\documentclass[12pt]{article}

\newcommand{\VersionInformation}{}  
\InputIfFileExists{Debug}{}{}

\usepackage{amsmath}
\usepackage{amsthm}
\usepackage{amsfonts}          
\usepackage{amssymb}           
\usepackage[mathscr]{eucal}
\usepackage{graphicx}
\usepackage{color}
\usepackage{hypbmsec}
\usepackage{fancyvrb}
\usepackage{sepnum}

\usepackage{multirow}
\usepackage{rotating}
\usepackage{slashbox}
\usepackage[isu,bf]{caption}
\newlength{\xtrawidth}
\newlength{\xtraheight}
\usepackage[all]{xy}           

\ifx\NoBackreferences\EMPTYMACRO
  \usepackage[linktocpage,bookmarks]{hyperref}
\else
  \usepackage[linktocpage,bookmarks]{hyperref}
\fi

\ifx\DEBUG\EMPTYMACRO
\else
  \usepackage{showlabels}
\fi

\def\clap#1{\hbox to 0pt{\hss#1\hss}}

\makeatletter
  \def\adots{\mathinner{\mkern2mu\raise\p@\hbox{.}
      \mkern2mu\raise4\p@\hbox{.}\mkern1mu
      \raise7\p@\vbox{\kern7\p@\hbox{.}}\mkern1mu}}
\makeatother

\newcommand{\comma}[1]{\ensuremath{\sepnum{{.}}{{,}}{}{#1}}}

\newcommand{\eqdef}{=}
\newcommand{\Q}{\ensuremath{{\mathbb{Q}}}}
\newcommand{\R}{\ensuremath{{\mathbb{R}}}}
\newcommand{\C}{\ensuremath{{\mathbb{C}}}}

\newcommand{\Z}{\mathbb{Z}}
\newcommand{\CP}{\ensuremath{\mathop{\null {\mathbb{P}}}}\nolimits}

\newcommand{\ibar}{\ensuremath{{\bar{\text{\it\i\/}}}}}
\newcommand{\jbar}{\ensuremath{{\bar{\text{\it\j\/}}}}}

\newcommand{\betabar}{\ensuremath{{\bar{\beta}}}}
\newcommand{\sbar}{\ensuremath{{\bar{s}}}}
\newcommand{\zbar}{\ensuremath{{\bar{z}}}}
\newcommand{\partialbar}{\ensuremath{\overline{\partial}}}

\newcommand{\FS}{\ensuremath{{\text{FS}}}}

\DeclareMathOperator{\diff}{d\!}
\DeclareMathOperator{\Span}{span}

\DeclareMathOperator{\Mat}{Mat}

\DeclareMathOperator{\Aut}{Aut}
\DeclareMathOperator{\AutBar}{\overline{Aut}}
\DeclareMathOperator{\Sym}{Sym}

\DeclareMathOperator{\Vol}{Vol}
\DeclareMathOperator{\dVol}{dVol}

\newcommand{\Rep}[1]{\ensuremath{\mathbf{\underline{#1}}}}
\newcommand{\barRep}[1]{\ensuremath{\overline{\Rep{#1}}}}

\newcommand{\Xt}{{\ensuremath{\widetilde{X}}}}

\newcommand{\ZZZ}{\ensuremath{{\Z_3\times\Z_3}}}
\newcommand{\Lsheaf}{\ensuremath{\mathscr{L}}}
\newcommand{\Osheaf}{\ensuremath{\mathscr{O}}}

\newcommand{\Fsheaf}{\ensuremath{\mathscr{F}}}

\newcommand{\Kahler}{K\"ahler}

\newcommand{\Qt}{{\ensuremath{\widetilde{Q}}}}
\newcommand{\QtF}{{\ensuremath{\widetilde{Q}}_F}}

\newcommand{\Pt}{{\ensuremath{\widetilde{P}}}}
\newcommand{\Rt}{{\ensuremath{\widetilde{R}}}}

\newcommand{\CY}{\text{CY}}

\newcommand{\Konly}{\ensuremath{{k}}}
\newcommand{\Nonly}{\ensuremath{{n}}}
\newcommand{\Kh}{\ensuremath{{k_h}}}
\newcommand{\Kphi}{\ensuremath{{k_\phi}}}
\newcommand{\Nh}{\ensuremath{{n_h}}}
\newcommand{\Nphi}{\ensuremath{{n_\phi}}}

\newcommand{\lambdahat}{\ensuremath{\hat{\lambda}}}

\newcommand{\CC}{C\nolinebreak\hspace{-.05em}\raisebox{.4ex}{\tiny\bf +}\nolinebreak\hspace{-.10em}\raisebox{.4ex}{\tiny\bf +}}

\newenvironment{descriptionlist}{%
\begin{list}%
{}%
{}}%
{\end{list}}

{\everymath{\displaystyle\everymath{}}\array}%
{\endarray}

\begin{document}
\begin{titlepage}
  \vspace*{-2cm}
  \VersionInformation
  \hfill
   \parbox[c]{5cm}{
     \begin{flushright}
       arXiv:0805.3689 [hep-th]
       \\
       UPR 1197-T
     \end{flushright}
   }
  \vspace*{\stretch1}
  \begin{center}
     \Huge 
     Eigenvalues and Eigenfunctions of the Scalar Laplace Operator
     on Calabi-Yau Manifolds
  \end{center}
  \vspace*{\stretch2}
  \begin{center}
    \begin{minipage}{\textwidth}
      \begin{center}
        \large Volker Braun${}^1$, 
        Tamaz Brelidze${}^1$,
        Michael R.~Douglas${}^2$, and
        \\
        Burt A.~Ovrut${}^1$
      \end{center}
    \end{minipage}
  \end{center}
  \vspace*{1mm}
  \begin{center}
    \begin{minipage}{\textwidth}
      \begin{center}
        ${}^1$ 
        Department of Physics,
        University of Pennsylvania,        
        \\
        209 S. 33rd Street, 
        Philadelphia, PA 19104--6395, USA
      \end{center}
      \begin{center}
        ${}^2$ 
        Rutgers University,
        Department of Physics and Astronomy,
        \\
        136 Frelinghuysen Rd.,
        Piscataway, NJ 08854--8019, USA
      \end{center}
    \end{minipage}
  \end{center}
  \vspace*{\stretch1}
  \begin{abstract}
    \normalsize 
    A numerical algorithm for explicitly computing the spectrum of the
    Laplace-Beltrami operator on Calabi-Yau threefolds is presented.
    The requisite Ricci-flat metrics are calculated using a method
    introduced in previous papers. To illustrate our algorithm, the
    eigenvalues and eigenfunctions of the Laplacian are computed
    numerically on two different quintic hypersurfaces, some
    $\Z_5\times\Z_5$ quotients of quintics, and the Calabi-Yau
    threefold with $Z_3\times\Z_3$ fundamental group of a heterotic
    standard model. The multiplicities of the eigenvalues are
    explained in detail in terms of the irreducible representations of
    the finite isometry groups of the threefolds.
  \end{abstract}
  \vspace*{\stretch5}
  \begin{minipage}{\textwidth}
    \underline{\hspace{5cm}}
    \centering
    \\
    Email: 
    \texttt{vbraun}, \texttt{brelidze}, \texttt{ovrut@physics.upenn.edu};
    \texttt{mrd@physics.rutgers.edu}.
  \end{minipage}
\end{titlepage}

\tableofcontents

\section{Introduction}
\label{sec:Intro}

A central problem of string theory is to find compactifications whose
low-energy effective action reproduces the standard model of
elementary particle physics.  One of the most promising candidates for
this task is the compactification of heterotic string theory on a
Calabi-Yau manifold~\cite{Candelas:1985en}. In particular, the
so-called ``non-standard embedding'' of $E_8\times E_8$ heterotic
strings has been a very fruitful approach to string
phenomenology~\cite{Lukas:1997fg,Donagi:1998xe, Lukas:1998hk,
  Lukas:1998yy, Lukas:1998ew, Donagi:1999gc, Donagi:1999jp}.

For a number of reasons, the most successful models of this type to
date are based on non-simply connected Calabi-Yau threefolds. These
manifolds admit discrete Wilson lines which, together with a non-flat
vector bundle, play an important role in breaking the heterotic $E_8$
gauge theory down to the standard model~\cite{Donagi:1999ez,
  Donagi:2000zs, Buchbinder:2002pr, Donagi:2000zf, Donagi:2000fw,
  Ovrut:2003zj, Donagi:2004ia, Donagi:2004qk}. In addition, they
project out many unwanted fields which would otherwise give rise to
exotic matter representations and/or additional replicas of standard
model fields. In particular, one can use this mechanism to solve the
doublet-triplet splitting problem~\cite{Donagi:2004ub,
  Donagi:2004su}. Finally, the non-simply connected threefolds have
many fewer moduli as compared to their simply connected covering
spaces~\cite{Braun:2005fk}. In recent work~\cite{Bouchard:2005ag,
  Braun:2005bw, Braun:2005nv, Braun:2005ux}, three generation models
with a variety of desirable features were introduced. These are based
on a certain quotients of a Schoen Calabi-Yau threefold, yielding a
non-simply connected Calabi-Yau manifold.

The ultimate goal is to compute all of the observable quantities of
particle physics, in particular gauge and Yukawa couplings, from the
microscopic physics of string theory~\cite{Candelas:1987rx,
  Braun:2006me, Braun:2005xp, Bouchard:2006dn}. There are many issues
which must be addressed to achieve this goal. Physical Yukawa
couplings, for example, depend on both coefficients in the
superpotential and the explicit form of the \Kahler{} potential. In a
very limited number of specific geometries~\cite{Candelas:1987rx,
  Candelas:1990rm, Greene:1993vm, Donagi:2006yf}, the former can be
computed using sophisticated methods of algebraic geometry,
topological string theory and the like. For the latter, one is usually
limited to the qualitative statement that a coefficient is ``expected
to be of order one''.  Improving our computational abilities and
extending these calculations to non-standard embedding has been an
outstanding problem~\cite{Candelas:1985en}.

Recently~\cite{DonaldsonNumerical, Douglas:2006hz}, a plan has been
outlined to analyze these problems numerically, at least in the
classical limit. The essential point is that, today, there are good
enough algorithms and fast enough computers to calculate Ricci-flat
metrics and to solve the hermitian Yang-Mills equation for the gauge
connection directly.  Given this data, one can then find the correctly
normalized zero modes of fields, determine the coefficients in the
superpotential and compute the explicit form of the \Kahler{}
potential. Some progress in this direction was made
in~\cite{DonaldsonNumerical, Douglas:2006hz, Douglas:2006rr,
  MR2161248, MR1064867} and also~\cite{Headrick:2005ch, Doran:2007zn,
  MR2154820}. Making effective use of symmetries~\cite{MR1255980,
  Braun:2007sn}, one can significantly improve the computational
procedure to find Calabi-Yau metrics and further extend it to
non-simply connected manifolds.  In this work, we take one step
further in the numerical approach to string theory compactification
and present an explicit algorithm to numerically solve for the
eigenvalues and eigenfunctions of the scalar Laplace operator. We use
as one of the inputs the Calabi-Yau metrics computed using the
techniques developed in~\cite{Braun:2007sn}.

We start, in \autoref{sec:CP3}, by discussing the general idea of the
method and list the key steps of our algorithm. This algorithm is then
applied to the simplest compact threefold, the projective space
$\CP^3$. This threefold is, of course, not a Calabi-Yau
manifold. However, in has the advantage of being one of the few
manifolds where the Laplace equation can be solved analytically. We
compare the numerical results of this computation with the analytical
solution in order to verify that our implementation is correct and to
understand the sources of numerical errors. We note that the
multiplicities of the approximate eigenvalues are determined by the
dimensions of corresponding irreducible representations of the
symmetry group of the projective space, as expected from the
analytical solution. We conclude the section by investigating the
asymptotic behavior of the numerical solution and comparing it with
Weyl's formula.

Having gone through this illustrative example, we apply our numerical
procedure to Calabi-Yau quintic threefolds in
\autoref{sec:genquintic}. The eigenvalues and eigenfunctions are
explicitly computed for both a quintic at a random point in moduli
space as well as for the Fermat quintic. We can again explain the
multiplicities of eigenvalues on the Fermat quintic as arising from
its enhanced symmetry; here, however, being a finite isometry
group. The asymptotics of the numerical solution is verified using
Weyl's formula. Note that the eigenvalues and eigenfunctions are not
known analytically in the Calabi-Yau case, so our numerical algorithm
is essential for their calculation.  Recently, Donaldson has proposed
a different algorithm to solve for the spectrum\footnote{The spectrum
  of an operator is the set of eigenvalues.} of the scalar
Laplacian. At the end of the section, we use it to numerically compute
the eigenvalues and eigenfunctions on a random quintic and on the
Fermat quintic and compare these to our results. In
\autoref{sec:Moduli}, we consider non-simply connected Calabi-Yau
manifolds, namely $\Z_5\times\Z_5$ quotients of certain quintic
threefolds. The eigenvalues and eigenfunctions of the Laplacian are
numerically computed using our algorithm, exploiting the Hironaka
decomposition discussed in our previous paper~\cite{Braun:2007sn}. In
this case, the multiplicities of the eigenvalues are determined by
finite ``pseudo-symmetries''~\cite{Green:1987mn}. We work out the
necessary representation theory and again find perfect agreement with
the multiplicities predicted by our numerical computation of the
eigenvalues. We conclude this section by studying the moduli
dependence of the eigenvalues for a one-parameter families of quintic
quotients.

In \autoref{sec:Z3Z3}, we apply this machinery to the case of a
certain $\Z_3\times\Z_3$ quotient of a Schoen
threefold~\cite{Braun:2004xv, Candelas:2007ac}. This is the Calabi-Yau
threefold underlying the heterotic standard model constructed
in~\cite{Braun:2005bw, Braun:2005nv, Braun:2005ux}. The essential new
feature is the existence of non-trivial \Kahler{} moduli, not just the
overall volume of the threefold as in all previous sections. As an
explicit example, we numerically compute the eigenvalues of the
Laplacian at two different points in the \Kahler{} moduli space,
corresponding to distinct ``angular'' directions in the \Kahler{}
cone. The group representation theory associated with the covering
space and the quotient is discussed.

We conclude in \autoref{sec:Laplace} by considering some physical
applications of the eigenvalues of the scalar Laplacian on a
Calabi-Yau threefold. In particular, we consider string
compactifications on these backgrounds and study the effect of the
massive Kaluza-Klein modes on the static gravitational potential in
four-dimensions. We compute this potential in the case of the Fermat
quintic, and explicitly show how the potential changes as the radial
distance approaches, and passes through, the compactification
scale. We then give a geometrical interpretation to the eigenvalue of
the first excited state in terms of the diameter of the Calabi-Yau
manifold. Inverting this relationship allows us to calculate the
``shape'' of the Calabi-Yau threefold from the numerical knowledge of
its first non-trivial eigenvalue.

Additional information is provided in three appendices. We explicitly
determine the first massive eigenvalue for the Laplacian $\CP^3$ in
\autoref{sec:CP3normalization}. Some technical aspects of semidirect
products, which are useful in understanding \autoref{sec:genquintic},
are discussed in \autoref{sec:semidirectproduct}. Finally, in
\autoref{sec:DonaldsonNotes}, we explain a modification of Donaldson's
algorithm for the numerical computation of Calabi-Yau metrics on
quotients, which is used \autoref{sec:Moduli}.

\section{Solving the Laplace Equation}
\label{sec:Algorithm}

Consider any $d$-dimensional, real manifold $X$. We will only be
interested in closed manifolds; that is, compact and without
boundary. Given a Riemannian metric\footnote{We denote the real
  coordinate indices by $\mu$, $\nu$, $\dots$.} $g_{\mu\nu}$ on $X$, the
Laplace-Beltrami operator $\Delta$ is defined as
\begin{equation}
  \label{eq:LaplaceBeltrami}
  \Delta 
  =
  - \frac{1}{\sqrt{g}} \partial_\mu (g^{\mu\nu} \sqrt{g} \partial_\nu)
  = 
  - \delta \diff = - * \diff * \diff
 \; ,
\end{equation}
where $g=\det{g_{\mu\nu}}$. Since this acts on functions, $\Delta$ is
also called the scalar Laplace operator. We will always consider the
functions to be complex-valued. Since $\Delta$ commutes with complex
conjugation, the scalar Laplacian acting on real functions would
essentially be the same.

An important question is to determine the corresponding eigenvalues
$\lambda$ and the eigenfunctions $\phi$ defined by
\begin{equation}
  \label{eq:DeltaEigen}
  \Delta \phi = \lambda \phi
  .
\end{equation}
As is well-known, the Laplace operator is hermitian. Due to the last
equality in eq.~\eqref{eq:LaplaceBeltrami}, all eigenvalues are real
and non-negative. The goal of this paper is to find the eigenvalues
and eigenfunctions of the scalar Laplace operator on specific
manifolds $X$ with metrics $g_{\mu\nu}$.

Since $X$ is compact, the eigenvalues of the Laplace operator will be
discrete. Let us specify the $n$-th eigenvalue by
$\lambda_n$. Symmetries of the underlying manifold will, in general,
cause $\lambda_n$ to be degenerate; that is, to have multiple
eigenfunctions. We denote by $\mu_n$ the multiplicity at level
$n$. Each eigenvalue depends on the total volume of the manifold. To
see this, consider a linear rescaling of distances; that is, let
$g_{\mu\nu} \mapsto \rho^2 g_{\mu\nu}$. Clearly,
\begin{equation}
  \Vol\big(\rho^2 g_{\mu\nu}\big) = 
  \rho^{d} \Vol\big(g_{\mu\nu}\big)
  , \quad
  \lambda_n\big(\rho^2 g_{\mu\nu}\big) = 
  \rho^{-2} \lambda_n\big(g_{\mu\nu}\big)
  .
  \label{bird1}
\end{equation}
Therefore, each eigenvalue scales as
\begin{equation}
  \label{eq:lambdaVol}
  \lambda_n \sim \Vol^{-\frac{2}{d}}
  .
\end{equation}
In the following, we will always normalize the volume to unity when
computing eigenvalues.

Now consider the linear space of complex-valued functions on $X$ and
define an inner product by
\begin{equation}
  \langle e|f\rangle
  =
  \int_X \bar{e} f \; \sqrt{g} \; \mathrm{d}^d x
  ,\qquad
  e,f\in C^{\infty}(X,\C).
  \label{eq:burt1}
\end{equation}
Let $\{f_a\}$ be an arbitrary basis of the space of complex
functions. For reasons to become clear later on, we will primarily be
working with bases that are not orthonormal with respect to the inner
product eq.~\eqref{eq:burt1}. Be that as it may, for any complex
function $e$ one can always find a function $\tilde{e}$ so that
\begin{equation}
  e=
  \sum_{a} f_{a} \langle f_{a}|\tilde{e}\rangle
  .
  \label{eq:burt2}
\end{equation}
Given the basis of functions $\{f_a\}$, the matrix elements
$\Delta_{ab}$ of the Laplace operator are
\begin{equation}
  \label{eq:Laplacematrix}
  \begin{split}
    \Delta_{a b}
    \;&=
    \big\langle f_a \big| \Delta \big| f_b \big\rangle
    =
    \int_X \bar{f}_a \Delta f_b \; \sqrt{g} \mathrm{d}^d x
    =
    - \int_X \bar{f}_a \diff * \diff f_b
    =
    \int_X \big\langle \diff f_a \big| \diff f_b \big\rangle
    \\
    \;&=
    \int_X g^{\mu\nu} 
    \big( \partial_\mu \bar{f}_a   \big)
    \big( \partial_\nu f_b \big)
    \; \sqrt{g} \mathrm{d}^d x
    .
  \end{split}
\end{equation}

Thus far, we have considered arbitrary $d$-dimensional, real manifolds
$X$ and any Riemannian metric $g_{\mu\nu}$. Henceforth, however, we
restrict our attention to even dimensional manifolds that admit a
complex structure preserved by the metric. That is, we will assume
that $X$ is a $D=\frac{d}{2}$-dimensional complex manifold with an
hermitian\footnote{In particular, \Kahler{} metrics are hermitian.}
metric\footnote{We denote the holomorphic and anti-holomorphic indices
  by $i$, $\ibar$, $j$, $\jbar$, $\dots$.} $g_{i\jbar}$ defined by
\begin{equation}
  g_{\mu\nu} \diff x^\mu  \otimes \diff x^\nu 
  =
  \frac{1}{2}
  g_{i\jbar} 
  \big(
  \diff z^i \otimes \diff z^\jbar + 
  \diff z^\jbar \otimes \diff z^i
  \big)
  .
\end{equation}
With $X$ so restricted, it follows that
\begin{equation}
  g^{\mu\nu} 
  \partial_\mu \bar{f}_a  
  \;
  \partial_\nu f_b 
  =
  2 g^{\ibar j} 
  \Big( 
  \partialbar_\ibar \bar{f}_a 
  \;
  \partial_j f_b
  +
  \partial_j \bar{f}_a 
  \;
  \partialbar_\ibar f_b
  \Big)
\end{equation}
and, hence,
\begin{equation}
  \Delta_{a b}
  =
  2 
  \int_X
  g^{\ibar j} 
  \Big( 
  \partialbar_\ibar \bar{f}_a 
  \;
  \partial_j f_b
  +
  \partial_j \bar{f}_a 
  \;
  \partialbar_\ibar f_b
  \Big)
  \det(g)
  \;
  \left( \tfrac{i}{2} \right)^D 
  \prod_{r=1}^D \diff z^r \wedge \diff\bar{z}^{\bar{r}}
  .
\end{equation}
Using this and eq.~\eqref{eq:burt2} for each eigenfunction
$\phi_{n,i}$, eq.~\eqref{eq:DeltaEigen} becomes
\begin{equation}
  \label{eq:generalizedeigen}
  \sum_{b}
  \big\langle f_a \big| \Delta \big| f_b \big\rangle
  \langle f_b|\tilde{\phi}_{n,i}\rangle
  =
  \sum_{b}\lambda_n\langle f_a|f_b\rangle
  \langle f_b|\tilde{\phi}_{n,i}\rangle
  ,
  \qquad
  i=1,\dots,\mu_n
  .
\end{equation}
Thus, in the basis $\{f_{a}\}$, solving the Laplace eigenvalue
equation is equivalent to the generalized eigenvalue problem for the 
infinite dimensional matrix $\Delta_{ab}$, where the matrix
$\langle f_a|f_b\rangle$ indicates the ``non-orthogonality'' of our
basis with respect to inner product eq.~\eqref{eq:burt1}.

In general, very little known about the exact eigenvalues and
eigenfunctions of the scalar Laplace operator on a closed Riemannian
manifold $X$, including those that are complex manifolds with
hermitian metrics. The universal exception are the zero modes, where
the multiplicity has a cohomological interpretation. Specifically, the
solutions to $\Delta \phi=0$ are precisely the locally constant
functions and, hence, the multiplicity of the zero eigenvalue is
\begin{equation}
  \mu_0(X) = h^0\big( X, \C \big) = \big| \pi_0(X) \big|
  ,
\end{equation}
the number of connected components of $X$. Furthermore, on symmetric
spaces $G/H$ one can completely determine the spectrum of the Laplace
operator in terms of the representation theory of the Lie groups $G$
and $H$. Indeed, in the next section we will discuss one such example
in detail. However, in general, and certainly for proper Calabi-Yau
threefolds, exact solutions of $\Delta\phi=\lambda\phi$ are unknown
and one must employ numerical methods to determine the eigenvalues and
eigenfunctions. The purpose of this paper is to present such a
numerical method, and to use it to determine the spectrum of $\Delta$
on physically relevant complex manifolds. Loosely speaking, the
algorithm is as follows.

First, we specify the complex manifold $X$ of interest as well as an
explicit hermitian metric. For \Kahler{} manifolds, the Fubini-Study
metric can always be constructed. However, this metric is never
Ricci-flat. To calculate the Ricci-flat Calabi-Yau metric, one can use
the algorithm presented in~\cite{DonaldsonNumerical, Douglas:2006rr}
and extended in~\cite{Braun:2007sn}. This allows a numerical
computation of the Calabi-Yau metric to any desired accuracy.  Giving
the explicit metric completely determines the Laplace operator
$\Delta$. Having done that, we specify a countably infinite set
$\{f_{a}\}$ that spans the space of complex functions. One can now
calculate any matrix element $\Delta_{ab}=\langle
f_a|\Delta|f_b\rangle$ and coefficient $\langle f_a| f_b \rangle$
using the scalar product specified in eq.~\eqref{eq:burt1} and
evaluated using numerical integration over $X$. As mentioned above,
the most convenient basis of functions $\{f_{a}\}$ will not be
orthonormal.  Clearly, calculating the infinite dimensional matrices
$\Delta_{ab}$ and $\langle f_a| f_b \rangle$, let alone solving for
the infinite number of eigenvalues and eigenfunctions, is not
possible. Instead, we greatly simplify the problem by choosing a
finite subset of slowly-varying functions as an approximate basis. For
simplicity of notation, let us take $\{f_a | a=1,\dots,k\}$ to be our
approximating basis. The $k\times k$ matrices $(\Delta_{ab})_{1\leq
  a,b \leq k}$ and $\langle f_a| f_b \rangle_{1\leq a,b \leq k}$ are
then finite dimensional and one can numerically solve
eq.~\eqref{eq:generalizedeigen} for the approximate eigenvalues and
eigenfunctions. It is important to note that this procedure
generically violates any underlying symmetries of the manifold and,
hence, each eigenvalue will be non-degenerate. Finally, we
successively improve the accuracy of the approximation in two ways: 1)
for fixed $k$ the numerical integration of the matrix elements is
improved by summing over more points and 2) we increase the dimension
$k$ of the truncated space of functions. In the limit where both the
numerical integration becomes exact and where $k \rightarrow \infty$,
the approximate eigenvalues $\lambda_n$ and eigenfunctions $\phi_n$
converge to the exact eigenvalues $\lambdahat_m$ and eigenfunctions
$\phi_{m,i}$ with multiplicity $\mu_m$. Inspired by our work on
Calabi-Yau threefolds, this algorithm to compute the spectrum of the
Laplacian was recently applied to elliptic curves
in~\cite{IuliuLazaroiu:2008pk}.


\section{\texorpdfstring{The Spectrum of  $\mathbf{\Delta}$ on $\mathbf{\CP^3}$}{The
    Spectrum of Delta on P3}}
\label{sec:CP3}

In this section, we use our numerical method to compute the
eigenvalues and eigenfunctions of $\Delta$ on the complex projective
threefold
\begin{equation}
  \label{eq:CP3def}
  \CP^3 = 
  S^7 \big/ U(1) =
  SU(4) \Big/ S\big( U(3) \times U(1) \big)  
\end{equation}
with a \Kahler{} metric proportional to the Fubini-Study metric,
rescaled so that the total volume is unity. As mentioned above,
since this is a symmetric space of the form $G/H$, the equation
$\Delta\phi=\lambda\phi$ can be solved analytically. The results were
presented in~\cite{MR510492}. Therefore, although $\CP^3$ is not a
phenomenologically realistic string vacuum, it is an instructive first
example since we can check our numerical algorithm against the exact
eigenvalues and eigenfunctions. Note that, in this case, the metric is
known analytically and does not need to be determined numerically.

\subsection{Analytic Results}
\label{sec:CP3analytic}

Let us begin by reviewing the known analytic
results~\cite{MR510492}. First, recall the Fubini-Study metric is
given by $g^\FS_{i\jbar}=\partial_i\bar\partial_\jbar K_\FS$ with
\begin{equation}
  K_\FS(z,\bar{z})
  =
  \frac{1}{\pi} 
  \ln
  \Big(|z_{0}|^{2}+|z_{1}|^{2}+|z_{2}|^{2}+|z_{3}|^{2} \Big)
  .
  \label{eq:K_FS}
\end{equation}
With respect to this metric the volume of $\CP^3$ is
\begin{equation}
  \Vol_\FS( \CP^3 ) 
  = 
  \int_{\CP^3} \det\big(g_{i\jbar}\big) \mathrm{d}^6 x
  =
  \int_{\CP^3} \frac{\omega_\FS^3}{3!} = \frac{1}{6}
  ,
  \label{eq:K_FSvol}
\end{equation}
where $\omega_\FS$ is the associated \Kahler{} $(1,1)$-form. However,
as discussed above, we find it convenient to choose the metric so as
to give $ \CP^3$ unit volume. It follows from eq.~\eqref{eq:K_FS}
and~\eqref{eq:K_FSvol} that one must rescale the \Kahler{} potential
to be
\begin{equation}
  K(z,\zbar) 
  = 
  {\sqrt[3]{6}} K_\FS(z,\zbar)
  = 
  \frac{\sqrt[3]{6}}{\pi} 
  \ln\Big( |z_0|^2 + |z_1|^2 + |z_2|^2 + |z_3|^2 \Big)
  .
  \label{eq:CP3_K}
\end{equation}
Then 
\begin{equation}
\Vol_{K}(\CP^3) = 1 \, , 
\label{do1}
\end{equation}
as desired. 

The complete set of eigenvalues of $\Delta$ on $\CP^3$ were found to
be~\cite{MR510492}
\begin{equation}
  \label{eq:CP3_lambda}
  \lambdahat_m 
  =
  \frac{4 \pi}{\sqrt[3]{6}} m (m+3)
  ,\quad 
  m=0,1,2,\dots
  ,
\end{equation}
where we determine the numerical coefficient, corresponding to our
volume normalization, in \autoref{sec:CP3normalization}. Furthermore,
it was shown in~\cite{MR510492} that the multiplicity of the $m$-th
eigenvalue is
\begin{equation}
  \label{eq:CP3_mu}
  \mu_m 
  =
  \binom{m+3}{m}^2 - \binom{m+2}{m-1}^2
  =
  \frac{1}{12} (m + 1)^2 (m + 2)^2 (2 m + 3)
  .
\end{equation}
This result for the multiplicity has a straightforward interpretation.
As is evident from the description of $\CP^3$ in
eq.~\eqref{eq:CP3def}, one can define an $SU(4)$ action on our
projective space. Thus the eigenstates of the Laplace operator
eq.~\eqref{eq:DeltaEigen} carry representations of $SU(4)$. In
general, any representation of $SU(4)$ is characterized by a three
dimensional weight lattice.  In particular, for each irreducible
representation there exists a highest weight
\begin{equation}
  w = m_1w_1+m_2w_2+m_3w_3
  ,
\end{equation}
where $w_1$, $w_2$, and $w_3$ are the fundamental weights and
$m_1,m_2,m_3\in\Z_{\geq 0}$. Starting with the highest weight, one can
generate all the states of the irreducible representation. It turns
out that multiplicity eq.~\eqref{eq:CP3_mu} is precisely the dimension
of the irreducible representation of $SU(4)$ generated by the highest
weight $m(w_1+w_3)=(m,0,m)$. Hence, the eigenspace associated with the
$m$-th eigenvalue $\lambdahat_m$ carries the irreducible representation
$(m,0,m)$ of $SU(4)$ for each non-negative integer $m$. For
convenience, we list the low-lying eigenvalues and their corresponding
multiplicities in \autoref{tab:CP3EV}.
\begin{table}[htbp]
  \renewcommand{\arraystretch}{1.3}
  \centering
  \begin{tabular}{c|cc}
    $m$ & $\mu_m$ & $\lambdahat_m$
    \\ \hline
    $0$ & $1$    & $0$ \\
    $1$ & $15$   & $\frac{16 \pi}{\sqrt[3]{6}} \simeq 27.662$ \\
    $2$ & $84$   & $\frac{40 \pi}{\sqrt[3]{6}} \simeq 69.155$ \\
    $3$ & $300$  & $\frac{72 \pi}{\sqrt[3]{6}} \simeq 124.48$ \\
    $4$ & $825$  & $\frac{112 \pi}{\sqrt[3]{6}} \simeq 193.64$ \\
    $5$ & $1911$ & $\frac{160 \pi}{\sqrt[3]{6}} \simeq 276.62$ \\
    $6$ & $3920$ & $\frac{216 \pi}{\sqrt[3]{6}} \simeq 373.44$ \\
    $7$ & $7344$ & $\frac{280 \pi}{\sqrt[3]{6}} \simeq 484.09$
  \end{tabular}
  \caption{Eigenvalues of $\Delta$ on $\CP^3$. 
    Each eigenvalue is listed with its multiplicity.}
  \label{tab:CP3EV}
\end{table}

The eigenfunctions of $\Delta$ on $\CP^3 = S^7\big/U(1)$ are the
$U(1)$-invariant spherical harmonics on $S^7$. In terms of homogeneous
coordinates $[z_0:z_1:z_2:z_3]$ on $\CP^3$, the eigenfunctions can be
realized as finite linear combinations of functions of the
form\footnote{We label the degree of the monomials here by $\Kphi$ to
  distinguish it from the degree $\Kh$ of polynomials in Donaldson's
  algorithm.}
\begin{equation}
  \frac{
    \Big(\text{degree $\Kphi$ monomial} \Big)
    \overline{\Big(\text{degree $\Kphi$ monomial} \Big)}
  }{
    \Big(|z_0|^2+|z_1|^2+|z_2|^2+|z_3|^2\Big)^\Kphi
  }
  .
  \label{eq:b}
\end{equation}
One can show this as follows. Let $\Rep{4}$ and $\barRep{4}$ be the
fundamental representations of $SU(4)$. Algebraically, one can show
that
\begin{equation}
  \Sym^\Kphi \Rep{4}
  \otimes 
  \Sym^\Kphi \barRep{4}
  = {\bigoplus}_{m=0}^\Kphi (m,0,m) 
  ,
  \label{eq:tonight1}
\end{equation}
where $(m,0,m)$ are the irreducible representations of $SU(4)$ defined
above. Now note that $\C[\vec{z}]_\Kphi$, the complex linear space
of degree-$\Kphi$ homogeneous polynomials in $z_0$, $z_1$, $z_2$,
$z_3$, naturally carries the $\Sym^\Kphi\Rep{4}$ reducible
representation of $SU(4)$. Similarly, $\C[\vec{\bar{z}}]_\Kphi$
carries the $\Sym^\Kphi \barRep{4}$ representation. Defining
\begin{equation}
 \Fsheaf_\Kphi 
 \eqdef 
 \frac{
   \C[z_0,z_1,z_2,z_3]_\Kphi
   \otimes
   \C[\zbar_0,\zbar_1,\zbar_2,\zbar_3]_\Kphi 
 }{ 
   \left( \sum_{j=0}^3 |z_j|^2 \right)^\Kphi 
 }  
\end{equation}
to be the space of functions spanned by the degree $\Kphi$ monomials,
then it follows from eq.~\eqref{eq:tonight1} that one must have the
decomposition
\begin{equation}
  \label{eq:harmfunc}
  \Fsheaf_\Kphi
  = 
  \bigoplus_{m=0}^\Kphi
  \Span\big\{ \phi_{m,1}, \dots, \phi_{m,\mu_m} \big\} \; ,
\end{equation}
where $\mu_m=\dim(m,0,m)$. Note the importance of the
$SU(4)$-invariant denominator, which ensures that the whole fraction
is of homogeneous degree zero, that is, a function on $\CP^3$.

To illustrate this decomposition, first consider the trivial case
where $\Kphi=0$. Noting that $\mu_0=1$, eq.~\eqref{eq:harmfunc} yields
\begin{equation}
  \phi_{0,1}=1
  ,
  \label{tonight2}
\end{equation}
corresponding to the trivial representation $\Rep{1}$ of $SU(4)$ and
the lowest eigenvalue $\lambda_0=0$. Now, let $\Kphi=1$. In this case
$\mu_0=1$ and $\mu_1=15$. It follows from eq.~\eqref{eq:harmfunc} that
there must exist a basis of $\Fsheaf_1$ composed of the eigenfunctions
of $\Delta$ in the $\Rep{1}$ and $\Rep{15}$ irreducible
representations of $SU(4)$ respectively. This is indeed the case. We
find that one such basis choice is
\begin{equation}
  \label{eq:CP3evec_0}
  \phi_{0,1} 
  = 
  \frac{
    |z_0|^2+|z_1|^2+|z_2|^2+|z_3|^2
  }{ \sum_{j=0}^3 |z_j|^2 }
  =
  1 
  ,
\end{equation}
corresponding to the lowest eigenvalue $\lambda_{0}=0$, and
\begin{equation}
  \label{eq:CP3evec_1}
  \begin{gathered}
    \begin{aligned}
      \phi_{1,1} =&~ 
      \textstyle
      z_0 \zbar_1 \Big/ \sum_{j=0}^3 |z_j|^2 
      & \qquad
      \phi_{1,2} =&~ 
      \textstyle
      z_1 \zbar_0 \Big/ \sum_{j=0}^3 |z_j|^2 
      \\
      \phi_{1,3} =&~ 
      \textstyle
      z_0 \zbar_2 \Big/ \sum_{j=0}^3 |z_j|^2 
      & \qquad
      \phi_{1,4} =&~ 
      \textstyle
      z_2 \zbar_0 \Big/ \sum_{j=0}^3 |z_j|^2 
      \\
      \phi_{1,5} =&~ 
      \textstyle
      z_0 \zbar_3 \Big/ \sum_{j=0}^3 |z_j|^2 
      & \qquad
      \phi_{1,6} =&~ 
      \textstyle
      z_3 \zbar_0 \Big/ \sum_{j=0}^3 |z_j|^2 
      \\
      \phi_{1,7} =&~ 
      \textstyle
      z_1 \zbar_2 \Big/ \sum_{j=0}^3 |z_j|^2 
      & \qquad
      \phi_{1,8} =&~ 
      \textstyle
      z_2 \zbar_1 \Big/ \sum_{j=0}^3 |z_j|^2 
      \\
      \phi_{1,9} =&~ 
      \textstyle
      z_1 \zbar_3 \Big/ \sum_{j=0}^3 |z_j|^2 
      & \qquad
      \phi_{1,10} =&~ 
      \textstyle
      z_3 \zbar_1 \Big/ \sum_{j=0}^3 |z_j|^2 
      \\
      \phi_{1,11} =&~ 
      \textstyle
      z_2 \zbar_3 \Big/ \sum_{j=0}^3 |z_j|^2 
      & \qquad
      \phi_{1,12} =&~ 
      \textstyle
      z_3 \zbar_2 \Big/ \sum_{j=0}^3 |z_j|^2 
    \end{aligned}
    \\
    \begin{aligned}
      \phi_{1,13} =&~ 
      \textstyle
      \Big( z_1 \zbar_1  - z_0\zbar_0 \Big) \Big/ \sum_{j=0}^3 |z_j|^2 
      \\
      \phi_{1,14} =&~ 
      \textstyle
      \Big( z_2 \zbar_2  - z_0\zbar_0 \Big) \Big/ \sum_{j=0}^3 |z_j|^2 
      \\
      \phi_{1,15} =&~ 
      \textstyle
      \Big( z_3 \zbar_3  - z_0\zbar_0 \Big) \Big/ \sum_{j=0}^3 |z_j|^2  \; ,
    \end{aligned}
  \end{gathered}
\end{equation}
corresponding to the first non-trivial eigenvalue $\lambda_1=\frac{16
  \pi}{\sqrt[3]{6}}$. Note that we recover the constant eigenfunction
for $\Kphi=0$ through the cancellation of the numerator in
eq.~\eqref{eq:CP3evec_0}. This pattern, where one recovers all the
lower eigenmodes through the factorization of the numerator in each
representation by an appropriate power of $\sum_{j=0}^3 |z_j|^2$,
continues for arbitrary $\Kphi$. In other words, there is a sequence
of inclusions
\begin{equation}
  \{1\} = \Fsheaf_0
  \subset
  \Fsheaf_1
  \subset 
  \Fsheaf_2
  \subset 
  \cdots
  \subset 
  C^\infty\big( \CP^3,\C \big)
  .
\end{equation}
Note that
\begin{equation}
  \dim \Fsheaf_\Kphi 
  = 
  \binom{\Kphi+3}{\Kphi}^2
  ,
\end{equation}
which, together with eq.~\eqref{eq:tonight1}, explains the
multiplicities given in eq.~\eqref{eq:CP3_mu}.

Although a basis of $\Fsheaf_\Kphi$ composed of eigenfunctions of
$\Delta$ would be the most natural, there is no need to go through the
exercise of decomposing the space into $SU(4)$-irreducible
representations. For numerical calculations, it is simpler to use the
equivalent basis
\begin{equation}
  \begin{split}
    \Fsheaf_\Kphi
    \;&=
    \Span
    \big\{ f_a \;\big|\; a= 0,\dots, \dim \Fsheaf_\Kphi -1 \big\}
    \\
    &= 
    \Span
    \left\{
      \Big(\text{degree $\Kphi$ monomial} \Big)
      \overline{\Big(\text{degree $\Kphi$ monomial} \Big)}
      \Bigg / \Big( \sum_{j=0}^3 |z_j|^2 \Big)^\Kphi
    \right\} 
  \end{split}
  \label{eq:CP3_f}
\end{equation}
for any finite value of $\Kphi$, even though these functions are
generically not themselves eigenfunctions of $\Delta$. In the limit
where $\Kphi \rightarrow \infty$, the basis eq.~\eqref{eq:CP3_f} spans the
complete space of eigenfunctions.

\subsection{Numerical Results}
\label{sec:CP3numeric}

Following the algorithm presented at the end of the
\autoref{sec:Algorithm}, we now numerically solve the eigenvalue
problem for the scalar Laplace operator $\Delta$ on $\CP^3$.  Unlike
more phenomenologically interesting Calabi-Yau threefolds, where one
must numerically compute the \Kahler{} metric using Donaldson's
method~\cite{DonaldsonNumerical, Douglas:2006rr, Braun:2007sn}, on
$\CP^3$ the \Kahler{} potential is given by eq.~\eqref{eq:CP3_K} and,
hence, the metric and $\Delta$ are known explicitly. This eliminates
the need for the first few steps of our algorithm, greatly simplifying
the calculations in this section. Furthermore, the $SU(4)$ action on
the eigenfunctions allows us to identify a complete basis for the
space of complex functions in terms of monomials of the form
eq.~\eqref{eq:b}. Since we know the exact eigenvalues and
eigenfunctions on $\CP^3$, this is an excellent venue for checking the
numerical accuracy of the remaining steps in our algorithm as well as
the correctness of our implementation.

Given the metric, $\Delta$ and the complete basis of
functions, the next step in our algorithm is to specify an
approximating basis for the linear space of complex functions. This is
easily accomplished by restricting to 
\begin{equation}
  \Fsheaf_\Kphi = 
  \Span
  \Big\{ 
  f_a \;\Big|\; 
  a= 0,\dots, \tbinom{\Kphi+3}{\Kphi}^2-1
  \Big\}
  ,
\end{equation}
see eq.~\eqref{eq:CP3_f}, for any finite value of $\Kphi$. Next, we
need to specify the volume measure in the integrals required to
evaluate the matrix elements $\langle f_a|\Delta|f_b\rangle$ and
$\langle f_a|f_b\rangle$. Each matrix element requires one 
integral over $\CP^3$, as in eq.~\eqref{eq:Laplacematrix}. The volume
form is completely determined by the metric to be
\begin{equation}
  \dVol_{K} = 
  \frac{1}{3!}\omega^3
,
\end{equation}
where $\omega$ is the \Kahler{} $(1,1)$-form given by the \Kahler{}
potential eq.~\eqref{eq:CP3_K}. Although $\CP^3$ is simple enough to
employ more elaborate techniques of integration, we will use the same
numerical integration algorithm as with Calabi-Yau threefolds later
on. That is, we approximate the integral by summing over $\Nphi$ random
points,
\begin{equation}
  \label{eq:numintegration}
  \frac{1}{\Nphi}
  \sum_{i=1}^{\Nphi} f(p_i) 
  \longrightarrow
  \int f \dVol
  ,
\end{equation}
where $f$ is an arbitrary function on $\CP^3$. The integration measure
$\dVol$ in eq.~\eqref{eq:numintegration} is determined by the
distribution of points. In other words, the random distribution of
points must be chosen carefully in order to approximate the integral
with our desired volume form $\dVol_{K}$. However, this can easily be
done: simply pick the points in an $SU(4)$-uniform
distribution. The corresponding integral measure is (up to overall
scale) the unique $SU(4)$-invariant volume form, the Fubini-Study
volume form. The normalization is fixed by our convention that
$\Vol_{K}(\CP^3)=1$.

The process of numerically evaluating integrals by summing over a
finite number $\Nphi$ of points has one straightforward
consequence. As discussed above, in the analytic solution the $m$-th
eigenvalue $\lambdahat_m$ is degenerate with multiplicity $\mu_m$
given in eq.~\eqref{eq:CP3_mu}. The reason for the degeneracy is that
the $m$-th eigenspace carries the $(m,0,m)$ highest weight
representation of $SU(4)$. However, even though the $\Nphi$ points
have an $SU(4)$-uniform distribution, the simple fact that they are
finite explicitly breaks the $SU(4)$ symmetry. The consequence of this
is that the degeneracy of each eigenvalue is completely broken. It
follows that in the numerical calculation, instead of one eigenvalue
$\lambdahat_m$ with multiplicity $\mu_m$, one will find $\mu_{m}$
non-degenerate eigenvalues $\lambda_n$. Only in the limit that $\Nphi
\rightarrow \infty$ will these converge to a single degenerate
eigenvalue as
\begin{equation}  
  \begin{array}{l@{~=~}l@{\quad\to\quad}l}
    \lambda_0 &
    \lambda_0,\dots,\lambda_{\mu_0-1} &
    \lambdahat_0=0
    ,
    \\
    \lambda_1,\dots,\lambda_{15} &
    \lambda_{\mu_0},\dots,\lambda_{\mu_0+\mu_1-1} &
    \lambdahat_1 = \frac{16 \pi}{\sqrt[3]{6}}
    ,
    \\
    \lambda_{16},\dots,\lambda_{99} &
    \lambda_{\mu_0+\mu_1},\dots,\lambda_{\mu_0+\mu_1+\mu_2-1} &
    \lambdahat_2 = \frac{40 \pi}{\sqrt[3]{6}}
    ,
    \\
    \multicolumn{3}{c}{\vdots}
  \end{array}
\end{equation}
\begin{figure}[htb]
  \centering
  \include{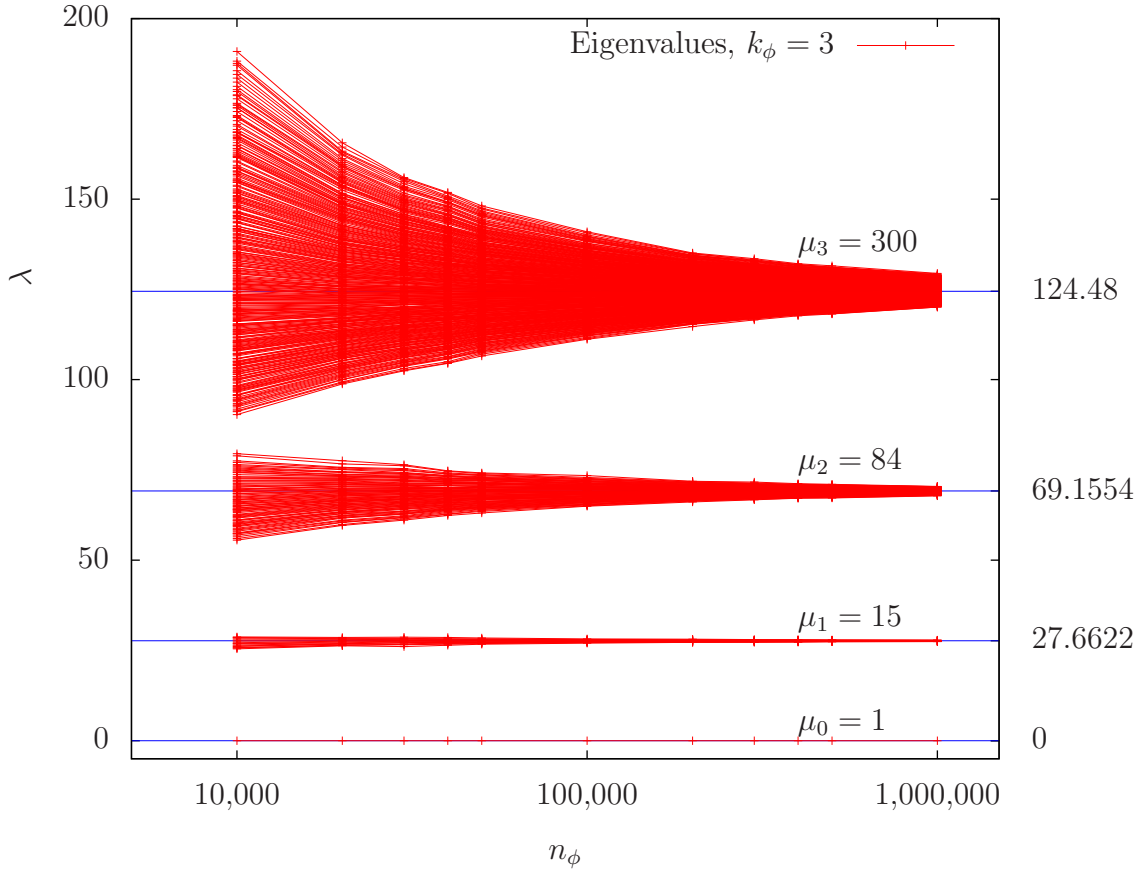}
  \caption{Spectrum of the scalar Laplacian on $\CP^3$ with the
    rescaled Fubini-Study metric. Here we fix the space of functions by
    choosing degree  $\Kphi=3$, and evaluate the Laplace operator at a varying
    number of points $\Nphi$.}
  \label{fig:SpecCP3Np}
\end{figure}

We are now ready to numerically compute the finite basis approximation
to the Laplace operator $\langle f_a|\Delta|f_b\rangle$ and the
coefficient matrix $\langle f_a|f_b\rangle$ for any fixed values of
$\Kphi$ and $\Nphi$. The coefficients do not form the unit matrix,
indicating that the approximating basis eq.~\eqref{eq:CP3_f} of $\Fsheaf_\Kphi$ is not  orthonormal. Even though one could orthonormalize the basis, this
would be numerically unsound and it is easier to directly solve the
generalized eigenvalue problem eq.~\eqref{eq:generalizedeigen}.  We
implemented this algorithm in \CC. In practice, the most
time-consuming part is the evaluation of the numerical integrals for
the matrix elements of the Laplace operator. We perform this step in
parallel on a $10$-node dual Opteron cluster, using
MPI~\cite{gabriel04:_open_mpi} for communication. Finally, we use
LAPACK~\cite{laug} to compute the eigenvalues and eigenvectors. Note
that the matrix eigenvectors are the coefficients $\langle f_a|
\tilde{\phi}\rangle$ and, hence, the corresponding eigenfunction is
\begin{equation}
  \phi=\sum_{a=0}^{\dim \Fsheaf_\Kphi-1} 
  f_a \langle f_a| \tilde{\phi}\rangle
  .
\end{equation}

We present our results in two ways. First fix $\Kphi$, thus
restricting the total number of non-degenerate eigenvalues $\lambda_n$
to $\dim \Fsheaf_\Kphi$. These eigenvalues are then plotted against the
number of points $\Nphi$ that we use to evaluate an integral. For
smaller values of $\Nphi$, the eigenvalues are fairly spread
out. However, as $\Nphi$ is increased the eigenvalues break into
distinct groups, each of which rapidly coalesces toward a unique
value. One can then compare the limiting value and multiplicity of
each group against the exact analytic result. We find perfect
agreement. To be concrete, let us present the numerical results for
the case $\Kphi=3$.  We plot these results in \autoref{fig:SpecCP3Np}.
As $\Nphi$ is increased from $\comma{10000}$ to $\comma{1000000}$, the
$\dim \Fsheaf_3 =400$ eigenvalues $\lambda_n$ cluster into $4$ distinct groups with
multiplicity $1$, $15$, $84$ and $300$. These clusters approach the
theoretical values of the first four eigenvalues respectively, as
expected. That is, the numerically calculated eigenvalues condense to
the analytic results for the eigenvalues and multiplicities listed in
\autoref{tab:CP3EV} on page~\pageref{tab:CP3EV}. At any $\Nphi$, the
eigenfunction $\phi_n$ associated with each $\lambda_n$ is evaluated
as a sum over the basis functions $\{ f_a | a=0,\dots,399 \}$. We do
not find it enlightening to present the numerical coefficients.
\begin{figure}[htb]
  \centering
  \include{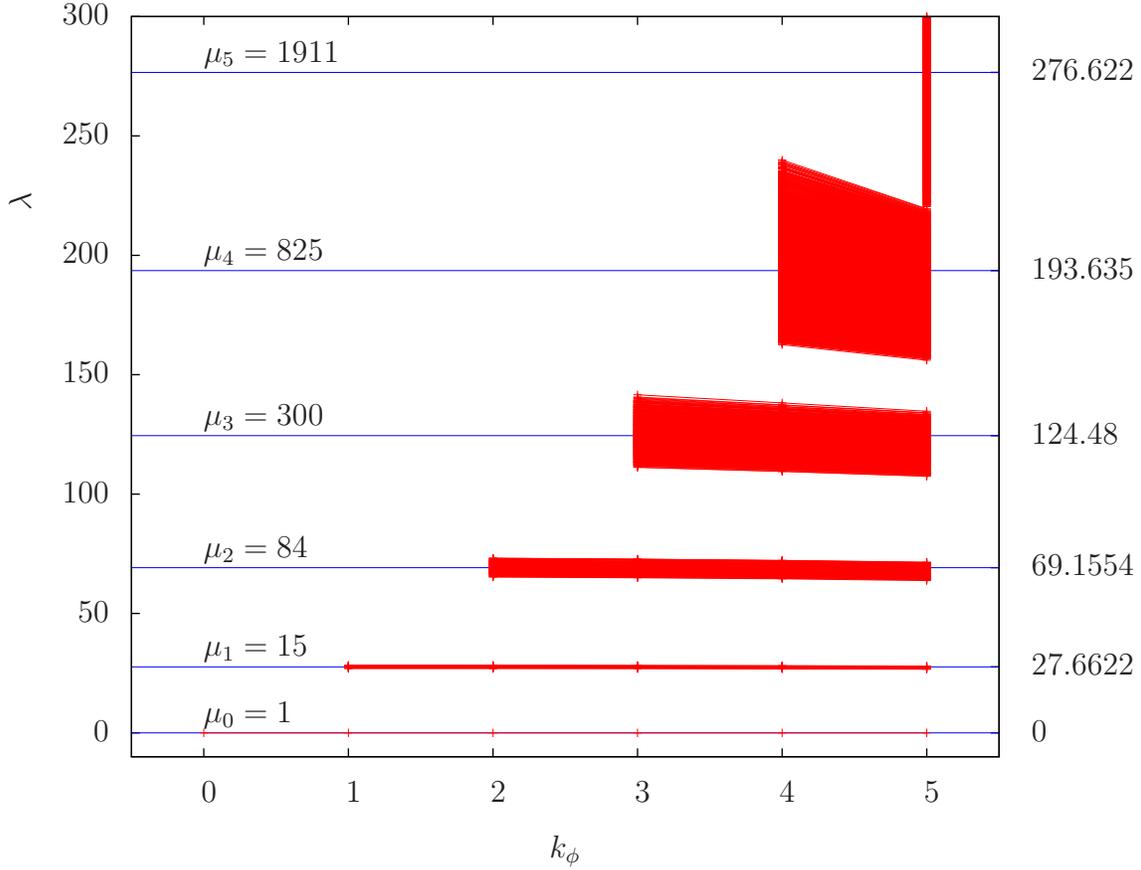}
  \caption{Spectrum of the scalar Laplacian on $\CP^3$ with the
    rescaled Fubini-Study metric. Here we evaluate the spectrum of the
    Laplace operator as a function of $\Kphi$, while keeping the
    number of points fixed at $\Nphi=\comma{100000}$.  Note that
    $\Kphi$ determines the dimension of the matrix approximation to
    the Laplace operator.}
  \label{fig:SpecCP3k}
\end{figure}

The second way to present our numerical results is to fix $\Nphi$ and
study the dependence of the eigenvalues on $\Kphi$. As was discussed
in \autoref{sec:CP3analytic}, since the eigenfunctions of the Laplace
operator are linear combinations of the elements of our basis, the
accuracy of $\lambda_n$ should not depend on $\Kphi$. However,
increasing $\Kphi$ does add higher-frequency functions to the
approximating space of functions. More explicitly, going from $\Kphi$ to
$\Konly_{\phi+1}$ will add an extra $\mu_{\Konly_{\phi+1}}$
eigenvalues to the numerical spectrum, corresponding to the dimension
of the $(\Konly_{\phi+1} ,0, \Konly_{\phi+1})$ irreducible
representation of $SU(4)$. This is exactly the behavior that we
observe in \autoref{fig:SpecCP3k}.

\subsection{Asymptotic Behaviour}
\label{sec:Weyl}

It is of interest to compare the asymptotic behaviour of the numerical solution
to the theoretical prediction of Weyl's formula, which determines
the asymptotic growth of the spectrum of the scalar Laplace operator.
Specifically, it asserts that on a Riemannian manifold $X$ of real
dimension $d$, the eigenvalues grow as $\lambda_n \sim
n^\frac{2}{d}$ for large $n$. Here it is important to keep track of multiplicities
by including the degenerate eigenvalue multiple times in the sequence
$\{\lambda_n\}$, as we do in our numerical calculations. The precise
statement of Weyl's formula is then that
\begin{equation}
  \label{eq:Weyl}
  \lim_{n\to \infty} 
  \frac{\lambda_n^{d/2}}{n}
  =
  \frac{
    (4 \pi)^\frac{d}{2} \Gamma\big(\tfrac{d}{2}+1\big)
  }{
    \Vol(X)
  }
  .
\end{equation}
Applying this to $\CP^3$, which has $d=6$ and the volume scaled to
$\Vol_{K}(\CP^3)=1$, we find that
\begin{equation}
  \label{eq:Weyl2}
  \lim_{n\to \infty} 
  \frac{\lambda_n^{3}}{n}
  = 384 \pi^3 
  .
\end{equation}
\begin{figure}[htbp]
  \centering
  \include{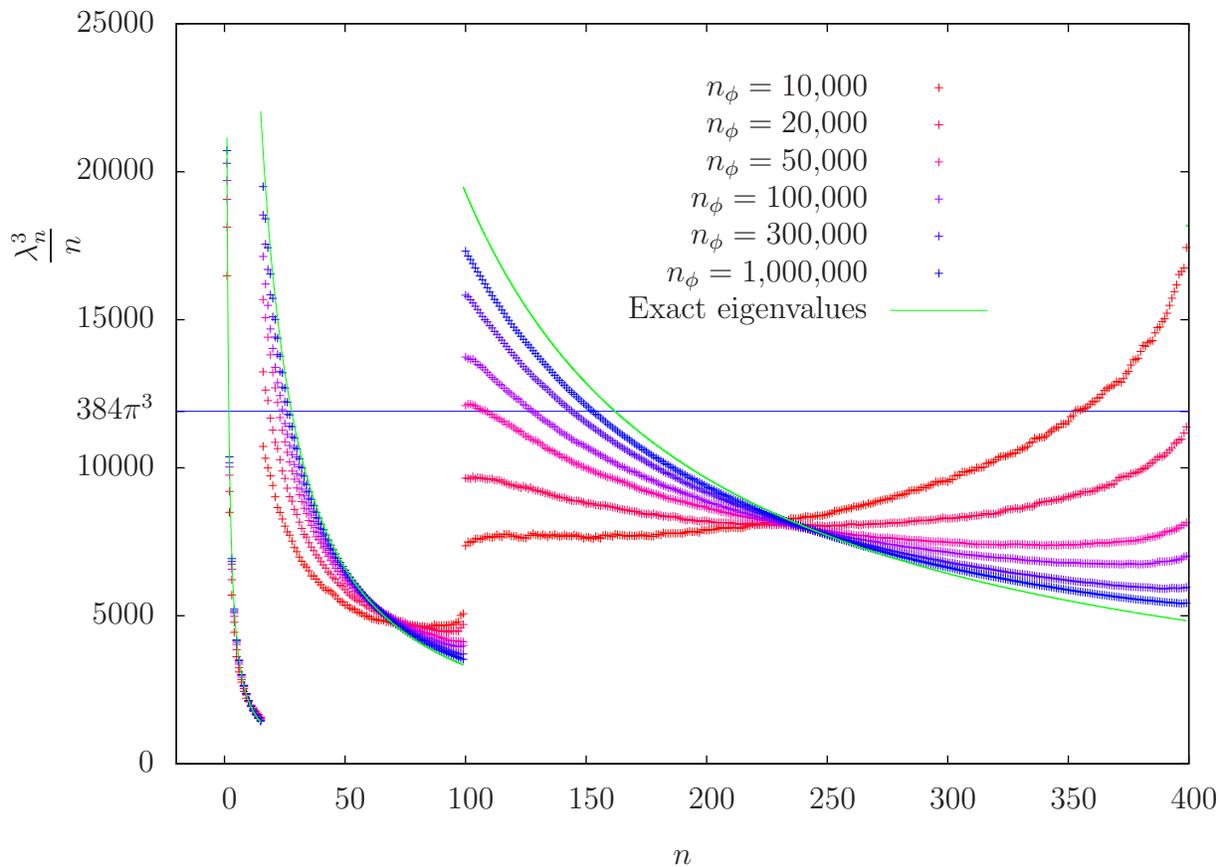}
  \caption{Check of Weyl's formula for the spectrum of the scalar
    Laplacian on $\CP^3$ with the rescaled Fubini-Study metric. We fix
    the space of functions by taking $\Kphi=3$ and evaluate
    $\frac{\lambda_n^3}{n}$ as a function of $n$ at a varying
    number of points $\Nphi$. Note that the data used for the eigenvalues is the same as for 
    $\Kphi=3$ in
    \autoref{fig:SpecCP3Np}.}
  \label{fig:SpecCP3Weyl}
\end{figure}
In \autoref{fig:SpecCP3Weyl} we choose $\Kphi=3$ and plot
$\frac{\lambda_n^3}{n}$ as a function of $n$ for the numerical values
of $\lambda_n$, as well as for the exact values listed in
\autoref{tab:CP3EV}. The numerical results are presented for six
different values of $\Nphi$. For each value of $\Nphi$, as well as for
the exact result, the $\frac{\lambda_n^3}{n}$ break into three groups,
corresponding to the first three massive levels with multiplicities
$15$, $84$, and $300$, respectively. Note that, as $\Nphi$ gets
larger, the numerical results converge to the exact result. That is,
each segment approaches a curve of the form
$\frac{\text{const.}}{n}$. Furthermore, as the number of eigenvalues
increase, the end-points of the curves asymptote toward the Weyl limit
$384 \pi^3$.

   
\section{ Quintic Calabi-Yau Threefolds}
\label{sec:genquintic}

Quintics are Calabi-Yau threefolds $\Qt \subset \CP^4$. Denote the
usual homogeneous coordinates on $\CP^4$ by
$z=[z_{0}:z_{1}:z_{2}:z_{3}:z_{4}]$. A hypersurface in $\CP^4$ is
Calabi-Yau if and only if it is the zero locus of a degree-5
homogeneous polynomial
\begin{equation}
  \tilde{Q}(z)=
     \sum_{n_0+n_1+n_2+n_3+n_4=5} 
     c_{(n_0,n_1,n_2,n_3,n_4)}
    z_0^{n_0}z_1^{n_1}z_2^{n_2}z_3^{n_3}z_4^{n_4}
    .
    \label{eq:ok1}
\end{equation}
By the usual abuse of notation, we denote both the defining polynomial
$\Qt(z)$ and the corresponding hypersurface $\{\Qt(z)=0 \} \subset
\CP^4$ by $\Qt$. There are $\binom{5+4-1}{4}=126$ degree-5 monomials,
leading to $126$ coefficients $c_{(n_0,n_1,n_2,n_3,n_4)} \in
\C$. These are not all independent complex structure parameters, since
the linear $GL(5,\C)$-action on the five homogeneous coordinates is
simply a choice of coordinates. Hence, the number of complex structure
moduli of a generic quintic $\Qt$ is $126-25=101$.

A natural choice of metric on $\CP^4$ is the Fubini-Study metric
$g_{i\jbar}=\partial_i \bar\partial_\jbar K_\FS$, where
\begin{equation}
  K_\FS=\frac{1}{\pi} \ln \sum_{i=0}^4 z_i \zbar_\ibar
  .
  \label{eq:cat1}
\end{equation}
This induces a metric on the hypersurface $\Qt$, whose \Kahler{}
potential is simply the restriction. Unfortunately, the restriction of
the Fubini-Study metric to the quintic is far from
Ricci-flat. Recently, however, Donaldson~\cite{DonaldsonNumerical}
presented an algorithm for numerically approximating Calabi-Yau
metrics to any desired accuracy. To do this in the quintic context,
one takes a suitable generalization, that is, one containing many more free
parameters, of the Fubini-Study metric. The parameters are then
numerically adjusted so as to approach the Calabi-Yau metric.

Explicitly, Donaldson's algorithm is the following. Pick a basis for
the quotient
\begin{equation}
  \label{eq:QuinticCoordRing}
  \C\left[ z_0, \dots, z_4 \right]_k
  \Big/
  \big\langle \Qt(z) \big\rangle
\end{equation} 
of the degree-$k$ polynomials on $\CP^4$ modulo the hypersurface
equation. Let us denote this basis by $s_\alpha$,
$\alpha=0,\dots,N(k)-1$ where
\begin{equation}
  \label{eq:N_k_equation}
  N(k) = 
  \begin{cases} 
    \binom{5+k-1}{k} 
    & 0 \leq k < 5 \\[1ex]
    \binom{5+k-1}{k} - \binom{k-1}{k-5} & k \geq 5 
    .
  \end{cases}
\end{equation}
For any given quintic polynomial $\Qt(z)$ and degree $k$, computing an
explicit polynomial basis $\{s_\alpha \}$ is straightforward. Now,
make the following ansatz
\begin{equation} 
  \label{eq:generalKahlerpotential}
  K_{h,k} 
  = 
  \frac{1}{k \pi} 
  \ln \sum_{\alpha ,\betabar=0}^{N(k)-1}
  h^{\alpha \betabar} s_\alpha \sbar_\betabar 
\end{equation}
for the \Kahler{} potential. The hermitian $N(k)\times N(k)$-matrix
$h^{\alpha \betabar}$ parametrizes the metric on $\Qt$ and is chosen to be the unique fixed point of the Donaldson T-operator
\begin{equation}
  \label{eq:T-operator}
  T(h)_{\alpha\betabar} =
  \frac{N(k)}{\Vol_\CY\big( \Qt \big)}
  \int_\Qt
  \frac{s_\alpha\sbar_{\bar\beta}}
  {\sum_{\gamma\bar\delta} h^{\gamma\bar\delta} s_\gamma \sbar_{\bar\delta}}
  \dVol_\CY
  ,
\end{equation}
where
\begin{equation}
  \dVol_\CY = \Omega \wedge \bar\Omega
\end{equation}
and $\Omega$ is the holomorphic volume form. The metric determined by
the fixed point of the T-operator is called ``balanced''. Hence, we
obtain for each integer $k \geq 1$ the balanced metric
\begin{equation}
  \label{eq:bmetric} 
  g_{i\jbar}^{(k)}
  =
  \frac{1}{k \pi} \partial_i \bar\partial_\jbar 
  \ln \sum_{\alpha ,\betabar=0}^{N(k)-1}
  h^{\alpha \betabar} s_\alpha \sbar_\betabar
  . 
\end{equation}
Note that they are formally defined on $\CP^4$ but restrict directly
to $\Qt$, by construction. One can show~\cite{MR1916953} that this
sequence
\begin{equation}
  g_{i\jbar}^{(k)} 
  \stackrel{
    k \to \infty
  }{
    \xrightarrow{\hspace{1cm}}
  }
  g^\CY_{i\jbar}
  \label{eq:cat2}
\end{equation}
of balanced metrics converges to the Calabi-Yau metric on $\Qt$.
 
It is important to have a measure of how closely the balanced metric
$g_{i\jbar}^{(k)}$ at a given value of $k$ approximates the exact
Calabi-Yau metric $ g^\CY_{i\jbar}$. One way to do this is the
following.  Let $g_{i\jbar}^{(k)}$ be a balanced metric, $\omega_k$
the associated $(1,1)$-form and denote by
\begin{equation}
  \Vol_K \big(\Qt,k\big)
  =
  \int_{\Qt}\frac{\omega_k^3}{3!} 
  , \qquad
  \Vol_\CY \big(\Qt\big)
  = 
  \int_\Qt \Omega\wedge\bar\Omega
  \label{eq:dog1}
\end{equation}
the volume of $\Qt$ evaluated with respect to $\omega_k$ and the
holomorphic volume form $\Omega$ respectively. Now note that the
integral
\begin{equation}
  \label{eq:sigmaQtDef}
  \sigma_k\big( \Qt \big) 
  = 
  \frac{1}{\Vol_\CY\big(\Qt\big)}
  \int_\Qt \left|
    1 - 
    \frac{
      \frac{\omega_k^3}{3!} \Big/ \Vol_K\big(\Qt,k\big)
    }{
      \Omega \wedge \bar\Omega \Big/ \Vol_\CY\big(\Qt\big)
    }
  \right| \dVol_\CY
\end{equation}
must vanish as $\omega_{k}$ approaches the Calabi-Yau \Kahler{}
form. That is
\begin{equation}
  \sigma_{k}  \stackrel{k \rightarrow \infty} {\longrightarrow} 0.
  \label{eq:dog2}
\end{equation}
Following~\cite{Douglas:2006rr}, we will use $\sigma_k$ as the error
measure for how far balanced metric $g_{i\jbar}^{(k)}$ is from being
Calabi-Yau. Finally, to implement our volume normalization we will
always scale the balanced metric so that
\begin{equation}
  \Vol_K\big(\Qt,k\big)=1
  \label{eq:VolKnorm}
\end{equation}
at each value of $k$.

\subsection{Non-Symmetric Quintic}
\label{sec:randomquintic}

In this subsection, we will pick random\footnote{To
  be precise, we pick uniformly distributed random numbers on the unit
  disk $\{z\in \C: |z|\leq 1\}$.} coefficients
$c_{(n_0,n_1,n_2,n_3,n_4)}$ for the $126$ different quintic monomials
in the $5$ homogeneous coordinates. An explicit example, which we use for 
the analysis in this section, is given by
\begin{multline}
  \Qt(z) =
  (-0.319235+0.709687 i) z_0^5
  +(-0.327948+0.811936 i) z_0^4 z_1
  \\
  +(0.242297+0.219818 i) z_0^4 z_2
  + \cdots +
  (-0.265416+0.122292 i) z_4^5
  .
  \label{eq:RandomQt}
\end{multline}
We refer to this as the ``random quintic''. Of course, any other
random choice of coefficients would lead to similar conclusions.  The
polynomial eq.~\eqref{eq:RandomQt} completely fixes the complex
structure. Furthermore, the single \Kahler{} modulus determines the
overall volume, which we set to unity.

Using Donaldson's algorithm~\cite{DonaldsonNumerical, Douglas:2006rr,
  Braun:2007sn} which we outlined above, one can compute an
approximation to the Calabi-Yau metric on the quintic defined by
eq.~\eqref{eq:RandomQt}. The accuracy of this approximation is
determined by
\begin{itemize}
\item The degree $k\in\Z_{\geq 0}$ of the homogeneous polynomials used
  in the ansatz eq.~\eqref{eq:generalKahlerpotential} for the
  \Kahler{} potential. To distinguish this degree from the one in the
  approximation to the Laplace operator, we denote them from now on by
  $\Kh$ and $\Kphi$, respectively. In this section, we will use
  \begin{equation}   
    \Kh = 8
    .
    \label{eq:Qt_Kh}
  \end{equation}
  Note that the choice of degree $\Kh$ determines the number of
  parameters
  \begin{equation}
    h^{\alpha\bar\beta}
    \in \Mat\big( N(\Kh)\times N(\Kh), \C \big)
  \end{equation}
  in the ansatz for the \Kahler{} potential,
  eq.~\eqref{eq:generalKahlerpotential}. This is why $\Kh$ is
  essentially limited by the available memory. We choose $\Kh=8$
  because it gives a good approximation to the Calabi-Yau metric, see
  below, without using a significant amount of computer memory
  ($\approx$ $7$ MiB).
\item The number of points used to numerically integrate within
  Donaldson's T-operator~\cite{Douglas:2006rr}. To distinguish this
  number from the number of points used to evaluate the Laplacian, we
  denote them by $\Nphi$ and $\Nh$ respectively. As argued
  in~\cite{Braun:2007sn}, to obtain a good approximation to the
  Ricci-flat metric one should choose $\Nh \gg N(\Kh)^2$, where $N(\Kh)$
  is the number of degree-$\Kh$ homogeneous monomials in the $5$
  homogeneous coordinates modulo the $\Qt(z)=0$ constraint, see
  eq.~\eqref{eq:N_k_equation}. In our computation, we will always take
  \begin{equation}
    \label{eq:Qt_Nh_formula}
    \Nh = 10 \cdot N(\Kh)^{2} + \comma{50000}.
  \end{equation}
  This rather arbitrary number is chosen for the following
  reasons. First, the leading term assures that $\Nh \gg N(\Kh)^2$ by an
  order of magnitude and, second, the addition of \comma{50000} points
  guarantees that the integrals are well-approximated even for small
  values of $\Kh$. It follows from eq.~\eqref{eq:N_k_equation} that for 
  $\Kh=8$ we will use
  \begin{equation}
    \Nh=\comma{2166000}
    \label{eq:Qt_Nh}   
  \end{equation}
  points in evaluating the T-operator.
\end{itemize}
Using the Donaldson algorithm with $\Kh$ and $\Nh$ given by
eqns.~\eqref{eq:Qt_Kh} and~\eqref{eq:Qt_Nh} respectively, one can now
compute a good approximation to the Calabi-Yau metric in a reasonable
amount of time\footnote{That is, within a few hours of ``wall''
  time.}. The expression for the metric itself is given as a sum over
monomials on $\Qt$ of degree $\Kh=8$ with numerically generated
complex coefficients. It is not enlightening to present it
here. However, it is useful to compute the error measure defined in
eq.~\eqref{eq:sigmaQtDef} for this metric. We find that
\begin{equation}
  \sigma_8\approx 5\times 10^{-2}
  ,
  \label{eq:parrot3}
\end{equation}
meaning that, on average, the approximate volume form
$\frac{\omega_8^3}{3!} $ and the exact Calabi-Yau volume form
$\Omega\wedge\bar\Omega$ agree to about $5\%$. Finally, having found
an approximation to the Ricci-flat metric, one can insert it into
eq.~\eqref{eq:LaplaceBeltrami} to determine the form of the scalar
Laplacian.

We can now compute the spectrum of the scalar Laplace operator as
discussed in the previous section. First, one must specify a
finite-dimensional approximation to the space of complex-valued
functions on $\Qt$. For any finite value of $\Kphi$, we choose
\begin{equation}
  \Fsheaf_\Kphi 
  =
  \Span
  \Bigg\{ 
    \frac{
      s_\alpha{\sbar}_{\bar\beta} 
    }{
      \big( \sum_{i=0}^4  |z_i|^2 \big)^\Kphi
    } 
  ~
  \Bigg|
  ~
  \alpha, \bar\beta =0,\dots,N(\Kphi)-1
  \Bigg\} 
  ,
  \label{eq:FkphiQt}
\end{equation}
where $\{s_\alpha|\alpha=0,\dots,N(\Kphi)-1\}$ are a basis
for the homogeneous polynomials modulo the hypersurface constraint
\begin{equation}
  \Span \{ s_{\alpha} \}=
  \C\left[ z_0, \dots, z_4 \right]_{\Kphi}
  \Big/
  \big\langle \Qt(z) \big\rangle 
  .
  \label{eq:parrot3b}
\end{equation} 
Such a basis was already determined during the Donaldson algorithm for
the metric, the only difference being that now the degree is $\Kphi$
instead of $\Kh$. The counting function $N(\Kphi)$ is given by
eq.~\eqref{eq:N_k_equation}. Clearly,
\begin{equation}
  \dim 
  \Fsheaf_\Kphi 
  = 
  N(\Kphi)^2
  .
  \label{eq:dimFk}
\end{equation}

Computing the matrix elements of the Laplace operator requires another
numerical integration which is completely independent of the one in
the T-operator. We denote the number of points in the matrix element
integration by $\Nphi$, as we did in the previous section. 
\begin{figure}[htbp]
  \centering
  \include{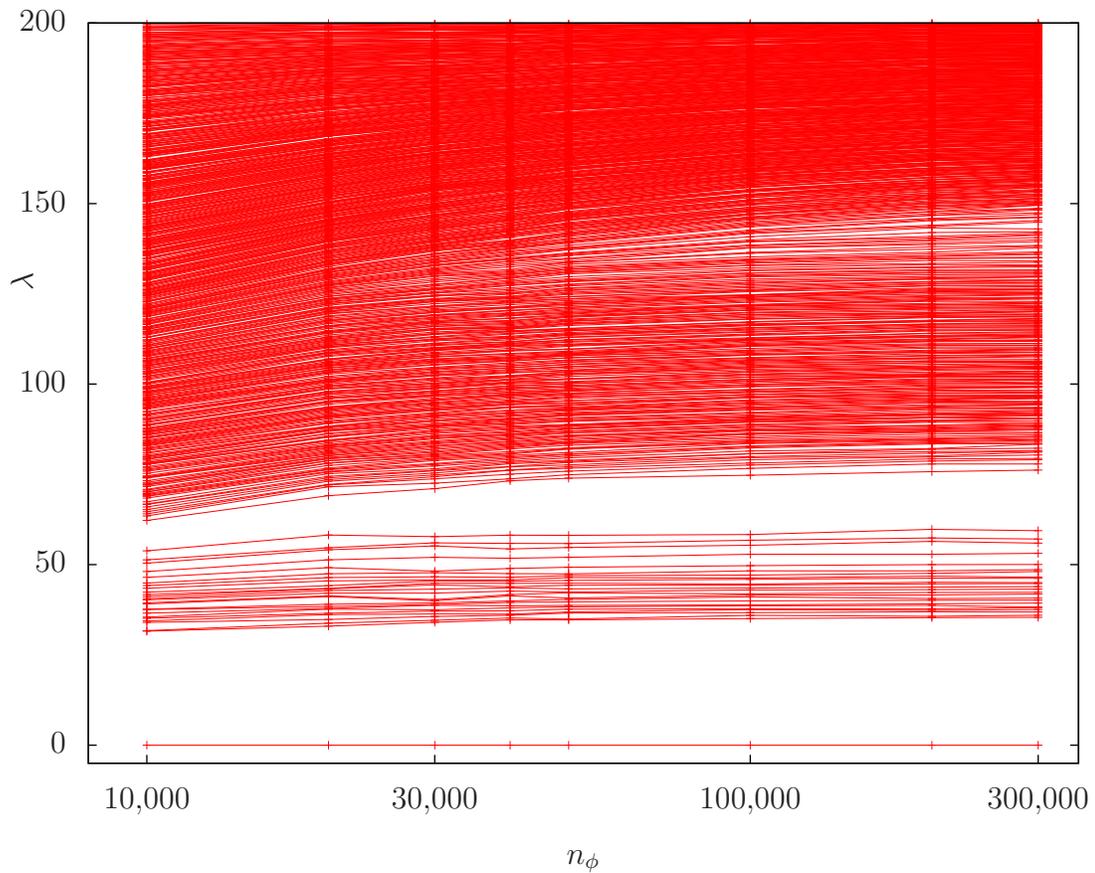}
  \caption{Eigenvalues of the scalar Laplace operator on the same
    ``random quintic'' defined in eq.~\eqref{eq:RandomQt}. The metric
    is computed at degree $\Kh=8$, using $\Nh=\comma{2166000}$
    points. The Laplace operator is evaluated at degree $\Kphi=3$ on a
    varying number $\Nphi$ of points.}
  \label{fig:SpecQtNp}
\end{figure}
We first present the resulting eigenvalue spectrum for fixed $\Kphi=3$
plotted against an increasing number of points $\Nphi$. Our results
are shown in \autoref{fig:SpecQtNp}. From eq.~\eqref{eq:N_k_equation}
we see that $N(3)= 35$ and, hence, there are $35^2=\comma{1225}$
non-degenerate eigenvalues $\lambda_0$, $\dots$,
$\lambda_{\comma{1224}}$. Note that for smaller values of $\Nphi$ the
eigenvalues are fairly spread out, and that they remain so as $\Nphi$
is increased. This reflects the fact that for any Calabi-Yau manifold
there is no continuous isometry, as there was for the
$\CP^{3}$. Furthermore, for the random quintic eq.~\eqref{eq:RandomQt}
there is no finite isometry group either.  Therefore, one expects each
eigenvalue to be non-degenerate, and our numerical results are clearly
consistent with this. At any $\Nphi$, the eigenfunctions $\phi_n$ are
a linear combination of the $\comma{1225}$ basis functions. We do not
find it enlightening to list the numerical coefficients explicitly.

Note that the accuracy of the numerical integration for the matrix
elements\footnote{Recall that $\Nh\to\infty$ is the continuum limit
  for the numerical integration in the T-operator, and
  $\Nphi\to\infty$ is the continuum limit for the numerical
  integration determining the matrix elements of the Laplace
  operator.} is not as crucial as in the T-operator, since we are
primarily interested in the low lying eigenvalues corresponding to
slowly-varying eigenfunctions. This is nicely illustrated by
\autoref{fig:SpecQtNp}, where the eigenvalues rather quickly approach
a constant value as we increase $\Nphi$, even though $\Nphi\ll
\Nh$. For this reason, $\Nphi=\comma{200000}$ gives a sufficiently good approximation 
and we will use this value for the reminder of this subsection.

\begin{figure}[htbp]
  \centering
  \include{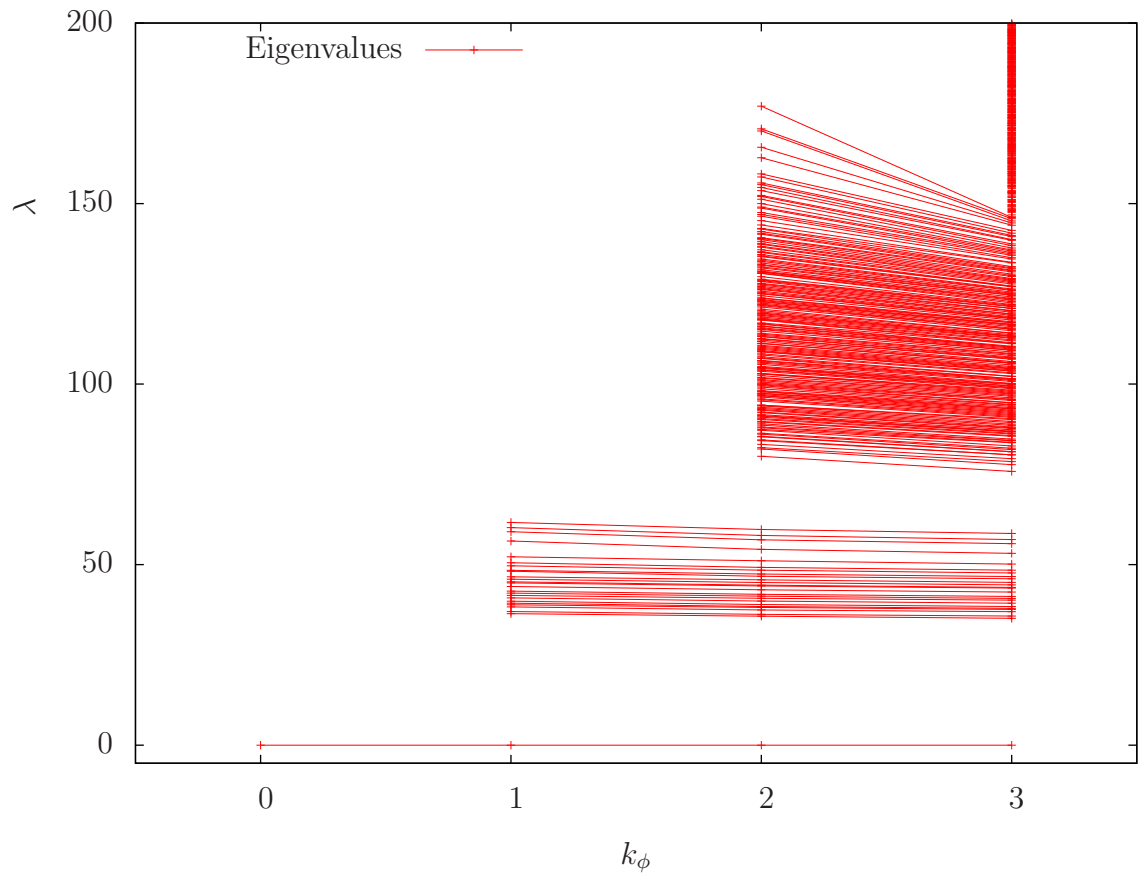}
  \caption{Eigenvalues of the scalar Laplace operator on a random
    quintic plotted against $\Kphi$. The metric is computed at degree $\Kh=8$, using
    $\Nh=\comma{2166000}$ points. The Laplace operator is then
    evaluated at $\Nphi=\comma{200000}$ points.}
  \label{fig:SpecQtk}
\end{figure}
A second way to present our numerical results is to fix $\Nphi$ and
study the dependence of the eigenvalues on $\Kphi$. This is presented
in \autoref{fig:SpecQtk}. We first note that the number of eigenvalues
indeed grows as $N(\Kphi)^2$, as it must. Second, as one expects, the
smaller eigenvalues do not change much as one increases $\Kphi$. The
higher eigenvalues, however, depend strongly on the truncation of the
space of functions, since their eigenfunctions vary quickly.

\begin{figure}[htbp]
  \centering
  \include{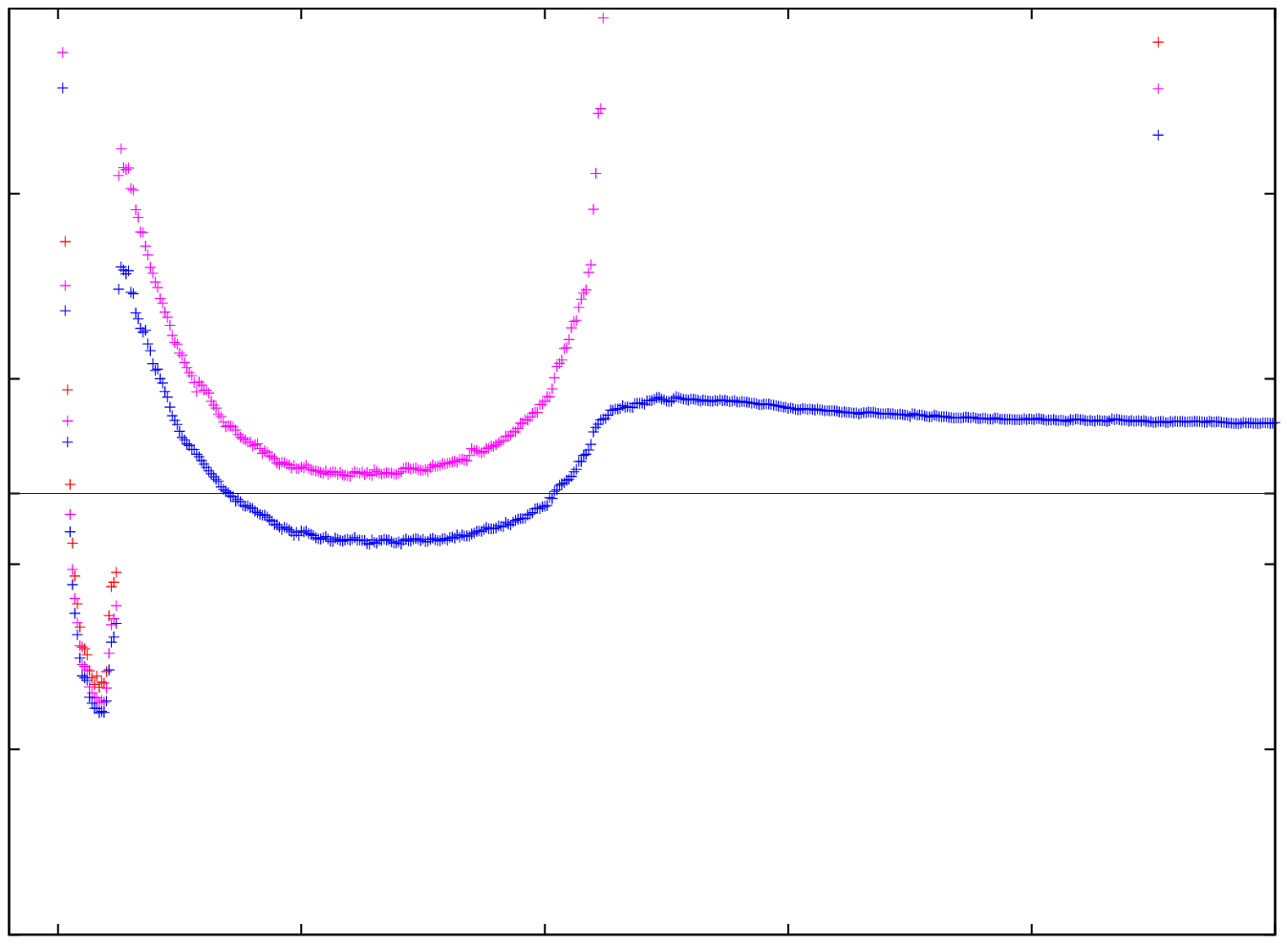}
  \caption{Check of Weyl's formula for the spectrum of the scalar
    Laplace operator on a random quintic. The metric is computed at
    degree $\Kh=8$, using $\Nh=\comma{2166000}$ points. The Laplace
    operator is evaluated at $\Nphi=\comma{200000}$ points and degrees
    $\Kphi=1,2,3$. Note that the data for the eigenvalues is the same as in
    \autoref{fig:SpecQtk}. According to
    Weyl's formula, the exact eigenvalues have to satisfy
    $\displaystyle \lim_{n\to \infty} \lambda_n^3/n = 384 \pi^3$.}
  \label{fig:SpecQWeyl}
\end{figure}
Finally, we plot $\lambda_n^3/n$ as a function of $n$ in
\autoref{fig:SpecQWeyl}. We see that this ratio does approach the
theoretical value of $384 \pi^3$ as $\Kphi$ and $n$ increase. This
confirms that the volume normalization in eq.~\eqref{eq:VolKnorm} is
being correctly implemented and that our numerical results are
consistent with Weyl's formula eq.~\eqref{eq:Weyl}.

\subsection{Fermat Quintic}
\label{sec:Fermat}

We repeat the analysis of the previous section for the Fermat
quintic defined by
\begin{equation}
  \label{eq:FermatQuintic}
  \QtF(z)=z_0^5+z_1^5+z_2^5+z_3^5+z_4^5
  .
\end{equation}
As before, the single \Kahler{} modulus is chosen so that the volume
of the Fermat quintic is unity. Now, however, we are at a 
different point in the complex structure moduli space,
eq.~\eqref{eq:FermatQuintic} instead of the random quintic
eq.~\eqref{eq:RandomQt}. Hence, we will perform the numerical
integrations now using points lying on a different hypersurface inside
$\CP^4$. Except for using different points, we compute the Calabi-Yau
metric on $\QtF$ using Donaldson's algorithm exactly as in the
previous subsection. In particular
\begin{itemize}
\item The degree $\Kh\in\Z_{\geq 0}$ of the homogeneous polynomials
  used in the ansatz eq.~\eqref{eq:generalKahlerpotential} for the
  \Kahler{} potential is chosen to be
  \begin{equation}   
    \Kh = 8
    .
    \label{eq:QtF_Kh}   
  \end{equation}
  This is the same degree as we used for the random quintic.
\item We take the number of points used to numerically integrate
  Donaldson's T-operator to be
 \begin{equation}
   \Nh=   10 \cdot N(8)^2 + \comma{50000} = 
   \comma{2166000}
   \label{eq:QtF_Nh}   
  \end{equation}
  This satisfies the condition that $\Nh \gg N(\Kh)^2$, ensuring that
  the numerical integration is sufficiently accurate.
\end{itemize}
Using $\Kh$ and $\Nh$ given by eqns.~\eqref{eq:QtF_Kh}
and~\eqref{eq:QtF_Nh} respectively, one can compute an approximation
to the Calabi-Yau metric using Donaldson's algorithm. The numerical
expression for the metric is tedious and will not be presented
here. The error measure eq.~\eqref{eq:sigmaQtDef} for this $\Kh=8$
balanced metric is
\begin{equation}
  \sigma_8\approx 5\times 10^{-2} 
  .
\end{equation}
Hence, the approximate volume form $\frac{\omega_8^3}{3!}$ and the
exact Calabi-Yau volume form $\Omega\wedge\bar\Omega$ agree to about
$5\%$. The metric determines the scalar Laplacian,
eq.~\eqref{eq:LaplaceBeltrami}. 

To determine the matrix elements of
the Laplace operator, one has to select an approximating basis for the
linear space of complex functions on $\QtF$,
eq.~\eqref{eq:FermatQuintic}. For any finite $\Kphi$, we again choose
the function space $\Fsheaf_\Kphi$ as in eqns.~\eqref{eq:FkphiQt}
and~\eqref{eq:parrot3b}. This basis was already determined during the
Donaldson algorithm for the metric. Computing the matrix elements of
the Laplace operator requires another numerical integration which is
completely independent of the one in the T-operator. As we did previously, we denote
the number of points in the matrix element integration by $\Nphi$.

\begin{figure}[htbp]
  \centering
  \include{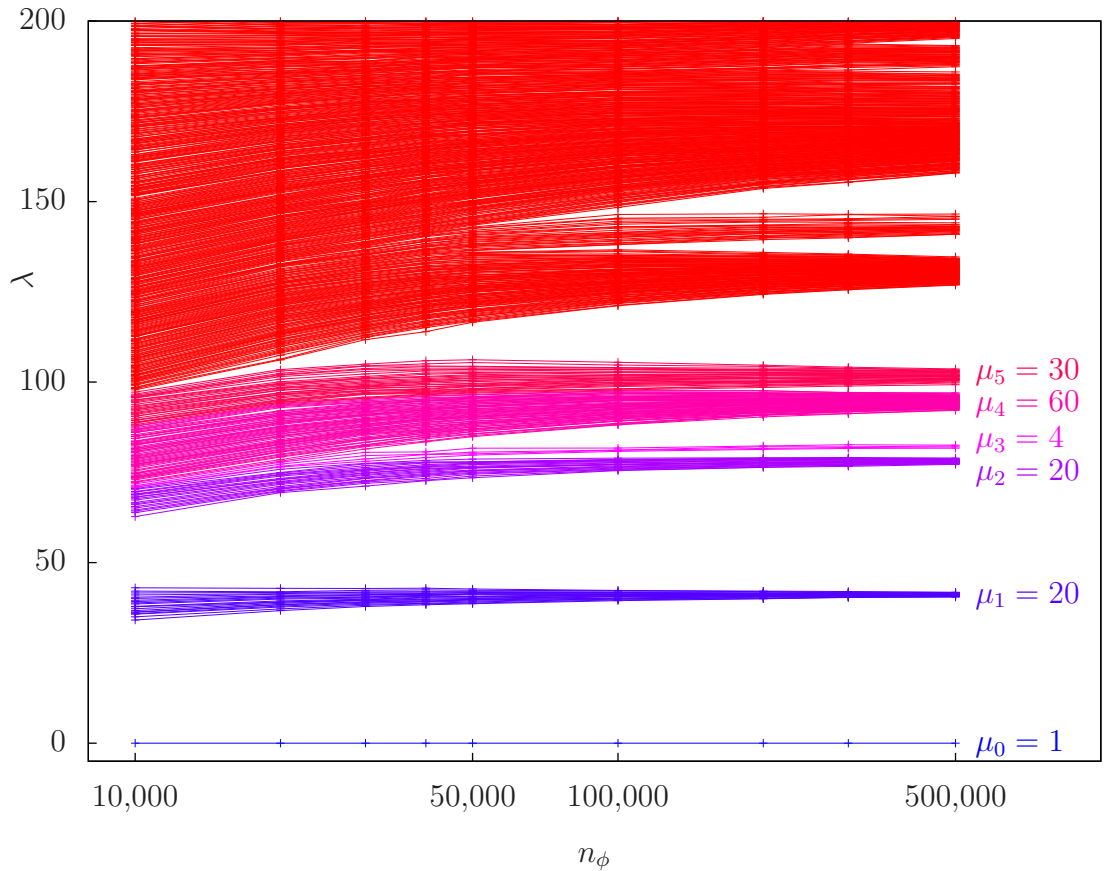}
  \caption{Eigenvalues of the scalar Laplace operator on the Fermat
    quintic. The metric is computed at degree $\Kh=8$, using
    $\Nh=\comma{2166000}$ points. The Laplace operator is evaluated at
    degree $\Kphi=3$ using a varying number $\Nphi$ of points.}
  \label{fig:SpecQtFNp}
\end{figure}
We first present the resulting eigenvalue spectrum for fixed $\Kphi=3$
plotted against an increasing number of points $\Nphi$. Our results
are shown in \autoref{fig:SpecQtFNp}. Note from eq.~\eqref{eq:dimFk}
that the total number of eigenvalues is given by
$\dim\Fsheaf_{3}=N(3)^2= \comma{1225}$. One immediately notices a
striking difference compared to the analogous graph for the random
quintic, \autoref{fig:SpecQtNp}. Here, the eigenvalues converge
towards degenerate levels. For smaller values of $\Nphi$, the
eigenvalues are fairly spread out. However, as $\Nphi$ is increased
the eigenvalues begin to condense into degenerate levels. Clearly,
this must be due to symmetries of the Fermat quintic.  As mentioned
above, no Calabi-Yau manifold has a continuous isometry. However,
unlike the random quintic, the Fermat quintic
eq.~\eqref{eq:FermatQuintic} does possess a finite isometry group,
which we will specify below in detail. Therefore, the exact
eigenvalues of $\Delta$ on $\QtF$ should be degenerate with
multiplicities given by the irreducible representations of this finite
group.  As we will see in \autoref{sec:fermatsymmetry}, the
numerically computed degeneracies of the eigenvalues exactly match the
irreducible representations of a this finite isometry group. Again, we
do not find it enlightening to present the numerical results for the
eigenfunctions. Moreover, as discussed previously, the accuracy of the
matrix element integration for low-lying eigenvalues need not be as
great as for the T-operator. As is evident from
\autoref{fig:SpecQtFNp}, a value of $\Nphi=\comma{500000}$ is already
highly accurate and we will use this value in the remainder of this
subsection.

\begin{figure}[htbp]
  \centering
  \include{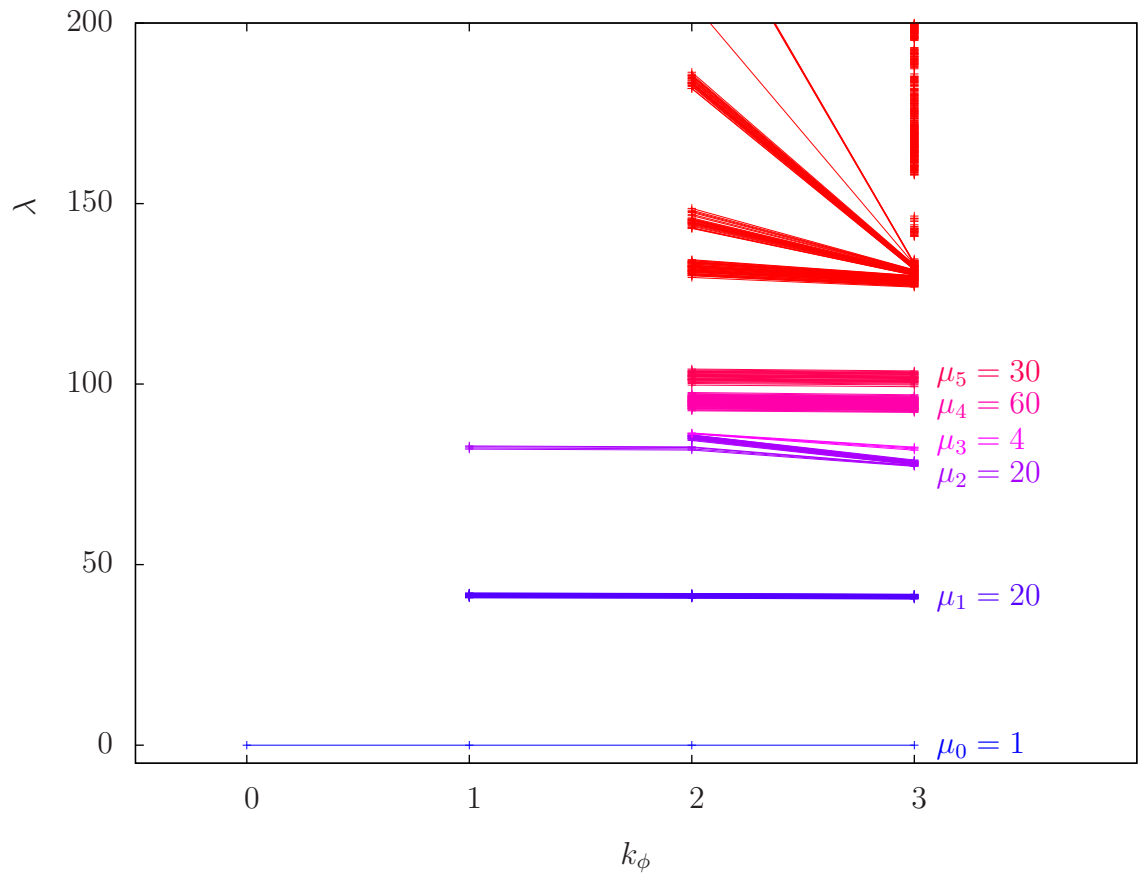}
  \caption{Eigenvalues of the scalar Laplace operator on the Fermat
    quintic. The metric is computed at degree $\Kh=8$, using
    $\Nh=\comma{2166000}$ points. The Laplace operator is evaluated at
    $\Nphi=\comma{500000}$ points with varying degrees $\Kphi$.}
  \label{fig:SpecQtFk}
\end{figure}
A second way to present our numerical results is to fix $\Nphi$ as in
the previous paragraph and study the dependence of the eigenvalues on
$\Kphi$. This is presented in \autoref{fig:SpecQtFk}. We first note
that the number of eigenvalues grows as $N(\Kphi)^2$, as it
must. Second, as one expects, the smaller eigenvalues do not change
much as one increases $\Kphi$, whereas the higher eigenvalues depend
strongly on the truncation of the space of functions. This is also to
be expected, since their eigenfunctions vary quickly.

\begin{figure}[htbp]
 \centering
 \include{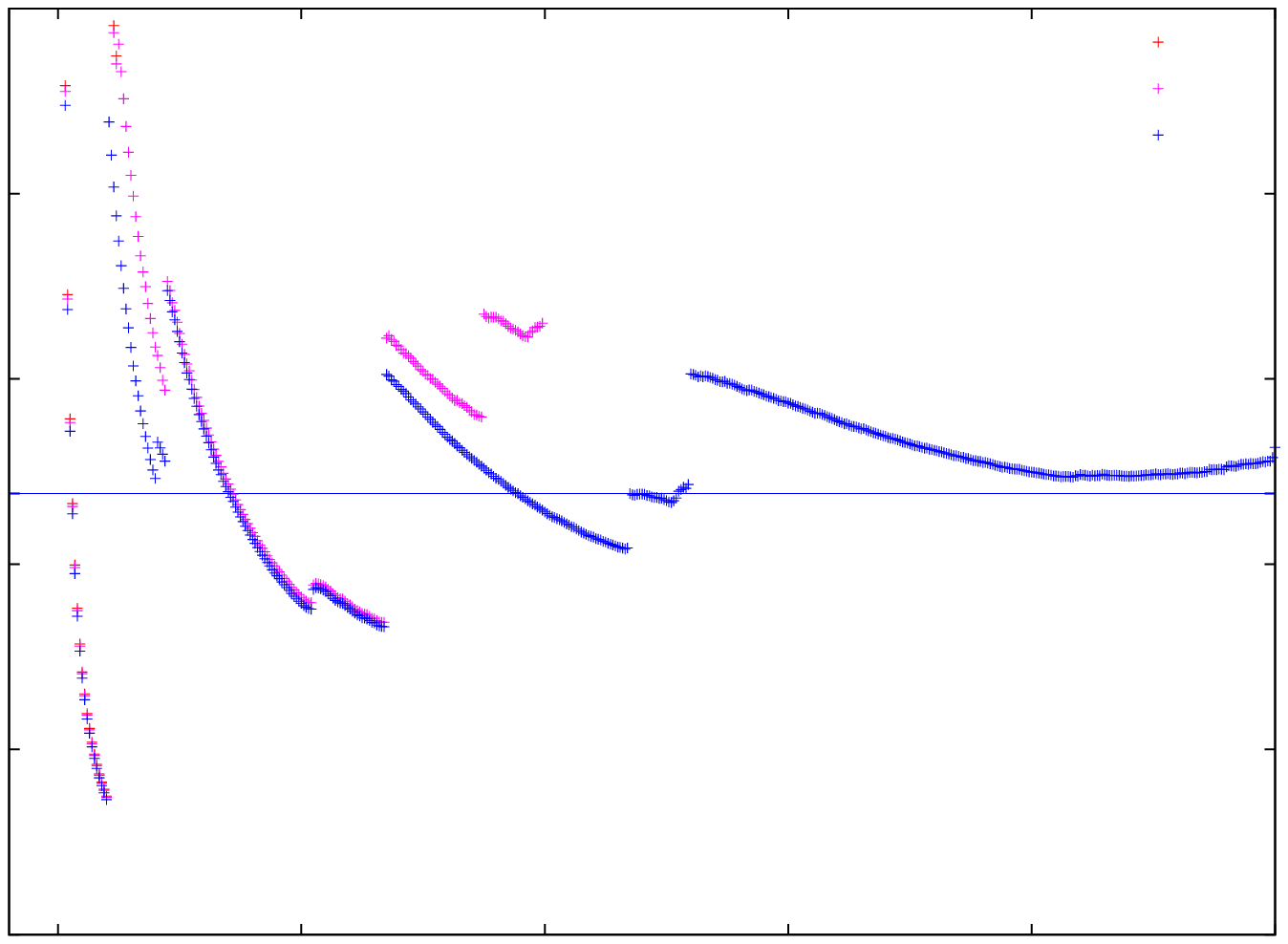}
 \caption{Check of Weyl's formula for the spectrum of the scalar
   Laplace operator on the Fermat quintic. The metric is computed at
   degree $\Kh=8$, using $\Nh=\comma{2166000}$ points.  The Laplace
   operator is evaluated at $\Nphi=\comma{500000}$ points and
   degrees $\Kphi=1,2,3$. Note that the data for the eigenvalues is the same as in
   \autoref{fig:SpecQtFk}. According to
   Weyl's formula, the exact eigenvalues have to satisfy
   $\displaystyle \lim_{n\to \infty} \lambda_n^3/n = 384 \pi^3$.}
 \label{fig:SpecQtFWeyl}
\end{figure}
Third, let us plot $\lambda_n^3/n$ as a function of $n$ in
\autoref{fig:SpecQtFWeyl}. This ratio approaches the theoretical value
of $384 \pi^3$ as $\Kphi$ and $n$ increase. This confirms that the
volume normalization in eq.~\eqref{eq:VolKnorm} is being correctly
implemented and that our numerical results are consistent with
Weyl's formula eq.~\eqref{eq:Weyl}.

\subsection{Symmetry Considerations}
\label{sec:fermatsymmetry}

Recall from \autoref{fig:SpecQtFNp} that the eigenvalues of the scalar
Laplace operator condense to a smaller number of degenerate levels as
$\Nphi\to\infty$, that is, in the limit where the numerical
integration becomes exact. The same phenomenon is clearly visible at
different values of $\Kphi$, see \autoref{fig:SpecQtFk}. Of course the
eigenvalues are never exactly degenerate due to numerical errors, but
counting the nearby eigenvalues allows one to determine the
multiplicities. Averaging over the eigenvalues in each cluster yields
an approximation to the associated degenerate eigenvalue. Using the
data from \autoref{fig:SpecQtFk}, we list the low-lying degenerate
eigenvalues and their multiplicities\footnote{Interestingly, the
  correct multiplicity $\mu_1=20$ was derived by a completely
 different  argument in~\cite{Kehagias:1999my}.} in
\autoref{tab:QtFresult}.
\begin{table}[htbp]
  \centering
  \renewcommand{\arraystretch}{1.3}
  \begin{tabular}{c|cccccc}
    $m$ & $0$ & $1$ & $2$ & $3$ & $4$ & $5$
    \\ \hline
    $\lambdahat_m$ &
    $1.18 \times 10^{-14}$ &
    $41.1 \pm 0.4$ &
    $78.1 \pm 0.5$ &
    $82.1 \pm 0.3$ &
    $94.5 \pm 1$ &
    $102  \pm 1$ 
    \\ \hline
    $\mu_m$ & $1$ &  $20$ & $20$ & $4$ & $60$ & $30$
  \end{tabular}
  \caption{The degenerate eigenvalues $\lambdahat_m$ 
    and their multiplicities $\mu_m$ 
    on the Fermat quintic, as computed from the numerical 
    values calculated with
    $\Kh=8$, $\Nh=\comma{2166000}$, 
    $\Kphi=3$, $\Nphi=\comma{500000}$. The errors are the standard
    deviation within the cluster of $\mu_n$ numerical eigenvalues.
  }
  \label{tab:QtFresult}
\end{table}
As discussed previously, multiplicities in the spectrum of the
Laplace-Beltrami operator results must follow from some symmetry. In
\autoref{sec:CP3}, we saw that the $SU(4)$ symmetry of $\CP^3$ leads to
degenerate eigenspaces of the scalar Laplacian. However, a proper
Calabi-Yau threefold never has continuous isometries, unlike
projective space. Nevertheless, a suitable non-Abelian\footnote{An
  Abelian symmetry group would only have one-dimensional
  representations and, hence, need not lead to degenerate
  eigenvalues. Note that any finite group has a finite number of
  irreducible representations and, therefore, one expects only a finite
  number of possible multiplicities for the eigenvalues of the Laplace
  operator. This is in contrast to the aforementioned $\CP^3$ case,
  where the multiplicities grow without bound.} {\it finite} group action is
possible and, in fact, explains the observed multiplicities, as we now
show.

First, note that for each distinct eigenvalue the corresponding space
of eigenfunctions must form a representation\footnote{An actual linear
  representation, \emph{not} just a representation up to phases
  (projective representation).} of the symmetry group. Clearly, the
degeneracies of the eigenvalues observed in \autoref{fig:SpecQtFNp}
and \autoref{fig:SpecQtFk} must arise from an isometry of $\QtF$.  In
fact, the Fermat quintic does have a large non-Abelian finite symmetry
group. To see this, note that the zero set of
eq.~\eqref{eq:FermatQuintic} is invariant under
\begin{itemize}
\item Multiplying a homogeneous coordinate by a fifth root of
  unity. However, not all $(\Z_5)^5$ phases act effectively because
  the projective coordinates are identified under the rescaling
  \begin{equation}
    \big[z_0:z_1:z_2:z_3:z_4\big]
    = 
    \big[
    \lambda z_0:\lambda z_1:\lambda z_2:\lambda z_3:\lambda z_4
    \big]
    .
  \end{equation}
  Only $(\Z_5)^5 \big/ \Z_5 \simeq (\Z_5)^4$ acts effectively.
\item Any permutation of the $5$ homogeneous coordinates. The
  symmetric group $S_5$ acts effectively.
\item Complex conjugation $\Z_2$.
\end{itemize}
The first two groups act by analytic maps, and together generate the
semidirect product
\begin{equation}
  \Aut\big( \QtF \big) = 
  S_5
  \ltimes
  \big(\Z_5\big)^4
\end{equation}
of order $75,000$. Our notation and the relevant group theory is discussed in \autoref{sec:semidirectproduct}. The full discrete symmetry group, including
the complex conjugation $\Z_2$, is
\begin{equation}
  \overline{\Aut}\big( \QtF \big) 
  = 
  \Z_2 
  \ltimes 
  \Aut\big( \QtF \big) 
  = 
  \big( S_5 \times \Z_2 \big)
  \ltimes
  \big(\Z_5\big)^4
  \label{ddd}
\end{equation}
and of order $\comma{150000}$. Note that even though the $\Z_2$ acts
as complex conjugation on the base space, the whole
$\overline{\Aut}(\QtF)$ acts linearly on the the basis of complex
functions on $\QtF$ and, hence, on the eigenfunctions. There are $80$
distinct irreducible representations occurring in $14$ different
dimensions, ranging from $1$ to $120$. We list them in
\autoref{tab:AutBarQF}.
\begin{table}
  \renewcommand{\arraystretch}{1.3}
  \centering
  \begin{tabular}{c|cccccccccccccc}
    $d$ &
    $1$ & $2$ & $4$ & 
    $5$ & $6$ & $8$ & 
    $10$ & $12$ & $20$ & 
    $30$ & $40$ & $60$ & 
    $80$ & $120$
    \\ \hline
    \parbox{24mm}{\centering\vspace{1ex} \# of irreps\\ in dim $d$}
    &
    $4$ & $4$ & $4$ & 
    $4$ & $2$ & $4$ &
    $4$ & $2$ & $8$ & 
    $8$ &$12$ & $18$ & 
    $4$ & $2$
  \end{tabular}
  \caption{Number of irreducible representations of 
    $\overline{\Aut}(\QtF) = \Z_2 \ltimes \Aut(\QtF)$ in each complex
    dimension.}
  \label{tab:AutBarQF}
\end{table}

We conclude by noting that the multiplicities listed in
\autoref{tab:QtFresult} also occur in \autoref{tab:AutBarQF}. That is,
the eigenspaces of the degenerate eigenvalues of the scalar Laplacian
on $\QtF$, computed using our numerical algorithm, indeed fall into
irreducible representations of the finite symmetry group $\big( S_5
\times \Z_2 \big) \ltimes \big(\Z_5\big)^4$, as they must. This gives
us further confidence that our numerical computation of the Laplacian
spectrum is reliable.

\subsection{Donaldson's Method}
\label{sec:Donaldson}

Donaldson~\cite{DonaldsonNumerical} conjectured a method to compute
the eigenvalues of the scalar Laplace operator that is completely
independent of our approach. His calculation of the spectrum of the
scalar Laplacian is very much tied into his algorithm for computing
balanced (Calabi-Yau) metrics. In our algorithm, on the other hand
side, any metric could be used and no particular simplifications arise
just because the metric happens to be balanced or Calabi-Yau. Because
they are so different, it is quite interesting to compare both
methods. We will now review his proposal, and then compare it with our
previous computation of the eigenvalues on the Fermat quintic as well
as the random quintic.

In this alternative approach to calculating the spectrum of the
Laplace-Beltrami operator, one first has to run through Donaldson's
algorithm for the metric. In particular, one had to choose a degree
$k$, fix a basis $\{s_\alpha|\alpha=0,\dots,N(k)-1\}$, and obtain the
balanced metric $h^{\alpha\bar\beta}$ as the fixed point of
Donaldson's T-operator. Let us write
\begin{equation}
  \big( s_\alpha, s_\beta ) = 
  \frac{
    s_\alpha\sbar_{\bar\beta}
  }{
    \sum_{\gamma\bar\delta} h^{\gamma\bar\delta} 
    s_\gamma
    \sbar_{\bar\delta}
  }
\end{equation}
for the integrand of the T-operator
eq.~\eqref{eq:T-operator}. Donaldson's alternative calculation of the
eigenvalues then hinges on the evaluation of the integral
\begin{equation}
  \label{eq:Qintegral}
  Q_{\alpha\bar\beta,\bar\gamma\delta}
  = 
  N(k)
  \int_X (s_\alpha,s_\beta) \overline{(s_\gamma,s_\delta)}
  \dVol_\CY
  ,
\end{equation}
where we again normalize $\Vol(X)=1$. One can think of $Q$ as a linear
operator on the space of functions\footnote{Note the similarity with
  the approximate space of functions $\Fsheaf_\Kphi$ used previously,
  eq.~\eqref{eq:FkphiQt}. When computing the matrix elements of the
  Laplace operator directly, the precise form of the denominator is
  not overly important as long as it has the correct homogeneous
  degree, and we always chose $(\sum|z_j|^2)^\Kphi$ for simplicity.}
\begin{equation}
  \Fsheaf^\text{D}_k
  = 
  \Span
  \Big\{ (s_\alpha,s_\beta) 
  ~\Big|~
  0 \leq \alpha, \bar\beta \leq N(k)-1
  \Big\}
  ,
\end{equation}
acting via
\begin{equation}
  Q:\Fsheaf^\text{D}_k\to\Fsheaf^\text{D}_k
  ,~
  (s_\alpha,s_\beta) \mapsto 
  \sum
  Q_{\alpha\bar\beta,\bar\gamma\delta}
  h^{\bar\gamma\sigma} h^{\bar\tau\delta}
  (s_\sigma, s_\tau )
  .
\end{equation}
In~\cite{DonaldsonNumerical}, Donaldson conjectures that
\begin{equation}
  \lim_{k\to\infty} Q 
  =
  e^{- \frac{\Delta}{4\pi \sqrt[3]{N(k)}}}
\end{equation}
as operators on
\begin{equation}
  \lim_{k\to\infty} \Fsheaf^\text{D}_k = C^\infty(X,\C)
  .
\end{equation}
For explicitness, let us look in more detail at the individual steps
as they apply to any quintic $X=\Qt\subset \CP^4$:
\begin{enumerate}
\item First, pick a degree $k$ and a basis
  $\big\{s_0,\dots,s_{N(k)-1}\big\}$ of degree-$k$ homogeneous
  polynomials modulo the hypersurface equation $\Qt=0$.
\item Compute the Calabi-Yau metric via Donaldson's algorithm. It is
  determined by the $N(k)\times N(k)$ hermitian matrix
  $h^{\alpha\betabar}$.
\item Compute the $N(k)^4$ scalar integrals in
  eq.~\eqref{eq:Qintegral}.  The numerical integration can be
  performed just as in Donaldson's T-operator, see
  \autoref{sec:genquintic}.
\item Compute the $N(k)^2\times N(k)^2$ matrix
  \begin{equation}
    Q{}_{N(k) \alpha + \bar\beta}{}^{N(k)\gamma+\bar\delta} 
    = 
    \sum_{\bar\sigma, \tau = 0}^{N(\Kh)-1}
    Q_{\alpha\bar\beta,\bar\sigma\tau}
    h^{\gamma\bar\sigma} h^{\tau\bar\delta}
  \end{equation}
  and find its eigenvalues $\Lambda_n$. Note that $Q_i^j$ is not
  hermitian\footnote{$Q_i^j$ is, however, conjugate to a hermitian
    matrix and hence has real eigenvalues.} and one should use the
  Schur factorization\footnote{Instead of the dqds algorithm we use
    for computing eigenvalues of hermitian matrices.} to compute
  eigenvalues.
\item Discard all $\Lambda_n\leq 0$, these correspond to high
  eigenvalues of the Laplacian that are not approximated well at the
  chosen degree $k$. The eigenvalues of the scalar Laplace operator
  are
  \begin{equation}
    \lambda_n = - 4\pi \sqrt[3]{N(k)} \; \ln \Lambda_n
    .
  \end{equation}
\end{enumerate}
We note that, in this approach to the spectrum of the Laplace-Beltrami
operator, there is only one degree $k$ that controls the accuracy of
the eigenvalues of the scalar Laplacian and at the same time the
accuracy of the Calabi-Yau metric. In fact, computing the integral
eq.~\eqref{eq:Qintegral} at degree $k$ is about as expensive as
computing Donaldson's T-operator at degree $2k$. In other words, a
general limitation of this approach is that one has to work with a
relatively low precision metric.

\begin{figure}[htbp]
  \centering
  \include{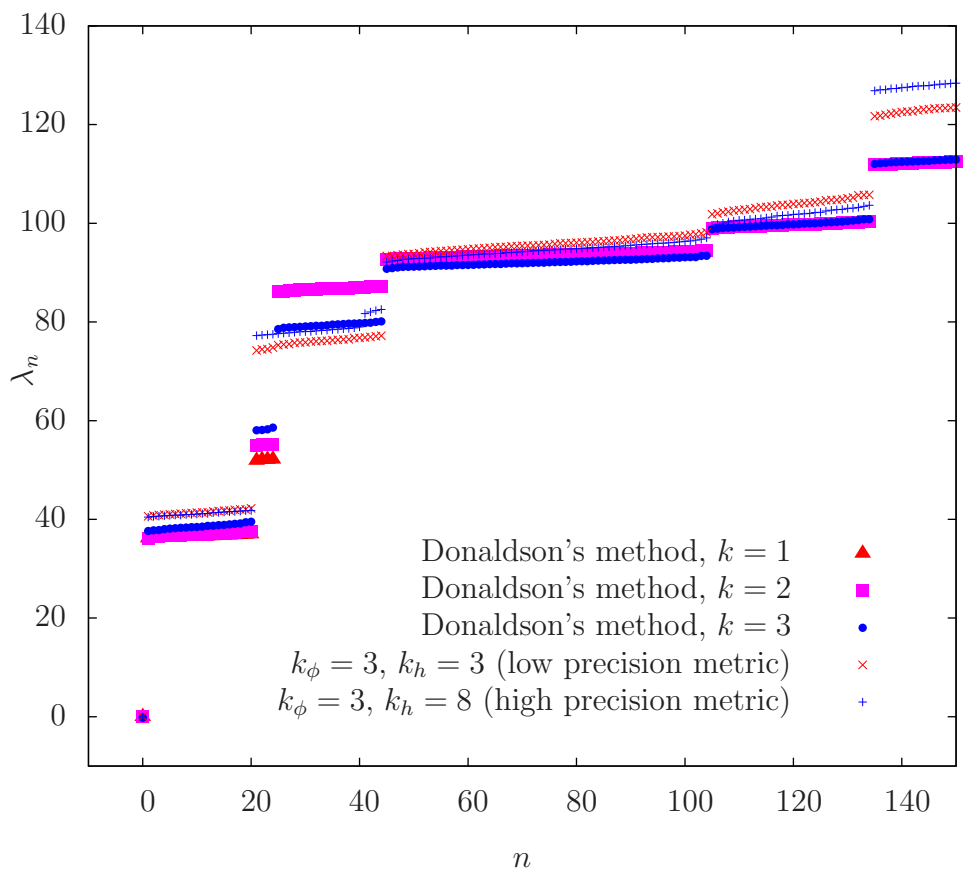}
  \caption{Donaldson's method of computing the spectrum (polygon
    symbols) of the scalar Laplace operator on the Fermat quintic
    compared to our direct computation (crosses). Note that the blue
    symbols are the highest-accuracy values, respectively. See
    \autoref{sec:Donaldson} for further discussion.}
  \label{fig:DonaldsonFermat}
\end{figure}

In \autoref{fig:DonaldsonFermat} we compare the two approaches for
computing the spectrum of the Laplace-Beltrami operator on the Fermat
quintic. We compute the eigenvalues using Donaldson's method at
degrees $k=1,2,3$ and evaluate the necessary integral
eq.~\eqref{eq:Qintegral} using $\Nonly=10 N(k)^{2}+\comma{100000}$
points. For comparison, we also plot the eigenvalues obtained by
directly computing the matrix elements of the Laplacian which we
always compute at degree $\Kphi=3$ using $\Nphi=\comma{500000}$
points. To estimate the effect of the metric on the eigenvalues, we
run our algorithm first with the metric obtain at degree $\Kh=3$
and\footnote{The number of points $\Nh$ is always obtained from the
  heuristic eq.~\eqref{eq:Qt_Nh_formula}.}  $\Nh=62250$ (bad
approximation to the Calabi-Yau metric, red diagonal crosses) as well
as with $\Kh=8$ and $\Nh=\comma{2166000}$ (good approximation to the
Calabi-Yau metric, blue upright crosses). We find that the eigenvalues
do not strongly depend on the details of the metric. Generally,
Donaldson's method and the direct computation yield very similar
results. There is a slight disagreement for the second and third
massive level, where the matrix element calculation points toward
$\mu_2=20, \mu_3=4$ while Donaldson's method suggests the opposite
order $\mu_3=4, \mu_4=20$. We suspect this is to be a numerical error
due to the finite degrees and it would be interesting to go to higher
degree in $k$, $\Kphi$, and $\Kh$.

\begin{figure}[htbp]
  \centering
  \include{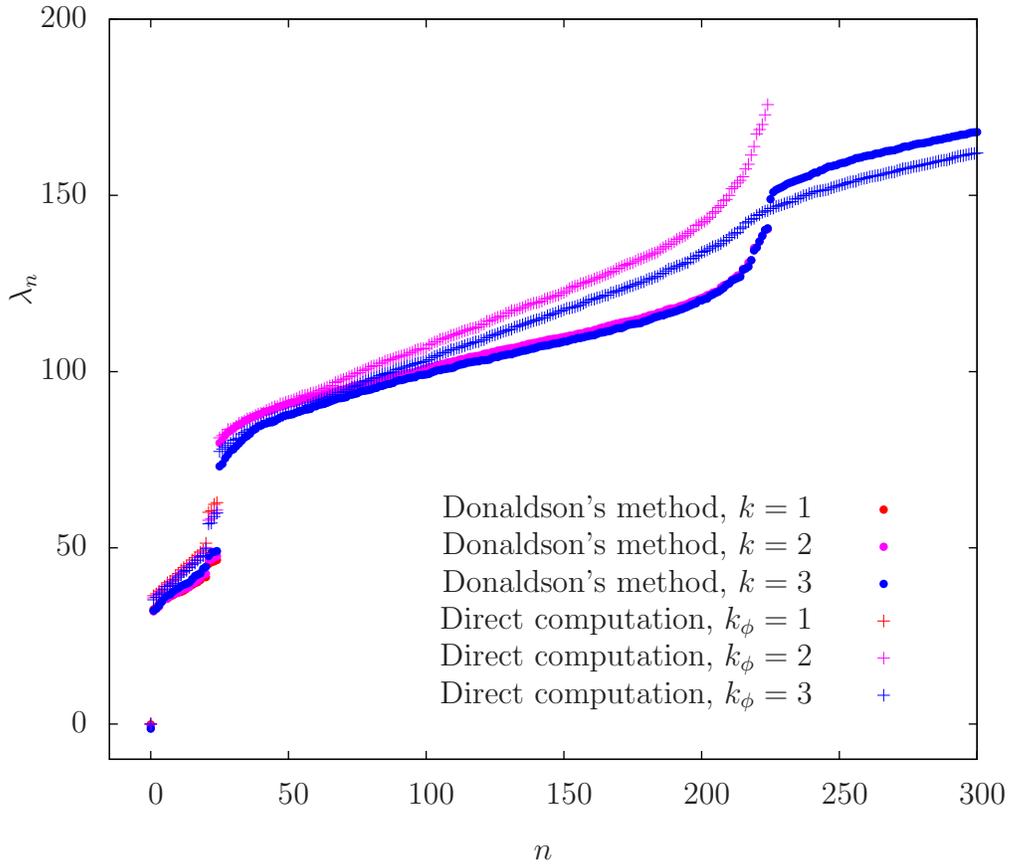}
  \caption{Donaldson's method of computing the spectrum (polygon
    symbols) of the scalar Laplace operator on the random quintic
    compared to our direct computation (crosses). Note that the blue
    symbols are the highest-accuracy values, respectively. In
    Donaldson's method the numerical integration was performed with
    $\Nonly=10 N(k)+\comma{100000}$ points. In the direct computation,
    the metric was approximated at degree $\Kh=8$ using
    $\Nh=\comma{2166000}$ points and the Laplace operator was
    evaluated at $\Nphi=\comma{500000}$ points.}
  \label{fig:DonaldsonRandom}
\end{figure}
Finally, in \autoref{fig:DonaldsonRandom} we repeat this comparison
for the quintic eq.~\eqref{eq:RandomQt} with random coefficients. In
this case, there are no discrete symmetries and one expects all
massive levels to be non-degenerate. We again find good agreement
between the two approaches towards solving the Laplace equation.

\section{\texorpdfstring{$\mathbf{\Z_5\times\Z_5}$ Quotients of
    Quintics}{Z5xZ5 Quotients of Quintics}}
\label{sec:Moduli}

Thus far, we have restricted our examples to quintic Calabi-Yau
threefolds $\Qt\subset\CP^4$. These manifolds are simply connected by
construction. However, for a wide range of applications in heterotic
string theory we are particularly interested in non-simply connected
manifolds where one can reduce the number of quark/lepton generations
as well as turn on discrete Wilson lines. Therefore, in this section
we will consider the free $\Z_5\times\Z_5$ quotient of quintic threefolds,
see~\cite{Braun:2007sn} for more details.

\subsection{\texorpdfstring{$\Z_5\times\Z_5$ Symmetric Quintics and
    their Metrics}{Z5xZ5 Symmetric Quintics and their Metrics}}
\label{sec:Z5Z5}
Explicitly, the group action
on the homogeneous coordinates $[z_0:\cdots:z_4]\in\CP^4$ is
\begin{equation}
 \label{eq:Z5xZ5act}
  \begin{split}
    g_1: 
    \big[ z_0:z_1:z_2:z_3:z_4 \big]
    \longrightarrow&~
    \big[ 
    z_0: 
    e^{\frac{2\pi i}{5}} z_1: 
    e^{2\frac{2\pi i}{5}} z_2:
    e^{3\frac{2\pi i}{5}} z_3: 
    e^{4\frac{2\pi i}{5}} z_4 \big]    
    ,
    \\
    g_2: 
    \big[ z_0:z_1:z_2:z_3:z_4 \big]
    \longrightarrow&~
    \big[ z_1:z_2:z_3:z_4:z_0 \big]
    .
  \end{split}
\end{equation}
As we discussed in \autoref{sec:genquintic}, a generic quintic is a
zero locus of a degree-$5$ polynomial containing $126$ complex
coefficients. However, only a small subset of these quintics is
invariant under the $\Z_5\times\Z_5$ action above. As we will show
below, the dimension of the space of invariant homogeneous degree-$5$
polynomials is $6$. Taking into account that one can always multiply
the defining equation by a constant, there are $5$ independent
parameters $\phi_1, \dots \phi_5\in\C$. Thus, the $\Z_5\times\Z_5$
symmetric quintics form a five parameter family which can be written
as
\begin{equation}
  \label{eq:QuinticZ5Z5}
  \begin{split}
    \Qt(z) 
    =&~
    \big( z_0^5+z_1^5+z_2^5+z_3^5+z_4^5 \big)    
    \\+&~ \phi_1
    \big( z_0 z_1 z_2 z_3 z_4 \big)
    \\+&~ \phi_2
    \big( 
    z_0^3 z_1 z_4+z_0 z_1^3 z_2+z_0 z_3 z_4^3+z_1 z_2^3 z_3+z_2
    z_3^3 z_4 
    \big)
    \\+&~ \phi_3
    \big( 
    z_0^2 z_1 z_2^2 + 
    z_1^2 z_2 z_3^2 + 
    z_2^2 z_3 z_4^2 + 
    z_3^2 z_4 z_0^2 + 
    z_4^2 z_0 z_1^2
    \big)
    \\+&~ \phi_4
    \big( 
    z_0^2 z_1^2 z_3 + 
    z_1^2 z_2^2 z_4 + 
    z_2^2 z_3^2 z_0 + 
    z_3^2 z_4^2 z_1 + 
    z_4^2 z_0^2 z_2
    \big)
    \\+&~ \phi_5
    \big( 
    z_0^3 z_2 z_3 + 
    z_1^3 z_3 z_4 + 
    z_2^3 z_4 z_0 + 
    z_3^3 z_0 z_1 + 
    z_4^3 z_1 z_2
    \big) 
    ,
  \end{split}
\end{equation}
where $\phi_1$, $\dots$, $\phi_5 \in\C$ are local coordinates on the
complex structure moduli space. From now on, $\Qt\subset \CP^4$ will
always refer to a quintic of this form.

For generic coefficients\footnote{For example, any sufficiently small
  neighbourhood of $(\phi_1,\dots,\phi_5)=(0,\dots,0)\in\C^5$. Note
  that setting all $\phi_i=0$ yields the Fermat quintic $\QtF$, see
  eq.~\eqref{eq:FermatQuintic}.} $\phi_i$, the hypersurface $\Qt$ is a
smooth Calabi-Yau threefold. Moreover, although the group action
eq.~\eqref{eq:Z5xZ5act} necessarily has fixed points in $\CP^4$, these
fixed points do not intersect a generic hypersurface $\Qt$. Thus the
quotient
\begin{equation}
  Q = \Qt \Big/ \big( \Z_5\times\Z_5 \big)
\end{equation}
is again a smooth Calabi-Yau threefold.
As a general principle, we will compute quantities on the quotient $Q$
by computing the corresponding invariant quantities on the covering
space $\Qt$. For example, the complex structure moduli space of $Q$ is
the moduli space of $\Z_5\times\Z_5$-invariant complex structures on
$\Qt$. Hence, its dimension is
\begin{equation}
  h^{2,1}(Q)
  = 
  \dim H^{2,1}\big(Q)
  = 
  \dim H^{2,1}\big(\Qt)^{\Z_5\times\Z_5}
  = 
  5
  ,
\end{equation}
corresponding to the $5$ independent parameters $\phi_1$, $\dots$,
$\phi_5$ in a $\Z_5\times\Z_5$-invariant quintic $\Qt(z)$.

In the same spirit, we will compute the Calabi-Yau metric on $Q$ by
performing the analogous computation on the covering space
$\Qt$. To begin, one must choose a degree $\Kh$ and determine a basis
$s_\alpha$ for the corresponding $\Z_5\times\Z_5$-invariant
homogeneous degree-$\Kh$ polynomials 
\begin{equation}
  \Span\{ s_\alpha \}
  =
  \C[ z_0, \dots, z_4 ]_\Kh^{\Z_5\times\Z_5}
  \Big/
  \big\langle \Qt(z) \big\rangle
  \label{train1}
\end{equation}
on $\Qt$. Note, however, that for any homogeneous degree-$\Kh$
polynomial $p_\Kh(z)$
\begin{equation}
  g_1g_2g_1^{-1}g_2^{-1} \Big( p_\Kh(z) \Big) 
  =
  e^{2\pi i\frac{\Kh}{5}}
  \; p_\Kh(z)
\end{equation}
and, hence, the two $\Z_5$ generators in eq.~\eqref{eq:Z5xZ5act} do
not always commute.  It follows that for a space of homogeneous
polynomials to carry a linear representation of $\Z_5\times\Z_5$, let
alone have an invariant subspace, their degree $\Kh$ must be divisible
by $5$; that is,
\begin{equation}
  \Kh \in 5\Z
  .
  \label{train2}
\end{equation}
This can be understood in various ways, and we refer
to~\cite{Braun:2007sn} for more details. Henceforth, we will assume
that eq.~\eqref{train2} is satisfied.

The first step in determining the basis of sections $\{s_\alpha\}$ on
$\Qt$ is to find a basis for the invariant polynomials $\C[ z_0,
\dots, z_4 ]_\Kh^{\Z_5\times\Z_5}$ on $\CP^4$. Such a basis is given
by the Hironaka decomposition
\begin{equation}
  \label{eq:HironakaZ5Z5}
  \C[z_0,z_1,z_2,z_3,z_4]^{\Z_5\times\Z_5}_\Kh=
  \bigoplus_{i=1}^{100}
  \eta_i 
  \, 
  \C[
  \theta_1,\theta_2,\theta_3,\theta_4,\theta_5 
  ]_{\Kh-\deg( \eta_i )}
  .
\end{equation}
Here, the $\theta_j=\theta_j(z)$ and $\eta_i=\eta_i(z)$ are themselves
homogeneous polynomials of various degrees\footnote{The degrees of the
  $\theta_j$, $\eta_i$ are multiples of $5$, of course.}. The
$\theta_1$, $\dots$, $\theta_5$ are called ``primary invariants'' and
the $\eta_1$, $\dots$, $\eta_{100}$ are called ``secondary
invariants''. The primary and secondary invariants are not unique, but
one minimal choice is~\cite{Braun:2007sn}
\begin{equation}
  \label{eq:thetadef}
  \begin{split}
    \theta_1 \eqdef&\; 
    z_0^5+z_1^5+z_2^5+z_3^5+z_4^5 
    \\
    \theta_2 \eqdef&\; 
    z_0 z_1 z_2 z_3 z_4 
    \\
    \theta_3 \eqdef&\; 
    z_0^3 z_1 z_4 +
    z_1^3 z_2 z_0 +
    z_2^3 z_3 z_1 +
    z_3^3 z_4 z_2 +
    z_4^3 z_0 z_3 
    \\
    \theta_4 \eqdef&\; 
    z_0^{10}+z_1^{10}+z_2^{10}+z_3^{10}+z_4^{10} 
    \\
    \theta_5 \eqdef&\; 
    z_0^8 z_2 z_3 +
    z_1^8 z_3 z_4 +
    z_2^8 z_4 z_0 +
    z_3^8 z_0 z_1 +
    z_4^8 z_1 z_2 
  \end{split}
\end{equation}
and
\begin{equation}
  \begin{split}
    \eta_1 \eqdef&\; 
    1
    ,\\ 
    \eta_2 \eqdef&\; 
    z_0^2 z_1 z_2^2 +
    z_1^2 z_2 z_3^2 +
    z_2^2 z_3 z_4^2 +
    z_3^2 z_4 z_0^2 +
    z_4^2 z_0 z_1^2
    ,\\
    \eta_3 \eqdef&\; 
    z_0^2 z_1^2 z_3 +
    z_1^2 z_2^2 z_4 +
    z_2^2 z_3^2 z_0 +
    z_3^2 z_4^2 z_1 +
    z_4^2 z_0^2 z_2
    ,\\
    \eta_4 \eqdef&\; 
    z_0^3 z_2 z_3 +
    z_1^3 z_3 z_4 +
    z_2^3 z_4 z_0 +
    z_3^3 z_0 z_1 +
    z_4^3 z_1 z_2
    ,\\ 
    \eta_5 \eqdef&\; 
    z_0^5 z_2^5 +
    z_1^5 z_3^5 +
    z_2^5 z_4^5 +
    z_3^5 z_0^5 +
    z_4^5 z_1^5 
    ,\\
    \vdots& \\
    \eta_{100} \eqdef&\;
    z_0^{30} + z_1^{30} + z_2^{30} + z_3^{30} + z_4^{30}
    .
  \end{split}
\end{equation}
For example, the $6$-dimensional space of invariant degree-$5$
homogeneous polynomials on $\CP^4$ is
\begin{equation}
  \begin{split}
    \C[z_0,z_1,z_2,z_3,z_4]^{\Z_5\times\Z_5}_5
    =&\;
    \bigoplus_{i=1}^{100}
    \eta_i 
    \, 
    \C[
    \theta_1,\theta_2,\theta_3,\theta_4,\theta_5 
    ]_{5-\deg( \eta_i )}
    \\
    =&\;
    \eta_1 \theta_1 
    \C \oplus
    \eta_1 \theta_2 
    \C \oplus
    \eta_1 \theta_3
    \C \oplus
    \eta_2
    \C \oplus
    \eta_3
    \C \oplus
    \eta_4
    \C
    ,
  \end{split}
\end{equation}
thus proving  eq.~\eqref{eq:QuinticZ5Z5}.

Using the Hironaka decomposition, we can now determining the basis
$s_\alpha$ in eq.~\eqref{train1} by modding out the equation
$\Qt(z)=0$ which defines the covering space.  This was discussed
in~\cite{Braun:2007sn}. The result is that one can simply eliminate
the first primary invariant using
\begin{equation}
  \theta_{1}
  =
  -\phi_{1}\theta_{2}-\phi_{2}\theta_{3}
  -\phi_{3}\eta_{2}-\phi_{4}\eta_{3}-\phi_{5}\eta_{4}
  ,
  \label{home1}
\end{equation}
yielding
\begin{equation}
\Span\{s_{\alpha}\}= \bigoplus_{i=1}^{100}
  \eta_i 
  \, 
  \C[
  \theta_2,\theta_3,\theta_4,\theta_5 
  ]_{\Kh-\deg( \eta_i )}
\end{equation}
where $\alpha=0,\dots,N^{\Z_5\times\Z_5}(\Kh)-1$. The number
$N^{\Z_5\times Z_5}(\Kh)$ of $\Z_5\times\Z_5$-invariant homogeneous
degree-$\Kh$ polynomials modulo $\Qt=0$ was tabulated
in~\cite{Braun:2007sn}. In particular, the first three values are
\begin{equation}
  N^{\Z_5\times\Z_5}(0)=1
  , \quad 
  N^{\Z_5\times\Z_5}(5)=5
  , \quad 
  N^{\Z_5\times\Z_5}(10)=35
  ,
  \label{train3}
\end{equation}
which we will use below.

We now have everything in place to compute the metric on $Q$. First,
one specifies the five complex structure parameters $\phi_{i}$ which
define the $\Z_5\times\Z_5$-symmetric covering space $\Qt$. Then, all
one has to do is to replace the homogeneous polynomials in the
procedure outlined in \autoref{sec:genquintic} by
$\Z_5\times\Z_5$-invariant homogeneous polynomials. Donaldson's
algorithm then calculates the Calabi-Yau metric on the
$\Z_5\times\Z_5$-symmetric quintic $\Qt$ and, hence, the metric on the
quotient $Q=\Qt\big/(\Z_5\times\Z_5)$. In fact, we use a refinement of
this method which is even more efficient, that is, achieves higher
numerical accuracy in less computing time. As it is not relevant to
the spectrum of the Laplace operator, we relegate the details to
\autoref{sec:DonaldsonNotes}. Henceforth, we will always use the
following parameters in the computation of the metric.
\begin{itemize}
\item The degree of the invariant homogeneous polynomials for the
  \Kahler{} potential is taken to be
  \begin{equation}
    \Kh=10
    .
    \label{aaa}
  \end{equation}
\item The number of points used to evaluate the T-operator is
  \begin{equation}
    \Nh= 10 \times
    \Big(
    \text{\# of independent entries in $h^{\alpha\betabar}$} 
    \Big) +
    \comma{100000} = 
    \comma{406250}
    .
    \label{bbb}
  \end{equation}
  Note that $h^{\alpha\bar\beta}$, the matrix of free parameters in
  Donaldson's ansatz for the metric, is block diagonal in
  \autoref{sec:DonaldsonNotes}. Therefore, the total number of
  independent entries is in fact \comma{30625} and not simply
  $N^{\Z_5\times\Z_5}(10)^2=\comma{1225}$.
\end{itemize}
As always, it is unenlightening to present the numerical result for
the approximation to the Calabi-Yau metric. It is useful, however, to
consider the error measure $\sigma_{10}$. As an important example, let
us choose as our Calabi-Yau manifold the $\Z_5\times\Z_5$ quotient of
the Fermat quintic $\QtF$. The computation of the metric takes about
half an hour of wall time, with the resulting error measure of
$\sigma_{10} = 2.8\times 10^{-2}$.

\subsection{The Laplacian on the Quotient}
\label{sec:DeltaQ}

Having computed the Calabi-Yau metric on the quotient
$Q=\Qt\big/(\Z_5\times\Z_5)$, we now turn to the calculation of the
spectrum of the Laplace-Beltrami operator $\Delta$. To begin, one must
specify a finite-dimensional approximation to the space of complex
valued functions on $Q$. Note, however, that the scalar functions on
$Q$ are precisely the invariant functions on the covering space
$\Qt$. More formally, an invariant function on $\Qt$ is of the form
$q^* f= f\circ q$, where $f$ is a function on the quotient $Q$ and
$q:\Qt\to Q$ is the quotient map. Hence, we will specify a
finite-dimensional approximation to the space of complex-valued
$\Z_5\times\Z_5$-invariant functions on $\Qt$. For any finite value of
$\Kphi$, we choose
\begin{equation}
  \Fsheaf^{\Z_5\times\Z_5}_\Kphi 
  =
  \Span
  \Bigg\{ 
    \frac{
      s_\alpha{\sbar}_{\bar\beta} 
    }{
      \big( \sum_{i=0}^4  |z_i|^2 \big)^\Kphi
    } 
  ~
  \Bigg|
  ~
  \alpha, \bar\beta =0,\dots,N^{\Z_5\times\Z_5}(\Kphi)-1
  \Bigg\} 
  ,
  \label{eq:FkphiQt2}
\end{equation}
where $\{s_\alpha\}$ is a basis for the invariant homogeneous
polynomials modulo the hypersurface constraint
\begin{equation}
  \Span \{ s_{\alpha} \}=
  \C\left[ z_0, \dots, z_4 \right]^{\Z_5\times\Z_5}_{\Kphi}
  \Big/
  \big\langle \Qt(z) \big\rangle 
  .
  \label{eq:parrot3b2}
\end{equation} 
We already had to determine such a basis while applying Donaldson's
algorithm for the metric, the only difference now is that the degree
is $\Kphi$ instead of $\Kh$. The counting function
$N^{\Z_5\times\Z_5}(\Kphi)$ is the same, and some of its values were
given in eq.~\eqref{train3}. Clearly,
\begin{equation}
  \dim 
  \Fsheaf^{\Z_5\times\Z_5}_\Kphi 
  = 
  \Big( N^{\Z_5\times\Z_5}(\Kphi) \Big)^2
  .
  \label{eq:dimFk2}
\end{equation}

Having specified $ \Fsheaf^{\Z_5\times\Z_5}_\Kphi $, we can now
calculate any matrix element on $Q$ simply by replacing the
approximating space of functions on $Q$ by the invariant functions on
$\Qt$ and integrating over $\Qt$. For example, the matrix elements of
the Laplacian on $Q$ are
\begin{equation}
  \Delta_{a b}
  =
  \big\langle f_a \big| \Delta \big| f_b \big\rangle
  =
  \int_Q \bar{f}_a \Delta  f_b \dVol(Q)
  =
  \frac{1}{\big|\Z_5\times\Z_5\big|}
  \int_\Qt (q^\ast \bar{f}_a) \Delta (q^\ast f_b)
  \dVol(\Qt)
  .
\end{equation}
Computing the matrix elements requires another numerical integration
that is completely independent of the one in the T-operator. As
previously, we denote the number of points in the matrix element
integration by $\Nphi$.

Having evaluated the matrix elements, one can now numerically solve
the matrix eigenvalue equation eq.~\eqref{eq:generalizedeigen} for the
eigenvalues and eigenfunctions of the Laplacian.  Note that the
factors of $\frac{1}{|\Z_5\times\Z_5|}$ cancel out of this equation,
leaving identical eigenvalues and eigenfunctions on $\Qt$ and $Q$,
respectively. Since the functions in $\Fsheaf^{\Z_5\times\Z_5}_\Kphi$
live on the covering space, we are actually solving
\begin{equation}
  \Delta_\Qt \phi_n = \lambda_n^{\Z_5\times\Z_5} \phi_n
  ,\qquad
  \phi_n \in C^\infty(\Qt,\C)^{\Z_5\times\Z_5}
\end{equation}
on $\Qt$. Note that, as always, the volume measure of the integrals is
chosen so that $\Vol(\Qt)=1$.  For the reasons stated above, the
invariant eigenfunctions on $\Qt$ can be identified with the
eigenfunctions of the Laplacian on $Q$ at the same eigenvalue, but
with $\Vol(Q)=\tfrac{1}{|\Z_5\times\Z_5|}=\tfrac{1}{25}$. However,
since we want to adhere to our convention of normalizing $\Vol(Q)=1$,
we have to rescale the volume and hence the eigenvalues
$\lambda_n^{\Z_5\times\Z_5}$. Using eqns.~\eqref{bird1}
and~\eqref{eq:lambdaVol}, the eigenvalues $\lambda_n$ on $Q$ are
\begin{equation}
  \lambda_n = \frac{\lambda_n^{\Z_5\times\Z_5}}{\sqrt[3]{25}}
  .
  \label{ccc}
\end{equation}
Using this method, one can compute the spectrum of the
Laplace-Beltrami operator on the quotient of any $\Z_5\times\Z_5$
symmetric quintic.

\subsection{Quotient of the Fermat Quintic}
\label{sec:QFQ}

As an explicit example, let us consider the quotient of the Fermat
quintic,
\begin{figure}[htbp]
  \centering
  \include{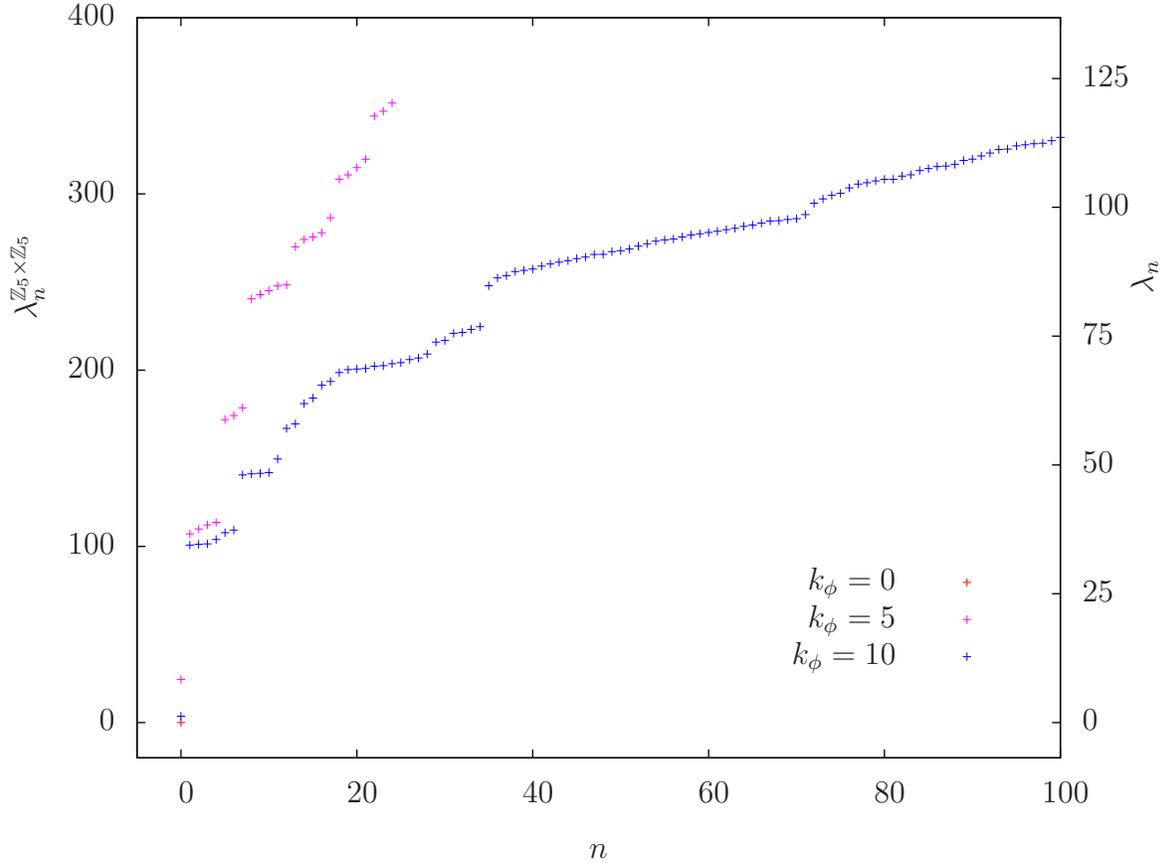}
  \caption{Eigenvalues $\lambda_n^{\Z_5\times\Z_5}$ of the scalar
    Laplace operator on the Fermat quintic $\QtF$ acting on
    $\Z_5\times \Z_5$-invariant eigenfunctions. Up to an overall
    factor due to our volume normalization, these are the same as the
    eigenvalues $\lambda_n$ of the scalar Laplace operator on the
    quotient $Q_F=\QtF\big/ (\Z_5\times\Z_5)$. The metric is computed
    at degree $\Kh=10$ and $\Nh=\comma{406250}$ points. The Laplace
    operator is evaluated using $\Nphi=\comma{100000}$ points.}
  \label{fig:SpecFourGenk}
\end{figure}
\begin{table}[htbp]
  \centering
  \renewcommand{\arraystretch}{1.3}
  \begin{tabular}{|c|cc|cc|}
    \hline
    $n$ & 
    $\lambda_n^{\Z_5\times\Z_5}$ &
    \vphantom{$\displaystyle \frac{A^1}{B_2}$}
    $\lambda_n = \frac{\lambda_n^{\Z_5\times\Z_5}}{\sqrt[3]{25}}$ &
    $\lambdahat$ & 
    $\mu$ 
    \\ \hline\hline
    $0$ & 
    $3.586$ &
    $1.226$ &
    $\lambdahat_0=1.23$ & 
    $\mu_0=1$ 
    \\ \hline
    $1$ & 
    $100.7$ &
    $34.45$ &
    \multirow{4}{*}{$\lambdahat_1=34.8\pm 0.5$} & 
    \multirow{4}{*}{$\mu_1=4$} 
    \\
    $2$ & 
    $101.2$ &
    $34.61$ & 
    &
    \\
    $3$ & 
    $101.4$ &
    $34.68$ &
    &
    \\
    $4$ & 
    $103.9$ &
    $35.53$ &
    &
    \\ \hline
    $5$ & 
    $107.8$ &
    $36.86$ &
    \multirow{2}{*}{$\lambdahat_2=37.1\pm 0.4$} & 
    \multirow{2}{*}{$\mu_2=2$}
    \\
    $6$ & 
    $109.2$ &
    $37.36$ &
    & 
    \\ \hline
    $7$ & 
    $140.50$ &
    $48.05$ &
    \multirow{4}{*}{$\lambdahat_3=48.3\pm 0.2$} & 
    \multirow{4}{*}{$\mu_3=4$}
    \\
    $8$ & 
    $141.16$ &
    $48.28$ &
    & 
    \\
    $9$ & 
    $141.47$ &
    $48.38$ &
    & 
    \\
    $10$ & 
    $141.78$ &
    $48.49$ &
    &
    \\ \hline
    $11$ & 
    $149.57$ &	
    $51.15$ &
    \multirow{1}{*}{$\lambdahat_4=51.2$} & 
    \multirow{1}{*}{$\mu_4=1$}
    \\ \hline
    $12$ & 
    $166.91$ &
    $57.08$ &
    \multirow{2}{*}{$\lambdahat_5=57.5\pm 0.6$} & 
    \multirow{2}{*}{$\mu_5=2$}
    \\
    $13$ & 
    $169.48$ &	
    $57.96$ &
    & 
    \\ \hline
    $14$ & 
    $181.00$ &
    $61.90$ &
    \multirow{2}{*}{$\lambdahat_6=62.4\pm 0.8$} & 
    \multirow{2}{*}{$\mu_6=2$}
    \\
    $15$ & 
    $184.15$ &
    $62.98$ &
    & 
    \\ \hline
    $16$ & 
    $191.49$ &
    $65.48$ &
    & 
    \\
    $17$ & 
    $193.55$ &
    $66.19$ &
    & 
    \\
    $18$ & 
    $198.65$ &
    $67.94$ &
    &
    \\ 
    $\vdots$ & $\vdots$ & $\vdots$ &
    $\vdots$ & $\vdots$
  \end{tabular}
  \caption{Low-lying eigenvalues of the scalar 
    Laplace operator on $Q_F$, the $\Z_5\times\Z_5$-quotient of the Fermat 
    quintic, computed with $\Kh=\Kphi=10$, $\Nh=\comma{406250}$,
    $\Nphi=\comma{100000}$. 
    The first two columns are the numerical results. 
    The third column specifies $\lambdahat$, the average    
    over the eigenvalues that are converging to a single degenerate 
    level. The final column counts the 
    multiplicities of each such level.}
  \label{tab:evQF}
\end{table}
\begin{equation}
  Q_F 
  = 
  \QtF
  \Big/ \big( Z_5\times\Z_5 \big)
  .
\end{equation}
We numerically computed the spectrum of the scalar Laplace operator for
each of the three values $\Kphi=0,5,10$ using eq.~\eqref{train3}. The
resulting eigenvalues are shown in \autoref{fig:SpecFourGenk}. Note
that we present both the eigenvalues $\lambda_n^{\Z_5\times\Z_5}$ on
$\Qt$ as well as the normalized eigenvalues $\lambda_n$ on $Q$ defined
by eq.~\eqref{ccc}.

We list the numerical values of the first few eigenvalues in
\autoref{tab:evQF} and make the following two observations. First, the
lowest eigenvalue $\lambda_0$ is no longer zero up to machine
precision, as it was in \autoref{tab:QtFresult}. This is so because
the constant function is not part of the approximate space of
functions at $\Kphi=10$ and, therefore, the lowest eigenvalue
$\lambda_0$ only approaches zero as $\Kphi$ increases. The actual
numerical value $\lambda_0\approx 1.2$ gives us an estimate of the
error introduced by truncating the space of functions.  Second, the
low-lying eigenvalues clearly form degenerate levels. As usual, the
numerical error caused by the truncation of the space of functions
increases as we go to higher eigenvalues. However, the first $16$
eigenvalues are sufficiently well separated that we can conjecture the
underlying multiplicities $\mu$. We list these degeneracies together
with the best approximation to the true eigenvalue $\lambdahat$ in
\autoref{tab:evQF}. Clearly, the degeneracies in the spectrum
strongly hint at an underlying symmetry. We will discuss the
associated isometry group in the following subsection.

\subsection{Group Theory and the Quotient Eigenmodes}
\label{sec:invHarmonics}

The free $\Z_5\times\Z_5$ action eq.~\eqref{eq:Z5xZ5act} is a subgroup
of the symmetries of the Fermat quintic,
\begin{equation}
  \Z_5\times\Z_5\subset \AutBar\big( \QtF \big)
  ,
\end{equation}
given in eq.~\eqref{ddd}. Naively, one now would like to form the
quotient to obtain the remaining symmetries on
$Q_F=\QtF/(\Z_5\times\Z_5)$. However, the $\Z_5\times\Z_5$ subgroup is
not normal, that is, not closed under conjugation. The only
possibility is to form the normal closure\footnote{Also called the
  conjugate closure.}
\begin{equation}
  \big<
  \Z_5 \times \Z_5
  \big>^{\AutBar(\QtF)}
  =
  \Big\{ 
  h^{-1} g h 
  ~\Big|~
  g \in \Z_5\times\Z_5
  ,~
  h \in \AutBar(\QtF)
  \Big\}
  .
\end{equation}
The quotient by the normal closure is well-defined, and we obtain 
\begin{equation}
  \AutBar(Q_F) 
  = 
  \AutBar\big(\QtF\big)
  \Big/ 
  \big<
  \Z_5 \times \Z_5
  \big>^{\AutBar(\QtF)}  
  =
  D_{20}
  ,
\end{equation}
the dihedral group with $20$ elements. However, just looking at the
representation theory of $\AutBar(Q_F)$ is insufficient to understand
the multiplicities of the eigenvalues of the Laplacian. Instead, one
must use all of $\AutBar(\QtF)$, even those elements (called
``pseudo-symmetries'' in~\cite{Green:1987mn}) that do not correspond to
symmetries of the quotient $Q_F$. On a practical level, we also note
that $D_{20}$ has only $1$- and $2$-dimensional irreducible
representations and could never explain the multiplicity
$\mu_1(Q_F)=4$, for example, listed in \autoref{tab:evQF}.

As we discussed in \autoref{sec:fermatsymmetry}, the symmetry group of
the Fermat quintic has $80$ distinct irreducible representations
occurring in $14$ different dimensions. Let us label them by
$\rho_{d,i}$, where $d$ is the complex dimension and $i=1,\dots,n_d$
distinguishes the $n_d$ different representations in dimension
$d$. Under the $\Z_5\times\Z_5$ quotient
\begin{equation}
  \QtF 
  \longrightarrow 
  Q_F = \QtF\big/(\Z_5\times\Z_5)
\end{equation}
all non-invariant eigenfunctions of the Laplacian are projected out
and each invariant eigenfunction descends to an eigenfunction on
$Q_F$. Hence, the degeneracies of the eigenvalues are counted by the
dimension
\begin{equation}
  \dim  \big( \rho_{d,i}^{\Z_5\times\Z_5} \big) 
\end{equation}
of the $\Z_5\times\Z_5$-invariant subspace. It turns out that, for the
chosen $\Z_5\times\Z_5 \subset \AutBar(\QtF)$, this dimension
depends only on $d$, and not on the index $i$. We denote the common
value by
\begin{equation}
  \label{eq:dimddef}
  \dim^{\Z_5\times\Z_5}_d
  \eqdef
  \dim \big( \rho_{d,1}^{\Z_5\times\Z_5} \big) 
  = \cdots = 
  \dim \big( \rho_{d,n_d}^{\Z_5\times\Z_5} \big) 
\end{equation}
and tabulate it in \autoref{tab:AutBarQFquot}.
\begin{table}
  \centering
  \renewcommand{\arraystretch}{1.5}
  \begin{tabular}{c|cccccccccccccc}
    $d$ &
    $1$ & $2$ & $4$ & 
    $5$ & $6$ & $8$ & 
    $10$ & $12$ & $20$ & 
    $30$ & $40$ & $60$ & 
    $80$ & $120$
    \\ \hline
    $n_d$
    &
    $4$ & $4$ & $4$ & 
    $4$ & $2$ & $4$ &
    $4$ & $2$ & $8$ & 
    $8$ &$12$ & $18$ & 
    $4$ & $2$
    \\ \hline
    $\dim^{\Z_5\times\Z_5}_d$
    &
    $1$ & $2$ & $0$ & 
    $1$ & $2$ & $0$ & 
    $2$ & $4$ & $0$ & 
    $2$ & $0$ & $4$ & 
    $0$ & $4$
  \end{tabular}
  \caption{Number $n_d$ of distinct irreducible representations of 
    $\AutBar(\QtF)$ in complex dimension $d$. We also list
    the dimension $\dim_d^{\Z_5\times\Z_5}$ of the
    $\Z_5\times\Z_5$-invariant 
    subspace for each representation, see eq.~\eqref{eq:dimddef}. 
    Note that it turns out to only
    depend on the dimension $d$ of the representation.
  }
  \label{tab:AutBarQFquot}
\end{table}

\begin{table}
  \centering
  \renewcommand{\arraystretch}{1.3}
  \begin{tabular}{c@{$\qquad\longrightarrow\qquad$}c}
    $\QtF$ & $Q_F$
    \\ \hline
    $\mu_0(\QtF)=1$  & $\mu_0(Q_F)=1$ \\ 
    $\mu_1(\QtF)=20$ & $0$ \\
    $\mu_2(\QtF)=20$ & $0$ \\ 
    $\mu_3(\QtF)=4$  & $0$ \\
    $\mu_4(\QtF)=60$ & $\mu_1(Q_F)=4$ \\ 
    $\mu_5(\QtF)=30$ & $\mu_2(Q_F)=2$
  \end{tabular}
  \caption{Projection of the multiplicity of eigenvalues on the 
    Fermat quintic $\QtF$ to the $\Z_5\times\Z_5$-quotient $Q_F$. }
  \label{tab:muQtFtoQF}
\end{table}
Using this and the multiplicities of the eigenvalues on the Fermat quintic $\QtF$ given in
\autoref{tab:QtFresult}, we can now perform the
$\Z_5\times\Z_5$-quotient and obtain the degeneracies of the scalar
Laplacian on the $Q_F$.
The results are listed in \autoref{tab:muQtFtoQF}. 
We find complete
agreement with the spectrum found by directly computing the eigenvalues on $Q_F$ given in 
\autoref{tab:evQF}.
Naturally, this comparison is limited by the number of eigenvalues we were able to compute on $\QtF$.
The agreement of the lower lying levels, however,  gives us confidence that the values of $\lambdahat_m$ and $\mu_{m}$ for $m=3,4,5,6,\dots$ given in \autoref{tab:evQF} are also a good approximation to the exact results on the quotient.

\subsection{Varying the Complex Structure}
\label{sec:family1}

To numerically compute any metric-dependent quantity on a
Calabi-Yau manifold, one has to fix the complex structure and \Kahler{}
moduli to specific values. This was done, for example, in~\autoref{sec:QFQ}, where the  moduli were 
chosen so that the covering space was the Fermat quintic with unit volume.
In this section, we will extend our results to the
one-parameter family of $\Z_5\times\Z_5$ symmetric quintics $\Qt_\psi$ defined by the vanishing
of the polynomial 
\begin{equation}
  \Qt_\psi =
  \sum z_i^5 - 5 \psi \prod z_i 
  .
\end{equation}
The Kahler modulus will always be fixed so that the volume of
$\Qt_\psi$ is unity. The complex structure parameter $\psi$ can, in
principle, take on any complex value. However, for simplicity, we will
only consider $\psi\in \R$ in this subsection. Note that each
$\Qt_\psi$ is indeed a quintic with the free $\Z_5\times\Z_5$ symmetry
in eq.~\eqref{eq:QuinticZ5Z5}. Hence, the quotient
\begin{equation}
  Q_\psi 
  =
  \Qt_\psi\Big/ \big(\Z_5\times\Z_5\big)
\end{equation}
is a smooth Calabi-Yau threefold. 

\begin{figure}[htbp]
  \centering
  \include{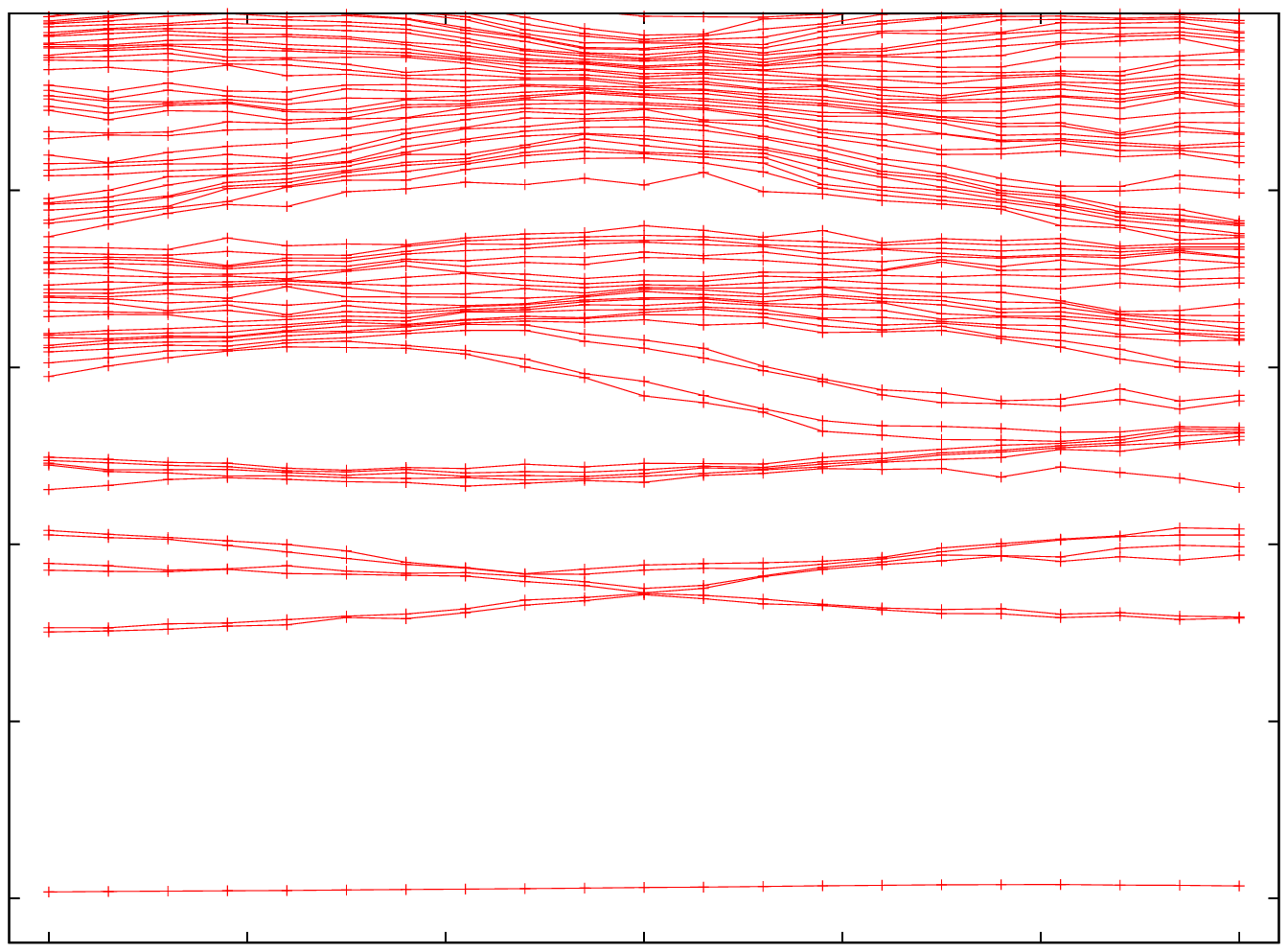}
  \caption{Spectrum of the scalar Laplace operator on the real
    $1$-parameter family $Q_\psi$ of quintic quotients. The metric is
    computed at degree $\Kh=10$ with $\Nh=\comma{406250}$. The Laplace
    operator is evaluated at $\Kphi=10$ and $\Nphi=\comma{50000}$
    points.}
  \label{fig:SpecFamily1}
\end{figure}

We have computed the spectrum of the scalar Laplace operator on this
one-parameter family of quotients for various values of $\psi$.  The
resulting $\psi$-dependent spectrum can be found in
\autoref{fig:SpecFamily1}.
Note that this one-parameter family of $\Z_5\times\Z_5$-symmetric
quintics passes through two special points,
\begin{descriptionlist}
\item[$\psi=0$:] Without the $\prod z_i$ term, $\Qt_{\psi=0} = \QtF$
  is exactly the Fermat quintic. We will investigate the symmetry
  enhancement at this point in the next subsection.
\item[$\psi=1$:] This is the so-called conifold point, where the
  quintic is singular. On the covering space $\Qt_{\psi=1} \subset
  \CP^4$, the singularity is at 
  \begin{equation}
    z_C = \big[ 1:1:1:1:1 ] 
  \end{equation}
  and its images under the $\Z_5\times\Z_5$ symmetry group. At these points the
  hypersurface equation fails to be transversal,
  \begin{equation}
    \frac{\partial \Qt_{\psi=1}}{\partial z_0} (z_C)
    = \dots = 
    \frac{\partial \Qt_{\psi=1}}{\partial z_4} (z_C)      
    =
    \Qt_{\psi=1}(z_C) = 
    0
    ,
  \end{equation}
causing the singularity.
\end{descriptionlist}
\noindent Perhaps surprisingly, the spectrum of the scalar Laplace operator
shows no trace of the conifold singularity at $\psi=1$. However, the reason for this is straightforward. The low-lying modes are
slowly-varying functions and, in particular, are almost
constant near any point-like singularity. For example, the first
massive eigenvalue is essentially determined by the diameter of the
manifold, see \autoref{sec:gap}, and does not depend on local details
of the metric.

\subsection{Branching Rules}
\label{sec:branching}

Let us return to spectrum of the Laplace-Beltrami operator in
\autoref{fig:SpecFamily1} and focus on the neighbourhood of
$\psi=0$. Clearly, $Q_{\psi=0}=Q_F$ is the quotient of the Fermat quintic, while
$Q_{\psi\not=0}$ is a deformation of the Fermat quotient that breaks
part of its discrete isometry group. In particular, note that for small non-zero values of $\psi$
\begin{itemize}
\item The first massive level $\mu_1(Q_F)=4$ splits into two pairs of
  eigenvalues.
\item The second massive level $\mu_2(Q_F)=2$ remains two-fold degenerate.
\end{itemize}
In this subsection, we will attempt to understand this from the
group-theoretical perspective. 

As discussed in \autoref{sec:invHarmonics}, the multiplicities of the
eigenvalues on the quotient $Q_\psi=\Qt_\psi/(\Z_5\times\Z_5)$ are
really determined by the representation theory of the symmetry group
of the covering space. We have to distinguish two cases.
\begin{descriptionlist}
\item[$\psi=0$:] This is the case of the Fermat quintic, whose
  symmetries we already discussed in \autoref{sec:fermatsymmetry},
  \begin{equation}
    \AutBar\big( \Qt_{\psi=0} \big) = 
    \AutBar\big( \QtF \big) =
    \big(  S_5 \times \Z_2  \big)  \ltimes  \big( \Z_5 \big)^4
    .
  \end{equation}
 The irreducible representations of  $\AutBar\big( \QtF \big)$ were presented in 
 \autoref{tab:AutBarQFquot}.  
\item[$\psi\not=0$:] In this case, the invariance of the $\prod z_i$
  monomial gives one further constraint on the $(\Z_5)^4$ phase
  rotations. In other words, turning on $\psi$ breaks the phase
  rotation symmetry to $(\Z_5)^3$. The remaining symmetry group
  is\footnote{Since we chose $\psi$ to be real, the complex conjugation
    $\Z_2$ remains unbroken.} 
  \begin{equation}
    \AutBar\big( \Qt_{\psi\not=0} \big) = 
    \big(  S_5 \times \Z_2  \big)  \ltimes  \big( \Z_5 \big)^3
    .
  \end{equation}
  The irreducible representations of  $\AutBar\big( \Qt_{\psi\not=0} \big)$ are given in 
  \autoref{tab:AutBarQpsiquot}.
  Note that, by construction, this group is a proper subgroup of the
  full symmetry group, both of which containing the free
  $\Z_5\times\Z_5$ action. That is,
  \begin{equation}
    \AutBar\big( \Qt_{\psi=0} \big)
    \quad\supset\quad
    \AutBar\big( \Qt_{\psi\not=0} \big)
    \quad\supset\quad
    \Z_5\times\Z_5
    .
  \end{equation}
\end{descriptionlist}
\begin{table}
  \centering
  \renewcommand{\arraystretch}{1.5}
  \begin{tabular}{c|cccccccccc}
    $d$ &
    $1$ & $4$ & 
    $5$ & $6$ &
    $20$ & $24$ & 
    $30$ & $40$ & 
    $48$ & $60$
    \\ \hline
    $n_d$
    &
    $4$ & $4$ &
    $4$ & $2$ &
    $8$ & $2$ &
    $8$ & $4$ &
    $2$ & $2$
    \\ \hline
    $\dim^{\Z_5\times\Z_5}_d$
    &
    $1$ & $0$ &
    $1$ & $2$ &
    $0$ & $4$ &
    $2$ & $0$ &
    $0$ & $4$
  \end{tabular}
  \caption{Number $n_d$ of distinct irreducible representations of 
    $\AutBar(\Qt_{\psi\not=0})$ in complex dimension $d$. We also list
    the dimension $\dim_d^{\Z_5\times\Z_5}$ of the
    $\Z_5\times\Z_5$-invariant 
    subspace for each representation.
    Note that it turns out to only
    depend on the dimension $d$ of the representation.
  }
  \label{tab:AutBarQpsiquot}
\end{table}
As one turns on the $\psi$-deformation, the eigenvalues must split
according to the group-theoretical branching rules. We list these in
\autoref{tab:psiBranch}. 
\begin{table}
  \centering
  \renewcommand{\arraystretch}{1.5}
  \begin{tabular}[t]{rcl}
    $\AutBar(\QtF)$ & $\supset$ & $\AutBar(\Qt_{\psi\not=0})$ 
    \\ \hline
    $\Rep{1}$   & $\longrightarrow$ & $\Rep{1}$ \\
    $\Rep{2}$   & $\longrightarrow$ & $\Rep{1} \oplus \Rep{1}$ \\
    $\Rep{4}$   & $\longrightarrow$ & $\Rep{4}$ \\
    $\Rep{5}$   & $\longrightarrow$ & $\Rep{5}$ \\
    $\Rep{6}$   & $\longrightarrow$ & $\Rep{6}$ \\
    $\Rep{8}$   & $\longrightarrow$ & $\Rep{4} \oplus \Rep{4}$ \\
    $\Rep{10}$   & $\longrightarrow$ & $\Rep{5} \oplus \Rep{5}$ \\
    $\Rep{12}$   & $\longrightarrow$ & $\Rep{6} \oplus \Rep{6}$ \\
  \end{tabular}
  $\quad$
  \renewcommand{\arraystretch}{1.5}
  \begin{tabular}[t]{rcl}
    $\AutBar\big(\QtF\big)$ & 
    $\supset$ & 
    $\AutBar\big(\Qt_{\psi\not=0}\big)$ 
    \\ \hline
    $\Rep{20}$   & $\longrightarrow$ & $\Rep{20}$ \\
    $\Rep{30}$   & $\longrightarrow$ & $\Rep{30}$ \\
    $\Rep{40}_1$ & $\longrightarrow$ & $\Rep{40}$ \\
    $\Rep{40}_2$ & $\longrightarrow$ & $\Rep{20}\oplus \Rep{20}$ \\
    $\Rep{60}_1$ & $\longrightarrow$ & $\Rep{60}$ \\
    $\Rep{60}_2$ & $\longrightarrow$ & $\Rep{30}\oplus \Rep{30}$ \\
    $\Rep{80}$   & $\longrightarrow$ & $\Rep{40} \oplus \Rep{40}$ \\
    $\Rep{120}$  & $\longrightarrow$ & 
    $\Rep{48} \oplus \Rep{48} \oplus \Rep{24}$ \\
  \end{tabular}
  \caption{Branching rules for the decomposition of the irreducible
    representations of $\AutBar(\QtF)$ into the
    irreducible representations of its subgroup $\AutBar(\Qt_{\psi\not=0})$.
    Note that there are always numerous distinct representations in
    each dimension, see \autoref{tab:AutBarQFquot} 
    and~\ref{tab:AutBarQpsiquot}. In particular, in dimension $60$ there
    are $10$ irreps of $\AutBar(\QtF)$, which we denote by $\Rep{60}_{1}$,
    that remain irreducible under $\AutBar(\Qt_{\psi\not=0})$ and $8$ irreps, denoted by 
    $\Rep{60}_{2}$, that branch
    into two distinct $30$-dimensional irreducible representations.
  }
  \label{tab:psiBranch}
\end{table}

Finally, we are really interested in the
eigenvalues on the quotient $Q_\psi$, which means that one must restrict
to the $\Z_5\times\Z_5$-invariants of each representation. For the
Fermat quintic, we listed the number and the dimension, $\dim_d^{\Z_5\times\Z_5}$, of
these invariants in \autoref{tab:AutBarQFquot}. We list the
analogous information for the $\Z_5\times\Z_5$-invariants within the irreducible representations of
$\AutBar(\Qt_{\psi\not=0})$ in \autoref{tab:AutBarQpsiquot}.
This allows us to compute the splitting of the eigenvalues on the
quotient $Q_\psi$. However, just knowing the multiplicities turns out
to be not quite enough since same-dimensional but different irreducible
representations can branch in different ways. In particular, the first
massive level on $Q_{\psi=0}$ comes from a $60$-dimensional
representation of $\Qt_{\psi=0}$, which can branch in two ways
according to \autoref{tab:psiBranch}. However, since we have seen in
\autoref{fig:SpecFamily1} that the eigenvalues do branch, this
$60$-dimensional representation must be of the type
$\Rep{60}_2$. 

To summarize, these group theoretical considerations are completely
compatible with the observed branching of the eigenvalues under the
complex structure deformation by $\psi$. The low-lying eigenvalues of
the scalar Laplacian on $Q_\psi$ split as
\begin{equation}
  \vcenter{\xymatrix@R=0mm@H=1ex@M=0mm{
      &
      \parbox{23mm}{\centering $\Z_5\times\Z_5$\\invariants} 
      \ar@{}[r]|-{\displaystyle \subset} &
      \parbox{23mm}{\centering $\AutBar\big(\Qt_{\psi=0}\big)$\\irreps} &
      \parbox{23mm}{\centering $\AutBar\big(\Qt_{\psi\not=0}\big)$\\irreps}
      \ar@{}[r]|-{\displaystyle \supset} & 
      \parbox{23mm}{\centering $\Z_5\times\Z_5$\\invariants}      
      \\ \\
      \ar@{=}[rrrrr] & & & & &
      \\ \\ \\
      &
      \mu_0\big( Q_{\psi=0} \big) = 1
     \ar@{}[r]|-{\displaystyle \subset} &
      \Rep{1}
      \ar[r] &
      \Rep{1}
      \ar@{}[r]|-{\displaystyle \supset} &
      \mu_0\big( Q_{\psi\not=0} \big) = 1    
      \\ \\
     &
     & & \Rep{30}       
     \ar@{}[r]|-{\displaystyle \supset} &
     \mu_2\big( Q_{\psi\not=0} \big) = 2 
     \\
     &
      \mu_1\big( Q_{\psi=0} \big) = 4
     \ar@{}[r]|-{\displaystyle \subset} & 
     \Rep{60}_2
     \ar[ru] \ar[rd] &
     \oplus
      \\
      &
      & & \Rep{30}
      \ar@{}[r]|-{\displaystyle \supset} &
      \mu_1\big( Q_{\psi\not=0} \big) = 2      
     \\ \\ \\
      &
      \mu_2\big( Q_{\psi=0} \big) = 2
      \ar@{}[r]|-{\displaystyle \subset} &
      \Rep{30}
      \ar[r] &
      \Rep{30}
     \ar@{}[r]|-{\displaystyle \supset} &
      \mu_3\big( Q_{\psi\not=0} \big) = 2
     .
    }}
 \end{equation}

\subsection{Another Family}
\label{sec:family2}

Finally, let us consider another family of complex structure
moduli. First, we deform the Fermat quintic to a generic
$\Z_5\times\Z_5$ invariant polynomial; that is, switch on all
coefficients in eq.~\eqref{eq:QuinticZ5Z5}. Then restrict to the real
one-parameter family of covering spaces defined by
\begin{equation}
  \begin{split}
    \Qt_\varphi =&\; 
    \sum z_i^5 
    + \varphi \prod z_i^5 
    + i \varphi \big( z_0^3 z_1 z_4 + \text{cyc} \big)
    \\ &\;
    + (1-i) \varphi \big( z_0^2 z_1 z_2^2 + \text{cyc} \big)
    - (1-2i) \varphi \big( z_0^2 z_1^2 z_3 + \text{cyc} \big)
    \\ &\;
    - (2-i)  \varphi \big( z_0^3 z_2 z_3 + \text{cyc} \big)
    \\
%
  \end{split}
\end{equation}  
and form the quotient spaces
\begin{equation}
    Q_\varphi =\; \Qt_\varphi\Big/ \big(\Z_5\times\Z_5\big)
    .
\label{pink1}
\end{equation}
For generic values of $\varphi$, this breaks all symmetries of the
Fermat quintic except for the free $\Z_5\times\Z_5$ that we are
dividing out. Consequently, we expect no degeneracies in the spectrum
of the Laplace-Beltrami operator. 
\begin{figure}[htbp]
  \centering
  \include{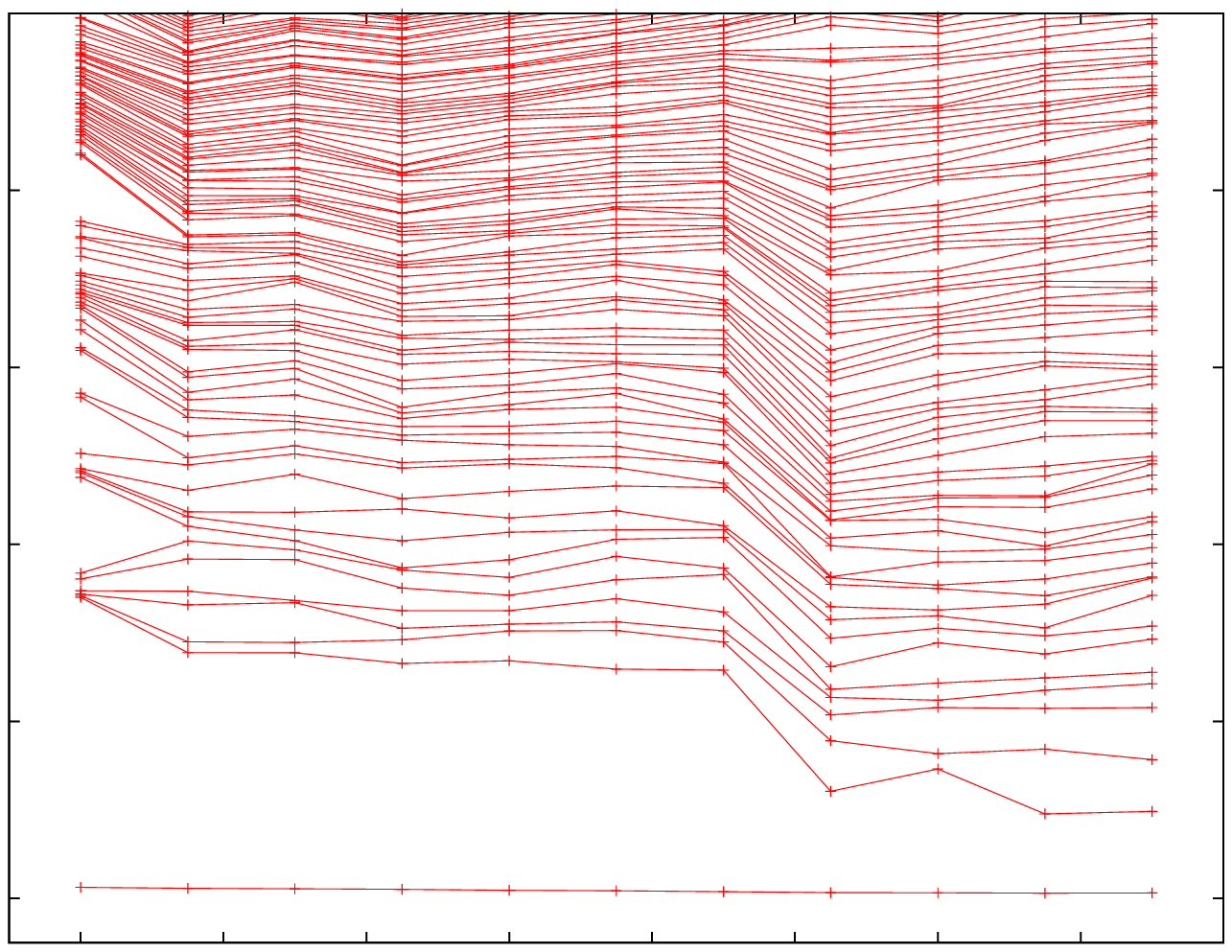}
  \caption{Spectrum of the scalar Laplace operator on the real
    $1$-parameter family $Q_\varphi$ of quintic quotients. The metric
    is computed at degree $\Kh=10$, $\Nh=\comma{406250}$ and the
    Laplace operator evaluated at $\Kphi=10$ and
    $\Nphi=\comma{50000}$ points.}
  \label{fig:SpecFamily2}
\end{figure}
In \autoref{fig:SpecFamily2}, we plot the spectrum of 
$\Delta$ and, indeed, observe that the degeneracies
of the eigenvalues on the Fermat quintic $Q_{\varphi=0}$ are broken as
$\varphi$ is turned on.


\section{A Heterotic Standard Model Manifold}
\label{sec:Z3Z3}

In this last section, we will compute the spectrum of the
Laplace-Beltrami operator on the torus-fibered Calabi-Yau threefold
$X$ with $\pi_1(X)=\ZZZ$ that was used in~\cite{Ovrut:2006yr} to
construct a heterotic standard model. The threefold $X$ is most easily
described in terms of its universal cover $\Xt$, which is the complete
intersection
\begin{equation}
  \Xt = 
  \Big\{ \Pt(x,t,y) = 0 = \Rt(x,t,y) \Big\}
  \subset 
  \CP^2_{[x_0:x_1:x_2]} \times \CP^1_{[t_0:t_1]} \times
  \CP^2_{[y_0:y_1:y_2]}
\end{equation}
defined by the degree-$(3,1,0)$ and $(0,1,3)$ polynomials
\begin{equation}
  \begin{split}
    \label{eq:manifold}
    \Pt(x,t,y) =&~ 
    t_0 
    \Big(x_0^3 + x_1^3 + x_2^3 + \lambda_1 x_0 x_1 x_2 \Big)
    + t_1 
    \lambda_3  \Big(  x_0^2 x_2 + x_1^2 x_0 + x_2^2 x_1  \Big) 
    \\
    \Rt(x,t,y) =&~ 
    t_1 
    \Big(
    y_0^3 + y_1^3 + y_2^3 + \lambda_2 y_0 y_1 y_2
    \Big)
    + t_0 
    \Big(
    y_0^2 y_1 + y_1^2 y_2 + y_2^2 y_0
    \Big)
    .
  \end{split}
\end{equation}
Note that $\lambda_1$, $\lambda_2$, $\lambda_3\in \C$ end up
parametrizing the complex structure of $X$. For generic $\lambda_i$,
the two maps
\begin{subequations}
  \begin{equation}
    \label{eq:gamma1}
    \gamma_1:
    \left\{
      \begin{array}{c@{~\mapsto~}cl}
        [x_0:x_1:x_2] &
        [x_0:\omega x_1:\omega^2 x_2]
        \\{}
        [t_0:t_1] &
        [t_0:\omega t_1] &
        \hphantom{\text{(no action)}} 
        \\{}
        [y_0:y_1:y_2] &
        [y_0:\omega y_1:\omega^2 y_2]
      \end{array}
    \right.
  \end{equation}
  and
  \begin{equation}
    \label{eq:gamma2}
    \gamma_2:  
    \left\{
      \begin{array}{c@{~\mapsto~}cl}
        [x_0:x_1:x_2] &
        [x_1:x_2:x_0]
        \\{}
        [t_0:t_1] &
        [t_0:t_1] &
        \\{}
        [y_0:y_1:y_2] &
        [y_1:y_2:y_0]
      \end{array}
    \right.
  \end{equation}
\end{subequations}
generate a free $\ZZZ$ group action on $\Xt$. Hence, the quotient
\begin{equation}
  X = \Xt \Big/ \big(\ZZZ\big)
\end{equation}
is a smooth Calabi-Yau threefold.
In addition to the $h^{2,1}(X)=3$ complex structure moduli of $X$,
there are also $h^{1,1}(X)=3$ \Kahler{} moduli. The \Kahler{} class on
the algebraic variety $X$ is determined by a line bundle $\Lsheaf$
whose first Chern class is represented by the \Kahler{} class,
\begin{equation}
  c_1(\Lsheaf) = [\omega_X] 
  \in 
  H^{1,1}(X,\Z) = H^{1,1}(X,\C) \cap H^2(X,\Z)
  .
\end{equation}
Pulling back to the covering space with the quotient map $q$, the \Kahler{} class is
equivalently encoded by an equivariant\footnote{As was shown
  in~\cite{Braun:2004xv, Braun:2007sn}, equivariance requires $a_1+a_2
  \equiv 0\mod 3$. We will always use the equivariant action
  specified by eqns.~\eqref{eq:gamma1} and~\eqref{eq:gamma2}.} line
bundle
\begin{equation}
  q^*\big(\Lsheaf\big) 
  =
  \Osheaf_\Xt(a_1,b,a_2) 
  ,
\end{equation}
which is determined by some $a_1,b,a_2 \in \Z_{>0}$. Note that, by
definition, the sections of $\Osheaf_\Xt(a_1,b,a_2)$ are the
homogeneous polynomials in $x$, $t$, and $y$ of multidegree
$(a_1,b,a_2)$. 

We now want to compute the Calabi-Yau metric
on the quotient $X$ using Donaldson's algorithm. However, as discussed in detail in the previous section, we will formulate everything in terms of
$\ZZZ$-invariant data on the covering space $\Xt$. First, one has to
pick a multidegree
\begin{equation}
  \Kh = \big(  a_1, b, a_2   \big)
  \quad
  \in \big( \Z_{>0} \big)^3
  ,~
  a_1+a_2\equiv 0\mod 3
\end{equation}
determining the \Kahler{} class of the metric. Then one has to find a
basis 
\begin{multline}
  \Span 
  \big\{ s_\alpha \big| \alpha=0,\dots,N(\Kh)-1 \big\}
  =\\=
  \C[x_0,x_1,x_2,t_0,t_1,y_0,y_1,y_2]_\Kh^\ZZZ
  \Big/ \big< \Rt(x,t,y), \Pt(x,t,y) \big>
\end{multline}
for the invariant sections of $\Osheaf_\Xt(a_1,b,a_2)$ modulo the
complete intersection equations, as described in detail
in~\cite{Braun:2007sn}. This is all the data needed to apply
Donaldson's algorithm and compute the approximate Calabi-Yau
metric. Note that, since we always normalize the volume to unity, the
exact Calabi-Yau metric only depends on the ray $\Q \Kh$ but not on
the ``radial'' distance $\gcd(a_1,b,a_2)$. However, the number of
sections $N(\Kh)$ and, therefore, the number of parameters in the
matrix $h^{\alpha\beta}$, does depend on $\Kh$ explicitly. Going from
$\Kh$ to $2\Kh$, $3\Kh$, $\dots$ increases the number of parameters
and subsequently improves the accuracy of the Calabi-Yau metric
computed through Donaldson's algorithm.

\subsection{\texorpdfstring{The Spectrum of the Laplacian on $X$}{The Spectrum of the Laplacian on X}}

Having determined the metric, we now turn towards the spectrum of the
Laplace-Beltrami operator. We do this again by computing the matrix
elements of the Laplacian on the covering in an approximate basis of
$\ZZZ$-invariant functions, completely analogous to
\autoref{sec:Z5Z5}. To specify the truncated space of invariant
functions on $\Xt$, fix a multidegree $\Kphi$ proportional to $\Kh$;
that is,
\begin{equation}
  \Kphi = (\Kphi_1,\Kphi_2,\Kphi_3)
  ~\in 
  \Q\Kh \cap \big(\Z_{\geq0}\big)^3
  .
\end{equation}
Then pick a basis $\big\{ s_\alpha \big| \alpha=0,\dots,N(\Kphi)-1
\big\}$ of degree-$\Kphi$ homogeneous, $\ZZZ$-invariant
polynomials. These define a finite-dimensional space of invariant
functions on $\Xt$ as
\begin{equation}
  \Fsheaf_\Kphi^\ZZZ
  =
  \left\{
    \frac{
      s_\alpha\sbar_{\bar\beta}
    }{
      \big( \sum |x_i|^2 \big)^{\Kphi_1} 
      \big( \sum |t_i|^2 \big)^{\Kphi_2} 
      \big( \sum |y_i|^2 \big)^{\Kphi_3}
    }
    ~\middle|~
    \alpha,\bar\beta =0,\dots, N(\Kphi)-1
  \right\}
  .
\end{equation}
By computing the matrix elements of the Laplacian and solving the
(generalized) matrix eigenvalue problem, we obtain the eigenvalues
$\lambda_n^\ZZZ$ of the Laplacian on the covering space $\Xt$ acting
on $\ZZZ$-invariant functions. These are identical to the eigenvalues
of the Laplacian on $X$, but with volume 
\begin{equation}
  \Vol(X) = \frac{1}{|\ZZZ|} \Vol(\Xt)
  .
\end{equation}
In the computation on $\Xt$ we normalized the volume to unity. Hence,
after rescaling the volume of $X$ back to one, the eigenvalues of the
scalar Laplacian on $X$ are
\begin{equation}
  \lambda_n = \frac{\lambda_n^\ZZZ}{\sqrt[3]{9}}
  .
\end{equation}

\begin{figure}[htbp]
  \centering
  \include{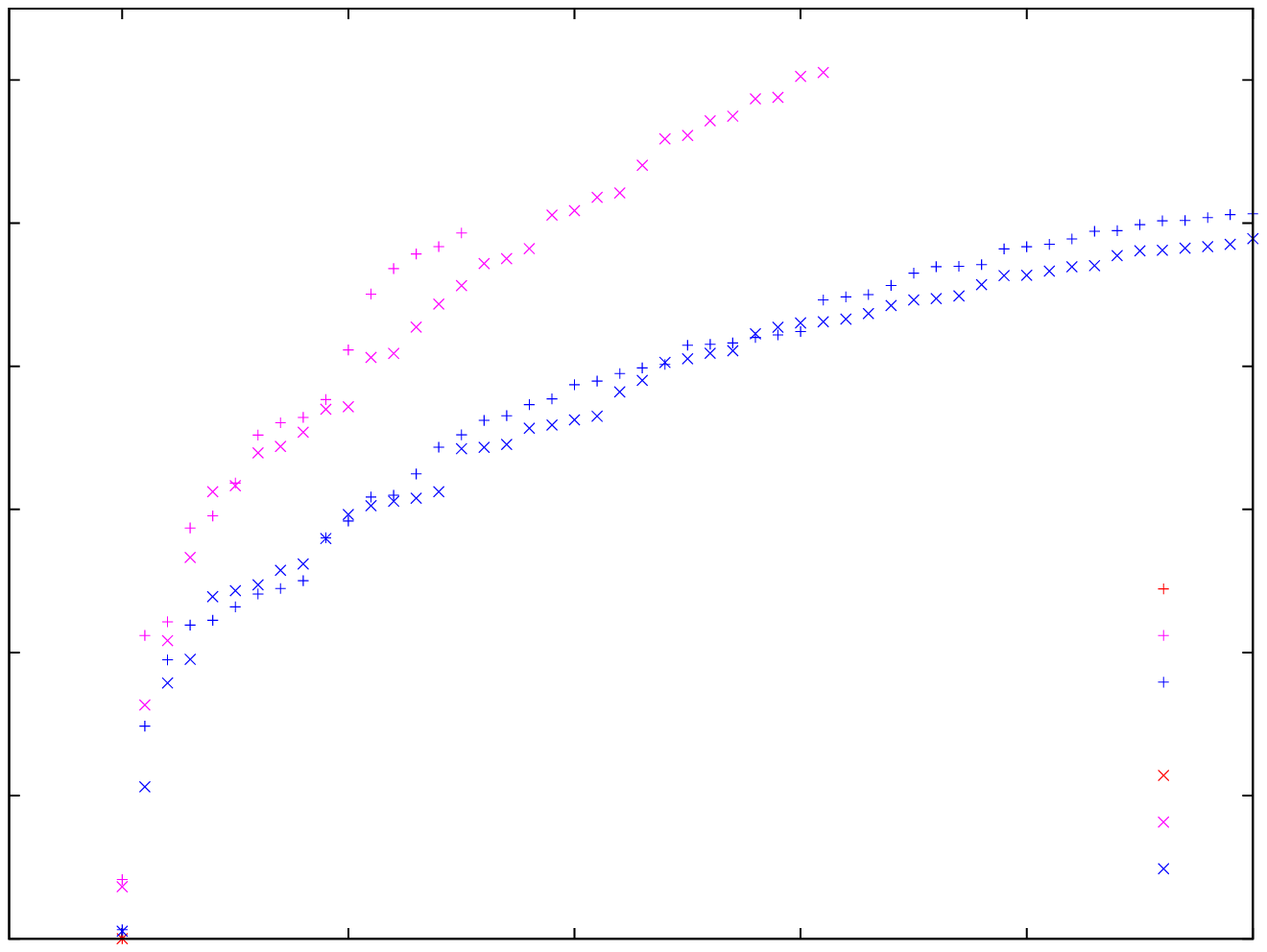}
  \caption{Eigenvalues of the scalar Laplace operator on the
    $\ZZZ$-threefold $X$ with complex structure
    $\lambda_1=0=\lambda_2$, $\lambda_3=1$ and at two distinct points
    in the \Kahler{} moduli space. The metric is computed at degree
    $\Kh=(6,3,3)$ and $\Nh=\comma{170560}$ points as well as degree
    $\Kh=(6,6,3)$ and $\Nh=\comma{290440}$ points, corresponding to
    the two different \Kahler{} moduli.  The matrix elements of the
    scalar Laplacian are always evaluated on $\Nphi=\comma{500000}$
    points. The blue pluses and crosses, corresponding in each case to
    $\Kphi$ with the largest radial distance, are the highest
    precision eigenvalues for the two metrics.}
  \label{fig:SpecZ3Z3}
\end{figure}
In \autoref{fig:SpecZ3Z3}, we compute the spectrum of the
Laplace-Beltrami operator on $X$ at two different points in the
\Kahler{} moduli space but with the same complex structure. Recall
that we always normalize the volume, corresponding to the ``radial''
distance in the \Kahler{} moduli space, to unity. The non-trivial
\Kahler{} moduli are the ``angular'' directions in the \Kahler{} cone,
and we consider the two different rays $\Q\cdot(2,1,1)$ and
$\Q\cdot(2,2,1)$. As expected, the actual eigenvalues do depend on the
\Kahler{} moduli, as is evident from
\autoref{fig:SpecZ3Z3}. 

Furthermore, note that there appear to be no multiplicities in the
spectrum. At first sight, this might be a surprise to the cognoscente,
as there \emph{is} a residual symmetry. By
construction~\cite{Braun:2004xv}, the covering space $\Xt$ comes with
a $(\Z_3)^4$ group action of which only a $\ZZZ$ subgroup acts freely
and can be divided out to obtain $X$. The remaining generators are
\begin{subequations}
  \begin{equation}
    \label{eq:gamma3}
    \gamma_3:  
    \left\{
      \begin{array}{c@{~\mapsto~}cl}
        [x_0:x_1:x_2] &
        [x_1:x_2:x_0] 
        \\{}
        [t_0:t_1] &
        [t_0:t_1] &
        \\{}
        [y_0:y_1:y_2] &
        [y_0:y_1:y_2] &
      \end{array}
    \right.
  \end{equation}
  and
  \begin{equation}
    \label{eq:gamma4}
    \gamma_4:  
    \left\{
      \begin{array}{c@{~\mapsto~}cl}
        [x_0:x_1:x_2] &
        [x_0:x_1:x_2] &
        \\{}
        [t_0:t_1] &
        [t_0:t_1]  &
        \\{}
        [y_0:y_1:y_2] &
        [y_1:y_2:y_0] 
      \end{array}
    \right.
  \end{equation}
\end{subequations}
in addition to $\gamma_1$ and $\gamma_2$, see eqns.~\eqref{eq:gamma1}
and~\eqref{eq:gamma2}. Moreover, we used the point
$\lambda_1=0=\lambda_2$, $\lambda_3=1$ where the polynomials
eq.~\eqref{eq:manifold} are also invariant under complex
conjugation. Hence, the symmetry group on the covering space is 
\begin{equation}
  \AutBar(\Xt)
  =
  \Z_2 \ltimes 
  \big( \Z_3 \big)^4
  = 
  D_6 \times \big(\Z_3\big)^3
  .
\end{equation}
To understand the latter identity, note the $Z_2$ action in the
semidirect product: 
\begin{itemize}
\item Complex conjugation commutes with $\gamma_2$, $\gamma_3$, and
  $\gamma_4$.
\item Complex conjugation does not commute with $\gamma_1$, but
  satisfies
  \begin{equation}
    \gamma_1\Big( 
    [\bar{x}_0:\bar{x}_1:\bar{x}_2],
    [\bar{t}_0:\bar{t}_1],
    [\bar{y}_0,\bar{y}_1,\bar{y}_2] 
    \Big) 
    =
    \overline{
      \gamma_1^2\Big( [x_0:x_1:x_2],[t_0:t_1],[y_0,y_1,y_2] \Big) 
    }
    .
  \end{equation}
  Hence, $\gamma_1$ together with complex conjugation generate $D_6$,
  the dihedral group with $6$ elements.
\end{itemize}
\begin{table}
  \centering
  \renewcommand{\arraystretch}{1.5}
  \begin{tabular}{c|ccc}
    $\AutBar(\Xt)$-Rep. & 
    $\rho_1,\dots,\rho_{36}$ & 
    $\rho_{37},\dots,\rho_{54}$ & 
    $\rho_{55},\dots,\rho_{81}$      
    \\ \hline
    $\dim(\rho)$ &
    $1$ & $1$ & $2$
    \\ 
    $\dim\big(\rho^\ZZZ\big)$
    &
    $0$ & $1$ & $0$
  \end{tabular}
  \caption{Number $n_d$ of distinct irreducible representations of 
    $\AutBar(\Xt)$ in complex dimension $d$. We also list
    the dimension $\dim_d^\ZZZ$ of the
    $\Z_3\times\Z_3$-invariant 
    subspace for each representation.
  }
  \label{tab:AutBarXtquot}
\end{table}
The group $\AutBar(\Xt)$ is of order $162=6\times 3^3$ and has one-
and two-dimensional representations due to the $D_6$ factor. As
discussed previously, the surviving eigenfunctions on the quotient $X$
are the $\ZZZ$-invariant eigenfunctions on the covering space $\Xt$.
Hence, we have to determine the subspace invariant under the freely
acting $\ZZZ$ inside of $\AutBar(\Xt)$. We list all this data in
\autoref{tab:AutBarXtquot}. We find that all the multiplicities on
$\Xt$ are, indeed, one.



\section{The Sound of Space-Time}
\label{sec:Laplace}

\subsection{Kaluza-Klein Modes of the Graviton}
\label{sec:KK}

Consider a $10$-dimensional spacetime of the form $\R^{3,1}\times Y$,
where $Y$ is some real, compact $6$-dimensional Calabi-Yau
manifold. Since $Y$ is compact, there is a scale associated with
it. Let us agree on a unit of length $L$ such that $\Vol(Y)=1 \cdot
L^6$. The gravitational interactions in this world are complicated,
but have two easy limiting cases. First, if the separation $r$ of two
probe masses $M_1$ and $M_2$ is large, then the gravitational
potential between them is given by Newton's law
\begin{equation}
  \label{eq:Newton}
  V(r\gg L) = - G_4 \frac{M_1 M_2}{r}
  .
\end{equation}
In the other extreme, when $r$ is very small, the potential becomes
the Green-Schwarz-Witten law
\begin{equation}
  V(r\ll L) = - G_{10} \frac{M_1 M_2}{r^7} 
  .
  \label{eq:10Dgravity}
\end{equation}
By dimensional analysis
\begin{equation}
  G_4 \sim 
  \frac{G_{10}}{L^6} \; ,
\end{equation}
with a constant of proportionality independent of $Y$ to be determined
below. In-between these two extremal limits for the separation $r$,
the gravitational potential is a complicated interpolation between
eq.~\eqref{eq:Newton} and eq.~\eqref{eq:10Dgravity}.

There are two alternative ways of describing fields on $\R^{3,1}\times
Y$. One can either directly use $10$-dimensional field theory, or work
with an infinite tower of massive Kaluza-Klein fields depending on
$\R^{3,1}$ only. Both methods are equivalent, but for the purposes of
this paper we only consider the Kaluza-Klein
compactification~\cite{Kaluza:1921tu, Klein:1926tv,
  ArkaniHamed:1998nn}. In this approach, the single $10$-dimensional
massless graviton $g^{(10D)}_{AB}$, $A,B=0,\dots,9$ is decomposed into
$4$-dimensional gravitons, vectors, and scalars. For simplicity, let
us only consider $4$-dimensional gravity, that is, $4$-d fields with
symmetrized indices $a,b=0,\dots,3$. Then
\begin{equation}
  g^{(10D)}_{ab}\big(x_0,\dots,x_3,y_1,\dots,y_6) = 
  \sum_{n=0}^\infty
  \phi_n(y_1,\dots,y_6) 
  \cdot
  g^{(4D),n}_{ab}(x_0,\dots,x_3)
  ,
\end{equation}
where the $(y_1,\dots,y_6)\in Y$-dependence of the $10$-dimensional
metric is now encoded in a basis of functions $\phi_n\in
\C^\infty(Y,\R)$. The most useful such basis consists of the solutions
to the equations of motion on $Y$, that is, the eigenfunctions of the
scalar Laplace operator
\begin{equation}
  \label{eq:DeltaY}
  \Delta_Y \phi_n(y_1,\dots,y_6) = \lambda_n \phi_n(y_1,\dots,y_6)
  ,
  \qquad
  \lambda_n \leq \lambda_{n+1}
  .
\end{equation}
The corresponding $4$-dimensional Lagrangian contains the infinite
tower of fields $g^{(4D),n}_{ab}$ of mass
\begin{equation}
  m_n =
  \sqrt{\lambda_n}
  ,\qquad
  n=0,\dots,\infty
  .
\end{equation}
As discussed previously, there is a unique zero mode $\lambda_0=0$
leading to a single massless graviton in $4$ dimensions. The
gravitational potential is then the sum of the potential due to the
massless graviton plus the Yukawa-interaction of the massive modes,
\begin{equation}
  \label{eq:Vr_Yuk}
  V(r) 
  = 
  - G_4 \frac{M_1 M_2}{r}
  \sum_{n=0}^\infty e^{-m_n r}
  = 
  - G_4 \frac{M_1 M_2}{r}
  \left( 
    1 + \sum_{n=1}^\infty e^{-m_n r}
  \right)
  .
\end{equation}
At distance scales $r\gg \tfrac{1}{m_1}$, only the massless graviton
propagates. This expected behaviour is clearly visible in the $r\gg
\tfrac{1}{m_1}$ limit of eq.~\eqref{eq:Vr_Yuk}, and one immediately
recovers eq.~\eqref{eq:Newton}. At distance scales $r\ll
\tfrac{1}{m_1}$, on the other hand, the massless graviton as well as
the infinite tower of massive spin-$2$ fields propagate. The
corresponding asymptotic behaviour of the gravitational potential is
less obvious. However, note that the asymptotic growth
\begin{equation}
  \lim_{n\to \infty} 
  \frac{\lambda_n^{3}}{n}
  =
  384 \pi^3 
  L^{-6}
  \quad\Leftrightarrow\quad
  m_n 
  \stackrel{n\to\infty}{\xrightarrow{\hspace{12mm}}} 
  2 \sqrt[6]{6} \sqrt{\pi} 
  \;
  n^{1/6}
  L^{-1} 
\end{equation}
of the Kaluza-Klein masses is known from Weyl's formula, see
\autoref{sec:Weyl}. Hence, the $r\ll \tfrac{1}{m_1}$ limit of
eq.~\eqref{eq:Vr_Yuk} is
\begin{equation}
  \begin{split}
    V(r) 
    =&\;
    - G_4 \frac{M_1 M_2}{r}
    \sum_{n=0}^\infty e^{-m_n r}
    \\
    \longrightarrow \; \sim&\;
    - G_4 \frac{M_1 M_2}{r}
    \int_{n=0}^\infty e^{- 2 \sqrt[6]{6}  \sqrt{\pi} n^{1/6} r/L} \diff n
    =
    - 
    \underbrace{
      \frac{15 G_4 L^6}{8 \pi^3} 
    }_{= G_{10}}
    \frac{M_1 M_2}{r^7}
    .
  \end{split}
\end{equation}
Again, this matches the expected behaviour eq.~\eqref{eq:10Dgravity}.

\begin{figure}[htbp]
  \centering
  \include{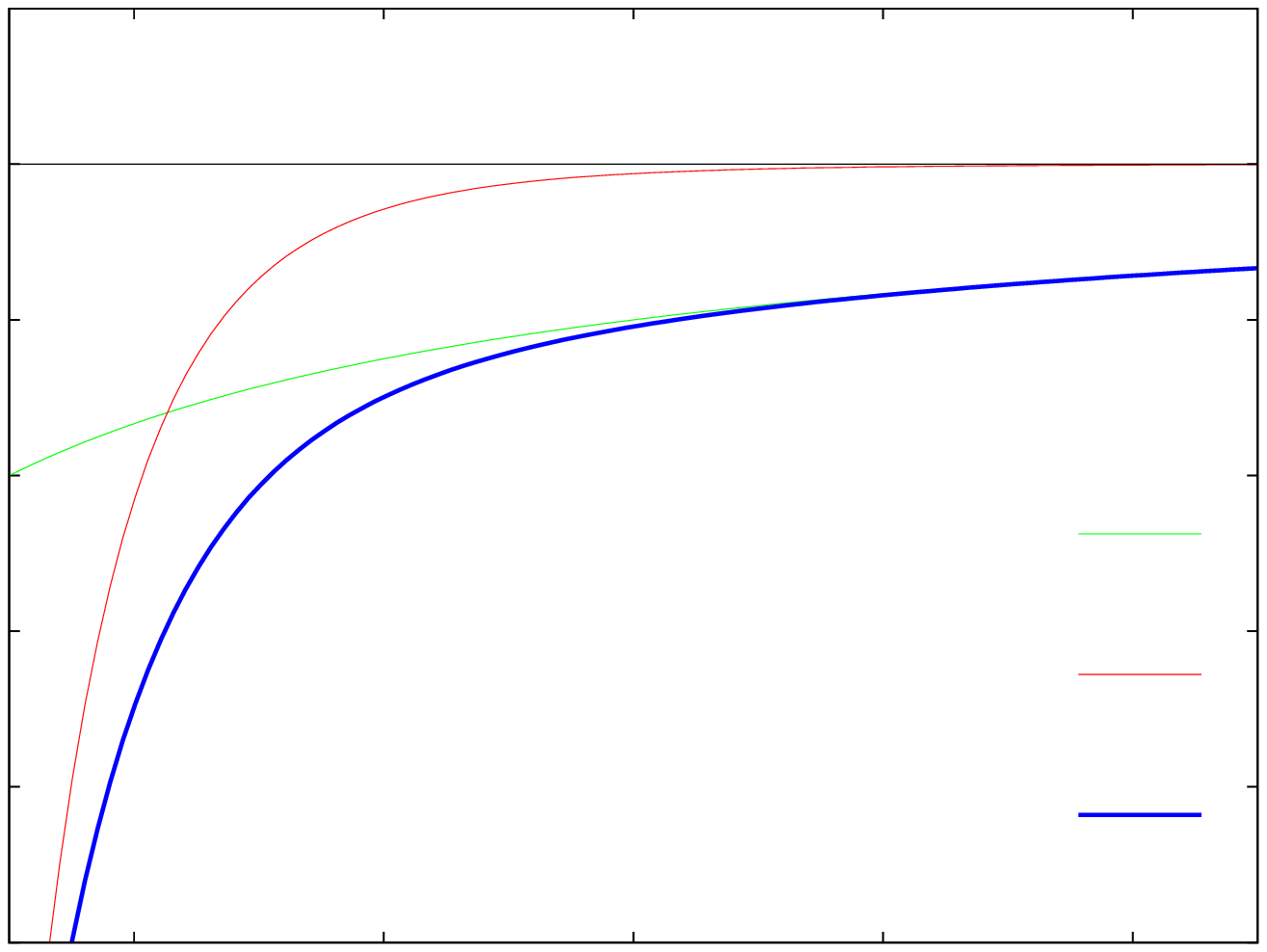}
  \caption{The gravitational potential $V(r)$ computed from
    eq.~\eqref{eq:Vr_Yuk} on $\R^{3,1}\times \QtF$, where $\QtF$ is
    the Fermat quintic with unit volume, $\Vol(\QtF)=1\cdot L^6$. The
    Kaluza-Klein masses $m_{n}=\sqrt{\lambda_{n}}$ are computed using
    the numerical results for $\lambda_{n}$ given in
    \autoref{sec:Fermat}.}
  \label{fig:Vr_QtF}
\end{figure}
The purpose of this section is to fill the gap between the extremal
limits and determine the gravitational potential at distances $r\simeq
L$. This explicitly depends on the details of the internal Calabi-Yau
threefold $Y$, and there is no way around solving
eq.~\eqref{eq:DeltaY}. The eigenvalues $\lambda_n$ and corresponding
eigenfunctions $\phi_n$ depend on the Calabi-Yau metric and can only
be computed numerically. We have presented a detailed algorithm for
calculating the spectrum of $\Delta$ in this paper, and given the
results for a number of different Calabi-Yau threefolds. As an
example, let us compute the gravitational potential $V(r)$ derived
from the numerical eigenvalues of the scalar Laplace operator on the
Fermat quintic discussed in \autoref{sec:Fermat}. The result is
plotted in \autoref{fig:Vr_QtF}.

\subsection{Spectral Gap}
\label{sec:gap}

As is evident from \autoref{fig:Vr_QtF}, deviations from the pure
$\frac{1}{r}$ (green line) and $\frac{1}{r^{7}}$ (red line) potentials
occur for $r$ in the region where these gravitational potentials have
a similar magnitude. In fact, these curves intersect at
\begin{equation}
  G_4 
  \frac{M_1 M_2}{r_{0}}
  =
  \frac{15 G_4 L^6}{8 \pi^3} 
  \frac{M_1 M_2}{r_{0}^7}
  \quad\Leftrightarrow\quad
  r_{0} = 
  \sqrt[6]{\frac{15}{8 \pi^3}} L
  \approx  0.627 L
  .
\end{equation}
Note that this point of intersection is independent of the Calabi-Yau
manifold and its geometry.  As will become clear below, for Calabi-Yau
threefolds which are relatively ``round'', such as the Fermat quintic,
$r_{0}$ is a good estimate for the point of substantial deviation from
the $\frac{1}{r}$ potential.  However, for geometries that are
stretched or develop a throat in at least one direction, this deviation
point is best determined by another scale, in principle independent of
the volume of the internal space. This other scale is the mass $m_1$
of the lightest Kaluza-Klein mode\footnote{The leading order
  correction to the gravitational potential is
  often~\cite{Leblond:2001ex, Kehagias:1999my} parametrized by the
  lowest Kaluza-Klein mass $m_1$ and its multiplicity $\mu_1$ as
  \begin{equation}
    V(r) 
    \approx
    - G_4 \frac{M_1 M_2}{r}
    \left( 1 + \mu_1 e^{-m_1 r} \right)
    .
  \end{equation}
  While this works well for symmetric spaces like spheres and tori
  with their large multiplicities and widely-separated eigenvalues,
  there are two issues when dealing with more general manifolds:
  \begin{itemize}
  \item The multiplicity is caused by symmetries, and tiny
    non-symmetric deformations can (and will) make the eigenvalues
    non-degenerate (see \autoref{sec:Moduli}).
  \item The separation between the zero mode and the first massive
    mode is, in general, much larger than the separation between the
    first and second mode. For example, on the non-symmetric ``random
    quintic'' Calabi-Yau threefold in \autoref{sec:randomquintic},
    \begin{equation}
      m_0 = 0
      ,\quad
      m_1 \approx 5.95
      ,\quad
      m_2 \approx 6.00
      .
    \end{equation}
  \end{itemize}
}, see eq.~\eqref{eq:Vr_Yuk}. For such manifolds, the spectral
gap\footnote{The first massive eigenvalue of the scalar Laplacian,
  $\lambda_1$, is also called the spectral gap since it is the gap
  between the unique zero mode $\lambda_0=0$ and the first massive
  mode.} $\lambda_1$ and, hence, the mass $m_1$ becomes
smaller. Eventually, the manifold may be sufficiently elongated that
$\frac{1}{m_1} \gg r_0$. In this case $\frac{1}{m_1}$ becomes the best
estimate of the point of deviation from the $\frac{1}{r}$ potential.

Of course, both the volume and $\lambda_1=m_1^2$ are determined by the
geometry of the internal Calabi-Yau manifold. However, what geometric
property really determines the spectral gap $\lambda_1$? In fact,
this is determined by the ``diameter'' of the manifold. Recall that
the diameter $D$ is defined to be the largest separation of any two
points, as measured by the shortest geodesic between them. Then, on an
arbitrary real $d$-dimensional manifold with non-negative scalar
curvature\footnote{In particular, a Calabi-Yau $\tfrac{d}{2}$-fold.},
the spectral gap is essentially determined by the diameter
via~\cite{MR794292, MR0378001, MR2002701}
\begin{equation}
  \label{eq:spectralgap}
  \frac{\pi^2}{D^2}  \leq  \lambda_1  \leq \frac{2d(d+4)}{D^2}
  \quad\Leftrightarrow\quad
  \frac{\pi}{D}  \leq  m_1  \leq \frac{\sqrt{2d(d+4)}}{D}
  .
\end{equation}
Clearly, in a compactification where all internal directions are
essentially of equal size, the diameter is of the order of $1\cdot
L$. However, as soon as there is even one elongated internal direction
or one long throat/spike develops, the diameter can be very
large. Hence, the spectral gap becomes very small and deviations from
$\frac{1}{r}$ gravity appear for relatively large values of $r\sim
\frac{1}{m_1}$.

The definition of the diameter $D$ is very impractical if one wants to
explicitly calculate it, since this would require global knowledge
about the shortest geodesics. However, to get a rough estimate of $D$,
one can reverse the inequalities eq.~\eqref{eq:spectralgap} and then
use the numerically computed value for $\lambda_1$. For example, on
the Fermat quintic our numerical computation in
\autoref{sec:fermatsymmetry} yielded $\lambda_1\approx
41.1$. Therefore, the diameter must be in the range
\begin{equation}
  0.490 \approx
  \frac{\pi}{\sqrt{\lambda_1}}
  \leq
  D
  \leq
  \frac{\sqrt{2\cdot 6(6+4)}}{\sqrt{\lambda_1}} 
  \approx 1.71
  .
  \label{finally1}
\end{equation}
Thus, computing the value of $\lambda_1$ numerically on a Calabi-Yau
threefold for specific values of its moduli gives us direct
information about the ``shape'' of the manifold; information that
would be hard to obtain by direct calculation of the diameter $D$. For
example, it follows from eq.~\eqref{finally1} that the Fermat quintic
is relatively ``round''.


\section*{Acknowledgments}

We are grateful to Evelyn Thomson for letting us use her 10 node
dual-core Opteron cluster.  This research was supported in part at
Rutgers by the U.~S.~Department of Energy grant DE-FG02-96ER40959, and
by the Department of Physics and the Math/Physics Research Group at
the University of Pennsylvania under cooperative research agreement
DE-FG02-95ER40893 with the U.~S.~Department of Energy, and an NSF
Focused Research Grant DMS0139799 for ``The Geometry of
Superstrings''.

\appendix
\makeatletter
\def\Hy@chapterstring{section}
\makeatother

\section{Spectrum of the Laplacian on Projective Space}
\label{sec:CP3normalization}

In this Appendix, we compute the lowest eigenvalue of the Laplace
operator on $\CP^3$ using the rescaled Fubini-Study \Kahler{} potential
eq.~\eqref{eq:CP3_K}. To do this, go to the coordinate patch where
$z_0=1$ and use $z_1$, $z_2$, $z_3$ as local coordinates. We find that
\begin{equation}
  \begin{gathered}
    g^{\ibar j}
    =
    \sqrt[3]{6} \pi \big( 1+|z_1|^2+|z_2|^2+|z_3|^2\big)
    \begin{pmatrix} 
      1+|z_1|^2   & z_2 \zbar_1 & z_3 \zbar_1 \\
      z_1 \zbar_2 & 1+|z_2|^2   & z_3 \zbar_2 \\
      z_1 \zbar_3 & z_2 \zbar_3 & 1+|z_3|^2 
    \end{pmatrix}
    ,
    \\
    \det \big( g_{i\jbar} \big) =
    \frac{6}{\big(1+|z_1|^2+|z_2|^2+|z_3|^2\big)^4 \pi^3}
  \end{gathered}
\end{equation}
and, hence,
\begin{equation}
  \Delta=
  2 \frac{1}{\det(g)} 
  \Big( \partialbar_\ibar g^{\ibar j} \det(g) \partial_i +
  \partial_j g^{\ibar j} \det(g) \partialbar_\jbar  \Big)
  .
  \label{buv1}
\end{equation}
One can now compute the eigenvalue corresponding to the eigenfunction
$\phi_{1,1}$ in eq.~\eqref{eq:CP3evec_1}. We find that
\begin{equation}
  \begin{split}
    \Delta \phi_{1,1} 
    =&\;
    2 \frac{1}{\det(g)} 
    \Big( 
    \partialbar_\ibar g^{\ibar j} \det(g) \partial_i 
    +
    \partial_j g^{\ibar j} \det(g) \partialbar_\jbar 
    \Big)
    \frac{\zbar_1}{1+|z_1|^2+|z_2|^2+|z_3|^2}
    \\
    =&\; 
   \left( \frac{16\pi}{\sqrt[3]{6}}\right)
    \frac{\zbar_1}{1+|z_1|^2+|z_2|^2+|z_3|^2}
    .
  \end{split}
\end{equation}
Hence, $\phi_{1,1}$ is indeed an eigenfunction of $\Delta$ with eigenvalue
\begin{equation}
  \lambda_1 = 
  \frac{16\pi}{\sqrt[3]{6}} = 
  \frac{4\pi}{\sqrt[3]{6}} \, \cdot 1 \cdot (1+3)
  .
\end{equation}
Hence, the numerical coefficient in eq.~\eqref{eq:CP3_lambda} is indeed
the correct one for our volume normalization $\Vol_{K}(\CP^3)=1$.

\section{Semidirect Products}
\label{sec:semidirectproduct}

Let $G$ and $N$ be two groups, and let 
\begin{equation}
  \psi: G \to \Aut(N)
\end{equation}
be a map from $G$ to the automorphisms of $N$. The semi-direct product
\begin{equation}
  G \;{}_\psi\!\!\ltimes N = 
  \Big\{ 
  (n,g)
  \;  \Big|  \;
  n\in N, g\in G
  \Big\}
\end{equation}
is defined to be the group consisting of pairs $(n,g)$ with the group
action
\begin{equation}
  (n_1, g_1) \cdot (n_2, g_2)
  = 
  \big( n_1 \cdot \psi(g_1)(n_2) ,~ g_1 \cdot g_2 \big)
  .
\end{equation}
Usually, one just writes $G\ltimes N$ with the map $\psi$ implied but
not explicitly named. Note that $G$ is a subgroup and $N$ is a normal
subgroup of the semidirect product.

For example, consider the semidirect product with $G=S_5$ and
$N=(\Z_5)^4$ used in \autoref{sec:fermatsymmetry}. These two groups
are acting on five homogeneous via permutations\footnote{$S_5$ is, by
  definition, the group of permutations of five objects.} and phase
rotations
\begin{equation}
  \Big( (n_1,n_2,n_3,n_4)
  ,~
  [z_0,z_1,z_2,z_3,z_4] \Big)
  \mapsto 
  \big[
  z_0,
  z_1 e^{\frac{2\pi i n_1}{5}},
  z_2 e^{\frac{2\pi i n_2}{5}},
  z_3 e^{\frac{2\pi i n_3}{5}},
  z_4 e^{\frac{2\pi i n_4}{5}}
  \big]
  ,
\end{equation}
respectively. The two group actions do not commute, and, therefore,
the total symmetry group is not simply the product $S_5\times
(\Z_5)^4$. The ``non-commutativity'' between $S_5$ and $(\Z_5)^4$ is
encoded in a map
\begin{equation}
  \psi: S_5 \to \Aut\Big( (\Z_5)^4 \Big)
  ,~
  \sigma \mapsto 
  \Big( \vec{n} \mapsto \sigma^{-1} \circ \vec{n} \circ \sigma 
  \Big)
  .
\end{equation}
To be completely explicit, note that the permutation group $S_5$ is
generated by the cyclic permutation $c$ and a transposition $t$,
acting as
\begin{equation}
  \begin{split}
    t:~&
    \big[ z_0, z_1, z_2, z_3, z_4 \big]
    \mapsto
    \big[ z_0, z_1, z_2, z_4, z_3 \big]
    ,
    \\
    c:~&
    \big[ z_0, z_1, z_2, z_3, z_4 \big]
    \mapsto
    \big[ z_1, z_2, z_3, z_4, z_0 \big]
    .
  \end{split}
\end{equation}
The generators $\langle c,t\rangle = S_5$ act, via $\psi$, on
$(\Z_5)^4$ as
\begin{equation}
  \begin{split}
    \psi(t):\;&
    (\Z_5)^4 \to (\Z_5)^4
    ,\quad
    (n_1,n_2,n_3,n_4) \mapsto (n_1,n_2,n_4,n_3)
    \\
    \psi(c):\;&
    (\Z_5)^4 \to (\Z_5)^4
    ,\quad
    (n_1,n_2,n_3,n_4) \mapsto (-n_4,n_1-n_4,n_2-n_4,n_3-n_4)
  \end{split}
\end{equation}
It is straightforward, if tedious, to show that $\psi$ is a group
homomorphism and that the total symmetry group generated by $S_5$ and
$(\Z_5)^4$ is, in fact, the semidirect product
\begin{equation}
  S_5 \;{}_\psi\!\!\ltimes (\Z_5)^4
  .
\end{equation}
By the usual abuse of notation, we always drop the subscript $\psi$ in
the main part of this paper.

\section{Notes on Donaldson's Algorithm on Quotients}
\label{sec:DonaldsonNotes}

For explicitness, let us consider the same setup as in
\autoref{sec:Z5Z5}, that is, $\Qt\subset\CP^4$ is a $\Z_5\times\Z_5$
symmetric quintic and we want to compute the metric on the quotient
$Q=\Qt \big/ (\Z_5\times\Z_5)$. To fix notation, let us denote the two
generators for the character ring of the group by
\begin{equation}
  \begin{aligned}
    \chi_1(g_1) =&\; e^{2\pi i /5}
    ,
    &\quad
    \chi_1(g_2) =&\; 1
    ,
    \\
    \chi_2(g_1) =&\; 1
    ,
    &\quad
    \chi_2(g_2) =&\; e^{2\pi i /5}
    .
  \end{aligned}
\end{equation}
We consider homogeneous polynomials in degrees $\Kh\in 5\Z$, so there
is a linear $\Z_5\times\Z_5$ group action. In
eq.~\eqref{eq:HironakaZ5Z5} we determined the invariant
polynomials. Now, let us slightly generalize this result and determine
``covariant polynomials'' transforming as some character $\chi$ of the
group,
\begin{equation}
  p \circ g(z) = \chi(g) p(z)
  \qquad
  g\in \Z_5\times\Z_5
  .
\end{equation}
These again form a linear space of $\chi$-covariant polynomials, which
we denote as
\begin{equation}
  \C[z_0,z_1,z_2,z_3,z_4]_\Kh^\chi 
  = 
  \Big\{
  p(z) 
  \Big|
  p \circ g(z) = \chi(g) p(z)
  \Big\}
  .
\end{equation}
Note that the covariant polynomials do not form a ring, but rather a
module over the invariant ring. Nevertheless, by a slight
generalization of the Hironaka decomposition, we can express the
covariants as a direct sum
\begin{equation}
  \label{eq:covHironakaZ5Z5}
  \C[z_0,z_1,z_2,z_3,z_4]^\chi_\Kh=
  \bigoplus_{i=1}^{100}
  \eta_i^\chi
  \, 
  \C[
  \theta_1,\theta_2,\theta_3,\theta_4,\theta_5 
  ]_{\Kh-\deg( \eta_i^\chi )}
  ,
\end{equation}
where the $\theta_1$, $\dots$, $\theta_5 \in
\C[z_0,z_1,z_2,z_3,z_4]^{\Z_5\times\Z_5}$ can be taken to be the
primary invariants of the original Hironaka decomposition
eq.~\eqref{eq:HironakaZ5Z5} and the ``secondary covariants''
$\eta_1^\chi$, $\dots$, $\eta_{100}^\chi$ are certain $\chi$-covariant
polynomials that need to be computed~\cite{GLS03}. For example, we
find
\begin{equation}
  \begin{split}
    \eta_1^{\chi_1} =&\; 
    z_0^4 z_1 +
    z_1^4 z_2 +
    z_2^4 z_3 +
    z_3^4 z_4 +
    z_4^4 z_0
    ,
    \\
    \eta_2^{\chi_1} =&\;
    z_0 z_1^3 z_3 + 
    z_1 z_2^3 z_4 + 
    z_2 z_3^3 z_0 + 
    z_3 z_4^3 z_1 + 
    z_4 z_0^3 z_2 
    ,~\dots
  \end{split}
\end{equation}
and
\begin{equation}
  \begin{split}
    \eta_1^{\chi_2} =&\; 
    z_0^5 + 
    e^{\frac{2 \pi i}{5}} z_1^5 + 
    e^{2\frac{2 \pi i}{5}} z_2^5 + 
    e^{3\frac{2 \pi i}{5}} z_3^5 + 
    e^{4\frac{2 \pi i}{5}} z_4^5
    ,
    \\
    \eta_2^{\chi_2} =&\;
    z_0 z_1^3 z_2 + 
    e^{\frac{2 \pi i}{5}} z_1 z_2^3 z_3 + 
    e^{2\frac{2 \pi i}{5}} z_2 z_3^3 z_4 + 
    e^{3\frac{2 \pi i}{5}} z_3 z_4^3 z_0 + 
    e^{4\frac{2 \pi i}{5}} z_4 z_0^3 z_1
    ,~\dots.
  \end{split}
\end{equation}
Note that we always take the defining quintic polynomial $\Qt(z)$ to
be completely\footnote{If $\Qt(z)$ were a $\chi$-covariant polynomial,
  it would still define a $\Z_5\times\Z_5$ invariant Calabi-Yau
  hypersurface. Everything in this paper would generalize
  straightforwardly, so we ignore this possibility to simplify
  notation.} invariant, see eq.~\eqref{eq:QuinticZ5Z5}. Restricting
everything to the hypersurface $\Qt(z)=0$, we get homogeneous
polynomials on the Calabi-Yau threefold. We pick bases
$\{s_\alpha^\chi\}$ for the $\chi$-covariant polynomials, that is,
\begin{equation}
  \begin{aligned}
    \chi = 1:&&\qquad
    \Span \big\{ s_\alpha^1 \big\}
    =&\;
    \C[z_0,z_1,z_2,z_3,z_4]_\Kh^{\Z_5\times \Z_5} \Big/ 
    \big< \Qt(z) \big>
    ,
    \\
    \chi \not= 1:&&\qquad
    \Span \big\{ s_\alpha^\chi \big\}
    =&\;
    \left(
      \C[z_0,z_1,z_2,z_3,z_4]_\Kh \Big/ 
      \big< \Qt(z) \big>
    \right)^\chi
    \\
    &&
    =&\;
    \C[z_0,z_1,z_2,z_3,z_4]_\Kh^\chi
    .
  \end{aligned}
\end{equation}

We now turn towards computing the metric on the quotient $Q$ or,
equivalently, computing the $\Z_5\times\Z_5$-invariant metric on the
covering space $\Qt$ by a variant of Donaldson's algorithm. For this,
we pick the ansatz
\begin{equation}
  K(z,\zbar) = 
  \frac{1}{\pi}
  \sum_{\chi = \chi_1^0 \chi_2^0}^{\chi_1^4\chi_2^4}
  \sum_{\alpha\bar\beta}
  h^{\chi \alpha \bar\beta} 
  s^\chi_\alpha \overline{s^\chi_\beta}
\end{equation}
for the Calabi-Yau metric. One can think of $h$ as a block-diagonal
matrix with blocks labelled by the characters $\chi$. The $T$-operator
is likewise block-diagonal, and therefore one obtains a balanced
metric as the fixed point of the iteration
\begin{equation}
  h^{\chi\alpha\bar\beta}_n
  \longrightarrow
  h^{\chi\alpha\bar\beta}_{n+1} = 
  T\big( h^{\chi\alpha\bar\beta}_n \big)^{-1} 
  .
\end{equation}
Note that this fixed point is the same\footnote{And different from the
  fixed point where one restricts to only the invariant sections. The
  latter is just the $\chi=1$ block.} as what one would obtain from
Donaldson's algorithm on the covering space $\Qt$ (without using any
symmetry). Only now the basis of sections is such that the impact of
the $\Z_5\times\Z_5$ symmetry is clearly visible: $h$ is
block-diagonal with blocks labelled by the characters $\chi$. 

As usual, the balanced metrics are better and better approximations to
the Calabi-Yau metric as one increases the degree $\Kh$. We find that
this method of computing the Calabi-Yau metric on the quotient $Q$ is
the most effective.

\bibliographystyle{utcaps} \renewcommand{\refname}{Bibliography}
\addcontentsline{toc}{section}{Bibliography} 

\bibliography{Volker,Metric}

\end{document}